\documentclass[twocolumn]{aastex62}

\pdfoutput=1

\usepackage{epsfig}
\usepackage{graphicx}
\usepackage{amsthm}
\usepackage{amssymb}
\usepackage{rotating}
\usepackage{multirow}
\usepackage{longtable}
\usepackage{savesym}
\savesymbol{tablenum}
\restoresymbol{SIX}{tablenum}
\usepackage{booktabs}

\received{September 30, 2018}
\revised{September 30, 2018}

\submitjournal{ApJ}

\shorttitle{Surveying G\ion{H}{2} Regions: I. W\,51\,A}
\shortauthors{Lim \& De\,Buizer}

\begin{document}

\title{Surveying the Giant \ion{H}{2} Regions of the Milky Way with \textit{SOFIA}: I. W\,51\,A}

\correspondingauthor{Wanggi Lim}
\email{wlim@usra.edu}

\author{Wanggi Lim and James M. De Buizer}
\affil{\footnotesize \textit{SOFIA}-USRA, NASA Ames Research Center, MS 232-12, Moffett Field, CA 94035, USA}

\begin{abstract}

 We discuss the first results from our mid-infrared imaging survey of Milky Way Giant \ion{H}{2} regions with our detailed analysis of W51A, which is one of the largest G\ion{H}{2} regions in our Galaxy. We used the FORCAST instrument on \textit{SOFIA} to obtain 20 and 37\,$\mu$m images of the central $10\arcmin\times20\arcmin$ area, which encompasses both of the G49.5-0.4 and G49.4-0.3 sub-regions. Based on these new data, and in conjunction with previous multi-wavelength observations, we conjecture on the physical nature of several individual sources and sub-components within W\,51\,A. We find that extinction seems to play an important role in the observed structures we see in the near- to mid-infrared, both globally and locally. We used the \textit{SOFIA} photometry combined with \textit{Spitzer}-IRAC and \textit{Herschel}-PACS photometry data to construct spectral energy distributions (SEDs) of sub-components and point sources detected in the \textit{SOFIA} images. We fit those SEDs with young stellar object models, and found 41 sources that are likely to be massive young stellar objects, many of which are identified as such in this work for the first time. Close to half of the massive young stellar objects do not have detectable radio continuum emission at cm wavelengths, implying a very young state of formation. We derived luminosity-to-mass ratio and virial parameters of the extended radio sub-regions of W51A to estimate their relative ages.

\end{abstract}


\section{Introduction} 

When a single massive star begins to form in a giant molecular cloud, it tends to be highly self-embedded and thus observable only in the mid-infrared (MIR) to sub-millimeter. At some point the central (proto-)star becomes hot enough that a substantial amount of Lyman continuum luminosity is produced. This ionizes the gas in its immediate surroundings, creating an \ion{H}{2} region that is bright in centimeter radio continuum emission. Initially this region is quite small ($\sim$0.01 pc), and thus is called a hyper-compact \ion{H}{2} (HC\ion{H}{2}) region  \citep{hoare07}. However, as the \ion{H}{2} region evolves and expands and more of the natal material becomes heated to higher temperatures, emission becomes observable at shorter and shorter infrared wavelengths. These ultra-compact \ion{H}{2} (UC\ion{H}{2}, $\sim$0.1pc) and compact \ion{H}{2} (C\ion{H}{2}, $\sim$1pc) phases can be quite bright at MIR wavelengths and sometimes even near-infrared (NIR) wavelengths  \citep{churchwell02}. This scenario holds for the formation of an individual massive star (or a tight multiple system of massive stars). However in the case of the most massive young stellar clusters in our Galaxy, there seems to be ongoing and/or sequential star formation, with the Lyman continuum emission from revealed massive stars as well as individual compact \ion{H}{2} regions combining to emit more than 10$^{50}$ LyC photons s$^{-1}$, and in the process create vast ionized regions within their host molecular clouds  \citep{1994ApJ...421..140V}. These large regions are called giant \ion{H}{2} (G\ion{H}{2}) regions, and typically have ionizing fluxes more than an order of magnitude larger than our nearest massive star-forming region, the Orion Nebula (i.e. M\,42). These objects tend to have angular sizes in the infrared of one to several arcminutes (given their typical $\sim$few kpc distances), and can be distinguished by their bright and optically thin radio continuum emission at cm wavelengths \citep{2004MNRAS.355..899C}. Also, such G\ion{H}{2} regions are a dominant source of emission contributing to the bolometric luminosity that we see from galaxies in general  \citep[e.g.][]{2008ApJ...672..214G}. Therefore, understanding the global and detailed properties of G\ion{H}{2} regions in our own Galaxy can be used as a template for interpreting what we observe in galaxies far away.

Our understanding of the formation of massive stars is not known to the same level of detail as stars like our own Sun. Discerning the similarities and differences of high-mass and low-mass star formation is essential to our fundamental understanding of star formation in general. Moreover, we know less about clustered star formation than isolated star formation. However, it is believed that the vast majority of all stars form within OB clusters \citep{1978PASP...90..506M}. G\ion{H}{2} regions are laboratories for the earliest stages of massive star formation and clustered star formation, and as such, a lot may be learned about the environments of forming OB clusters.

This is the first paper in a large-scale project with the goal of creating a 20 and 37\,$\mu$m imaging survey of all known G\ion{H}{2} regions within the Milky Way with the \textit{Stratospheric Observatory For Infrared Astronomy} (\textit{SOFIA}) and its mid-infrared instrument FORCAST. Though the \textit{Spitzer Space Telescope} and \textit{Wide-field Infrared Survey Explorer} (\textit{WISE}) satellite imaged these regions at comparable resolutions near 20\,$\mu$m, often the \textit{Spitzer} 24\,$\mu$m and \textit{WISE} 22\,$\mu$m images were severely saturated in the brightest areas. There also exist \textit{Midcourse Space Experiment} (\textit{MSX}) 21\,$\mu$m images of each of these regions, and while they are unsaturated, the resolution is $\sim$18$\arcsec$, or 7$\times$ worse than what we can achieve with \textit{SOFIA} at 20\,$\mu$m. Observing near 20\,$\mu$m is also possible from ground-based observatories, but from the ground the sky emission is much brighter than these sources and one must observe through a sky and background subtraction technique called ``chopping and nodding''. However, these regions are highly extended in emission and no ground-based observatory can chop larger than $\sim$1$\arcmin$. Furthermore, typical ground-based cameras also have small field-of-view ($<$1$\arcmin$). Both of these issues mean that images typically obtained with ground-based facilities can only target small subregions and the images they obtain are often contaminated with negative emission from the chop and nod reference beams (which can complicate flux calibration and accuracy, as well as artificially change the observed morphology and source structure). At 37\,$\mu$m, \textit{SOFIA}-FORCAST has unique wavelength coverage, allowing us to probe cooler dust (50--100K) and even more extinguished regions than is possible at 20\,$\mu$m with the best resolution ever achievable at that wavelength ($\sim$3.0$\arcsec$). 

\citet{2004MNRAS.355..899C} used a published 6\,cm all-sky survey along with data from the \textit{MSX} and \textit{IRAS} archives to identify 56 bona-fide G\ion{H}{2} regions. Observations of these targets are ongoing, and we aim to observe as many of these sources as we can with \textit{SOFIA} to understand their physical properties individually and as a population. In this paper, and several papers to follow, we will discuss individual G\ion{H}{2} regions, highlighting the properties of each region as determined from the \textit{SOFIA} data, and compare that data to other data in the literature. We plan to finish the series of G\ion{H}{2} region papers with one detailing the global properties of Milky Way G\ion{H}{2} regions as a population, with comparisons to extragalactic G\ion{H}{2} regions and starbursts.

 We start here with an in-depth look at our \textit{SOFIA} observations of the extensive W\,51\,A G\ion{H}{2} region. This source was one of the first observed for this program, and is one of the largest regions in our source list in terms of angular diameter. W\,51\,A is also one of the largest and brightest G\ion{H}{2} regions in our Galaxy, weighing in at 100 times the mass of Orion \citep[$\sim$1$\times10^5 M_\sun$ for W\,51\,A versus $\sim$1$\times10^3 M_\sun$ for M\,42;][]{2010ApJS..190...58K,2018MNRAS.473.4890S}, with an ionizing flux more than 100 times that of Orion \citep[$N^{H}_{\rm LyC} / s\gtrsim$6$\times10^{48}$ versus $\sim$1$\times10^{51}$ for M\,42 and W\,51\,A, respectively;][]{2004MNRAS.355..899C,1993A&AS..101..127F,1994ApJS...91..713M}. It is sufficiently large, complicated, and well-studied that we devote to it this entire first paper.

In the next section (\S\,\ref{sec:obs}), we will discuss the new {\it SOFIA} observations and give information on the data obtained of W\,51\,A. In \S\,\ref{sec:results1}, we will give more background on this region as we compare our new data to previous observations and discuss individual sources and regions in-depth. In \S\,\ref{sec:data}, we will discuss our data analysis, modeling, and derivation of physical parameters of sources and regions. Our conclusions are summarized in \S\,\ref{sec:conclusion}.

\begin{figure*}[htb!]
\figurenum{1}
\epsscale{1.14}
\plotone{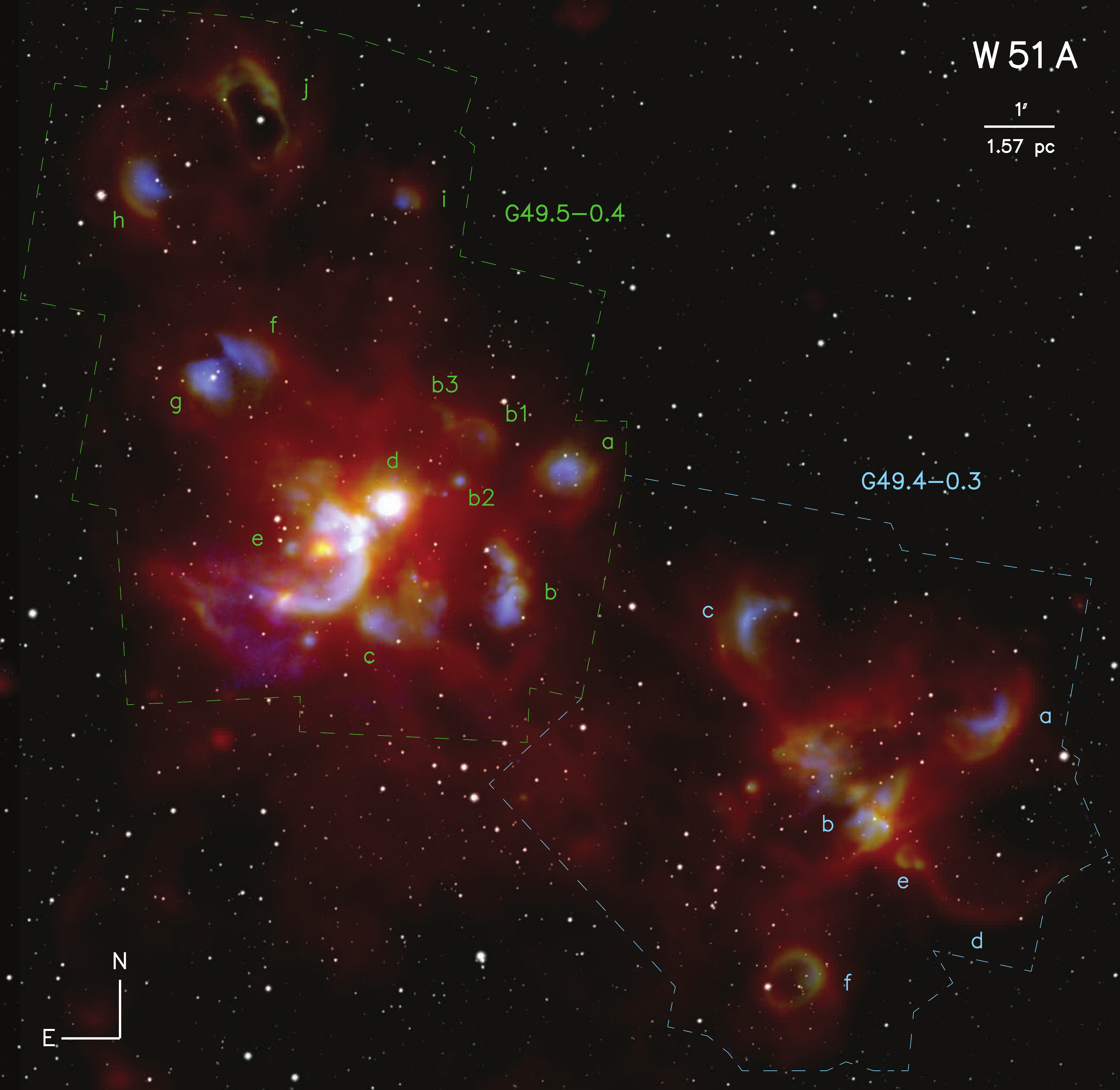}
\caption{A 3-color image of W\,51\,A. Blue is the \textit{SOFIA}-FORCAST 20\,$\mu$m image, green is the \textit{SOFIA}-FORCAST 37\,$\mu$m image, and red is the \textit{Herschel} 70\,$\mu$m image. Overlaid in white is the \textit{SDSS} z-band star field, which traces the revealed stars and field stars. The dashed contours show the boundaries of the \textit{SOFIA} image mosaic, and the area encompassed by the green dashed lines is the sub-component G49.5-0.4, and the area encompassed by the blue dashed lines is G49.4-0.3.\label{fig:f1}}
\end{figure*}

\section{Observations and Data Reduction} \label{sec:obs}

Data for this program have been collected over several \textit{SOFIA} Cycles dating back to Cycle 1 in 2013. All data were obtained using the FORCAST instrument \citep{2013PASP...125...1393H}. FORCAST is a dual-array mid-infrared camera capable of taking simultaneous images at two wavelengths. The short wavelength camera (SWC) is a 256$\times$256 pixel Si:As array optimized for 5–-25\,$\mu$m observations; the long wavelength camera (LWC) is a 256$\times$256 pixel Si:Sb array optimized for 25–-40\,$\mu$m observations. After correction for focal plane distortion, FORCAST effectively samples at 0$\farcs$768 pixel$^{-1}$, which yields a 3$\farcm$4$\times$3$\farcm$2 instantaneous field of view. Observations were obtained in the 20\,$\mu$m ($\lambda_{eff}$=19.7\,$\mu$m; $\Delta\lambda$=5.5\,$\mu$m) and 37\,$\mu$m ($\lambda_{eff}$=37.1\,$\mu$m; $\Delta\lambda$=3.3\,$\mu$m) filters simultaneously using an internal dichroic. 

All images were obtained by employing the standard chop-nod observing technique used in the thermal infrared, with chop and nod throws sufficiently large to sample clear off-source sky (typically $\sim$7$\arcmin$). We also dithered the observations  to help correct for any additional array artifacts (e.g. bad pixels) that are not removed via the chop and nod process.   As detailed in \citet{2013PASP...125...1393H}, this process does not always completely flatten the background of FORCAST data, leaving low-spatial frequency background variations that changes from exposure to exposure and which cannot be easily removed. Therefore, some significantly large areas of the images obtained can have slightly non-zero (including negative) backgrounds. Furthermore, the background around bright sources can be suppressed due to electronic crosstalk \citep[see again][]{2013PASP...125...1393H}, creating negative areas of background. 

The W\,51\,A G\ion{H}{2} region is much larger ($\sim$15$\arcmin\times$15$\arcmin$) than the FORCAST field of view, and thus had to be mapped using multiple pointings. Though the total exposure time for each pointing was planned to be the same (in order to yield a mosaicked image with relatively uniform signal-to-noise), in actuality the time varied due to changes in flight plans, losses of time in flight, or changes in observing efficiencies over the cycles. For W\,51\,A, we created a mosaic from 19 individual pointings, each composed of the coaddition of 9-10 dither images, with each final dither-coadded image having an average on-source exposure time of about 180s at both 20\,$\mu$m and 37\,$\mu$m. However, the exposure time in any given area could be different given that edges of the final images produced at each pointing after coadding the dithers have variable exposure times and each pointing had significant field overlap ($>$10\%) with adjacent pointings.  The overlapping areas can have factors of 2-4 larger exposure time than non-overlapping areas. The total fraction of overlapped area in the SOFIA maps are 24.6\,\% and 26.4\,\% for 20 and 37\,$\mu$m, respectively.

Flux calibration for each of the 19 individual pointings was created via the \textit{SOFIA} Data Cycle System (DCS) pipeline. The pipeline uses calibrators (stars and asteroids) observed over multiple flights to derive a calibration factor (Jy per raw data unit) for each image. These calibration factors take into account airmass and aircraft altitude of each observation, and once corrected for these conditions, these calibration factors show remarkably stable values across multiple flights, and thus are assumed to be reliable. The flux density calibration error of the W\,51\,A field is $\sim$3.3\% at 20\,$\mu$m and $\sim$8.0\% at 37\,$\mu$m.

Some of the images produced from the dither-coadded individual pointings had additional residual high-spatial frequency background noise due to imperfect nod subtraction. To remove this high frequency pattern noise (seen only in the 20\,$\mu$m images), the data were corrected using a custom-developed Interactive Data Language (IDL) software package built around its native Fast Fourier Transform code (\textit{fft.pro}). The noise was corrected by isolating it in Fourier space and removing it before transforming the data back into image space. We modified all raw data so that all 198 dither images were inspected and corrected by the IDL Fourier Transform code. The flux density difference before and after this correction are maintained under 2\% across all 20\,$\mu$m images. 

Another issue that had to be dealt with when mosaicking the individual pointings is the FORCAST array crosstalk mentioned above. This means that when there is a particularly bright source on the array (e.g. IRS\,1 and IRS\,2 regions), the array response can cause the images of adjacent pointings to have discontinuous backgrounds. Through trial and error we found that the best way to mosaic all of the data and minimize the effect of this was to use a combination of the \textit{SOFIA} Pipeline Software and custom mosaicking routines. We used the \textit{SOFIA} Pipeline Software to make three sub mosaics that showed smooth background over the sub-fields. We then used custom IDL routines to match the backgrounds of the three sub-fields with exposure time weighting to create the final W\,51\,A map. We tested the photometric variances among the final mosaic produced solely with the \textit{SOFIA} Pipeline, the IDL-corrected mosaic, and flux calibrated individual pointing images from the \textit{SOFIA} Pipeline prior to mosaicking. The intensities of individual sources in all three cases are in agreement to within better than 10\% which implies the background correction method does not substantially affect scientific results.  There still exist areas with slightly negative background intensities in the final 20 and 37\,$\mu$m mosaic maps, however as we will discuss in  \S\,\ref{sec:cps}, this in the end does not affect our compact source photometry since the issue is mitigated by applying proper background subtraction.

In addition to the FORCAST data, our analyses also utilize the \textit{Spitzer} IRAC 3.6, 4.5, 5.8 and 8.0\,$\mu$m data of the Galactic Legacy Infrared Mid-Plane Survey Extraordinaire survey (GLIMPSE, \citet{2009PASP..121..213C}) as well as the \textit{Herschel} PACS 70\,$\mu$m and 160\,$\mu$m and SPIRE 250, 350 and 500\,$\mu$m data of the \textit{Herschel} infrared Galactic Plane Survey (Hi-GAL, \citet{2010A&A...518L.100M}).

Because all FORCAST data were taken in the dichroic mode, one can determine precise relative astrometry of the two wavelength images that were obtained simultaneously. The relative astrometry between filters is known to better than 0.5 pixels ($\sim$0$\farcs$38). All images then had their astrometry absolutely calibrated using \textit{Spitzer} data by matching up the centroids of point sources in common between the \textit{Spitzer} and \textit{SOFIA} data. Absolute astrometry of the final \textit{SOFIA} images is assumed to be better than 1$\farcs$0.

In order to perform photometry on MIR point sources, we employed the aperture photometry program \textit{aper.pro}, which is part of the IDL DAOPHOT package available in The IDL Astronomy User's Library (http://idlastro.gsfc.nasa.gov).

\begin{figure*}[htb!]
\figurenum{2}
\epsscale{0.65}
\plotone{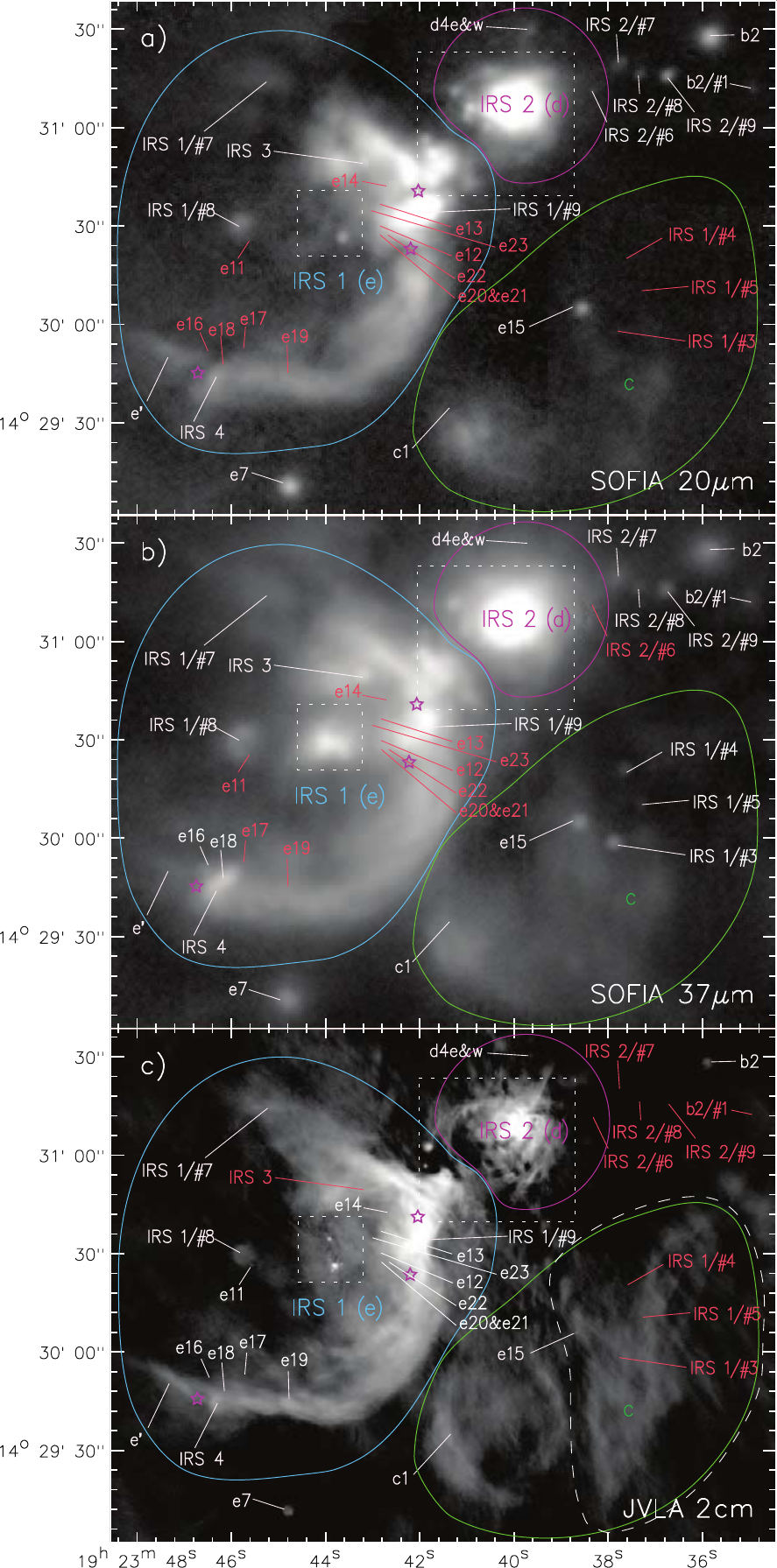}
\caption{IRS\,1, IRS\,2, IRS\,3, and c regions with all radio and infrared source positions labeled. Wavelength of each image is given in the lower right of each panel. Red labeled sources indicate a non-detection at that wavelength. The area encompassed by the blue lines identifies the region of IRS\,1 (a.k.a radio source e), the area encompassed by the purple lines identifies the region of IRS\,2 (a.k.a. radio source d), and the area encompassed by the green lines identifies the c region seen in radio and infrared. Three purple stars denote the approximate location of the mid-infrared dark lanes bisecting the e arc of emission. The smaller d0tted box identifies the area shown in Figure \ref{fig:f3}, and the larger dotted box shows the area shown in Figure \ref{fig:f5}. a) \textit{SOFIA} 20\,$\mu$m image. b) \textit{SOFIA} 37\,$\mu$m image. c) \textit{JVLA} radio continuum image at 2\,cm \citep{2016AnAp...595A..27G}. The area encompassed by dashed white lines designates the radio emitting area referred to as ``the arc-like area between regions b and c1'' by \citet{1994ApJS...91..713M} which we incorporate in this work into the extended c region.\label{fig:f2}}
\end{figure*}

\begin{figure*}[htb!]
\figurenum{3}
\epsscale{1.16}
\plotone{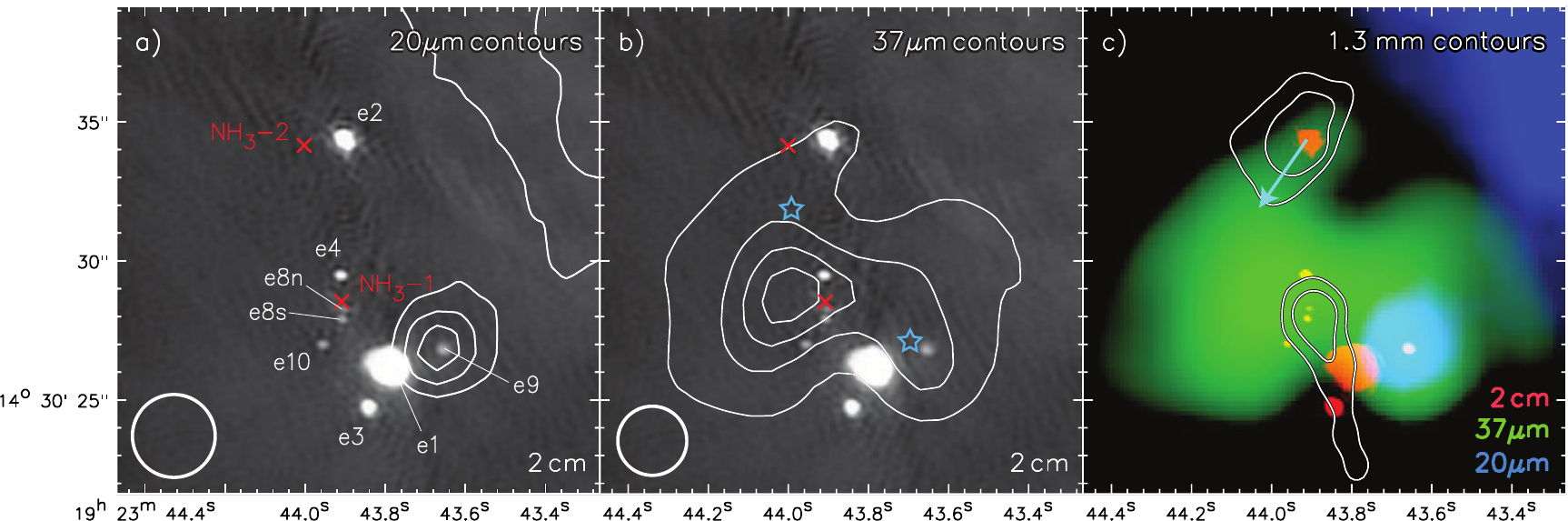}
\caption{The e1/e2 cluster. a) The \textit{SOFIA}-FORCAST 20\,$\mu$m contours are overlaid on the 2\,cm \textit{JVLA} contours from \citet{2016AnAp...595A..27G}. All of the radio continuum sources are labeled in white. The two red Xs mark the peak position of the ammonia clumps seen by \citet{1983ApJ...266..596H}. The resolution of the 20\,$\mu$m data is given by the circle in the lower right. b) The contours are the deconvolved \textit{SOFIA}-FORCAST 37\,$\mu$m data. The blue stars show the location of the point sources seen in the \textit{Spitzer} data by \citet{2016ApJ...825...54B}. c) An RGB composite image with the wavelengths for each color shown in the lower right corner. Overlaid are the smoothed 1.3\,mm data from \citet{2017ApJ...842...92G}. The blue arrow shows the direction of the blue-shifted outflow originating from e2 \citep{2010ApJ...718L.181S}. \label{fig:f3}}
\end{figure*}

\section{Comparing \textit{SOFIA} Images to Previous Imaging Observations} \label{sec:results1}

W\,51 was first detected as an \ion{H}{2} region by \citet{1958BAN....14..215W} through its free-free radio continuum emission. Over a decade later it was identified as a molecular cloud from its CO emission \citep{1971ApJ...165..229P}. The 430 MHz observations of \citet{1967AnAp...30...59K} were the first to resolve W\,51 into four large ($\sim$10--20$\arcmin$) radio components, which were labeled A through D, with W\,51\,A being the brightest among them. \citet{1970ApL.....5...99W}, were the first to further resolve W\,51\,A into two components labeled G49.5-0.4 and G49.4-0.3.  \citet{1972MNRAS.157...31M} observed W\,51\,A in centimeter continuum emission and further resolved G49.5-0.4 into eight regions named a through h, and G49.4-0.3 into three regions labeled a through c.  About two decades later, \citet{1994ApJS...91..713M} identified G49.5-0.4\,i and G49.4-3\,d, e, and f from Very Large Array (VLA) centimeter observations. G49.5-0.4\,j was first defined by \citet{2000ApJ...543..799O}. Peaks and compact sources within or near these regions are indexed with numbers. Our \textit{SOFIA} imaging data covers the entire W\,51\,A region, including both G49.5-0.4 and G49.4-0.3 (Figure \ref{fig:f1}).

\subsection{G49.5-0.4}

The strongest radio continuum emission regions in G49.5-0.4 are e and d. G49.5-0.4 was first mapped in the infrared by \citet{1974ApJ...187..473W}, where they identified two bright infrared components: IRS\,1, which was coincident with the radio source e; and IRS\,2, which was coincident with the radio source d. Both of these regions are well-studied and have garnered most of the observations trained on W\,51. We will discuss these two regions first before tackling the other regions of G49.5-0.4 below.

\subsubsection{The W\,51\,A IRS\,1 region (a.k.a. G49.5-0.4\,e)}

\textit{IRS\,1} --- The e region of G49.5-0.4 encompasses the entire 1$\farcm$5 arc-shaped IRS\,1 infrared region and its surroundings east of the d complex. The IRS\,1 arc as seen in the mid-infrared is bisected by dark lanes (Figure \ref{fig:f2}), first discussed by \citet{1994ApJ...433..164G} in their work with 2\,$\mu$m images of W\,51\,A. These dark lanes are centered at the locations shown with star symbols in Figure \ref{fig:f2}. \citet{1994ApJ...433..164G} point out several lines of evidence including the fact that the radio continuum maps show no gaps at these infrared-dark locations to conclude that the dark lanes are cold and dense dust filaments seen in absorption against the bright emission of the e arc.  The \textit{SOFIA} data show that these features are suppressed in their infrared emission even out to 37\,$\mu$m. If this suppression is due to extinction from a dense, cold dust filament, it would be expected that, at a long enough wavelengths, one would see the continuum emission from the cold dust concentrated in these dark lanes. Interestingly, there is no indication of concentrated emission from these dark lanes in the \textit{Herschel} 70 and 160\,$\mu$m data. In fact, the northernmost dark lane is clearly suppressed in emission out to 160\,$\mu$m. Perhaps more importantly, there are no indications of the northern two dark lanes having any enhanced emission in the 1.3\,mm ALMA continuum maps of \citet{2017ApJ...842...92G}. This may indicate that the gaps are not actually due to dense cold dust filaments, but may simply be areas with less dust, contrary to previous assessments. 

\begin{figure*}[htb!]
\figurenum{4}
\epsscale{1.15}
\plotone{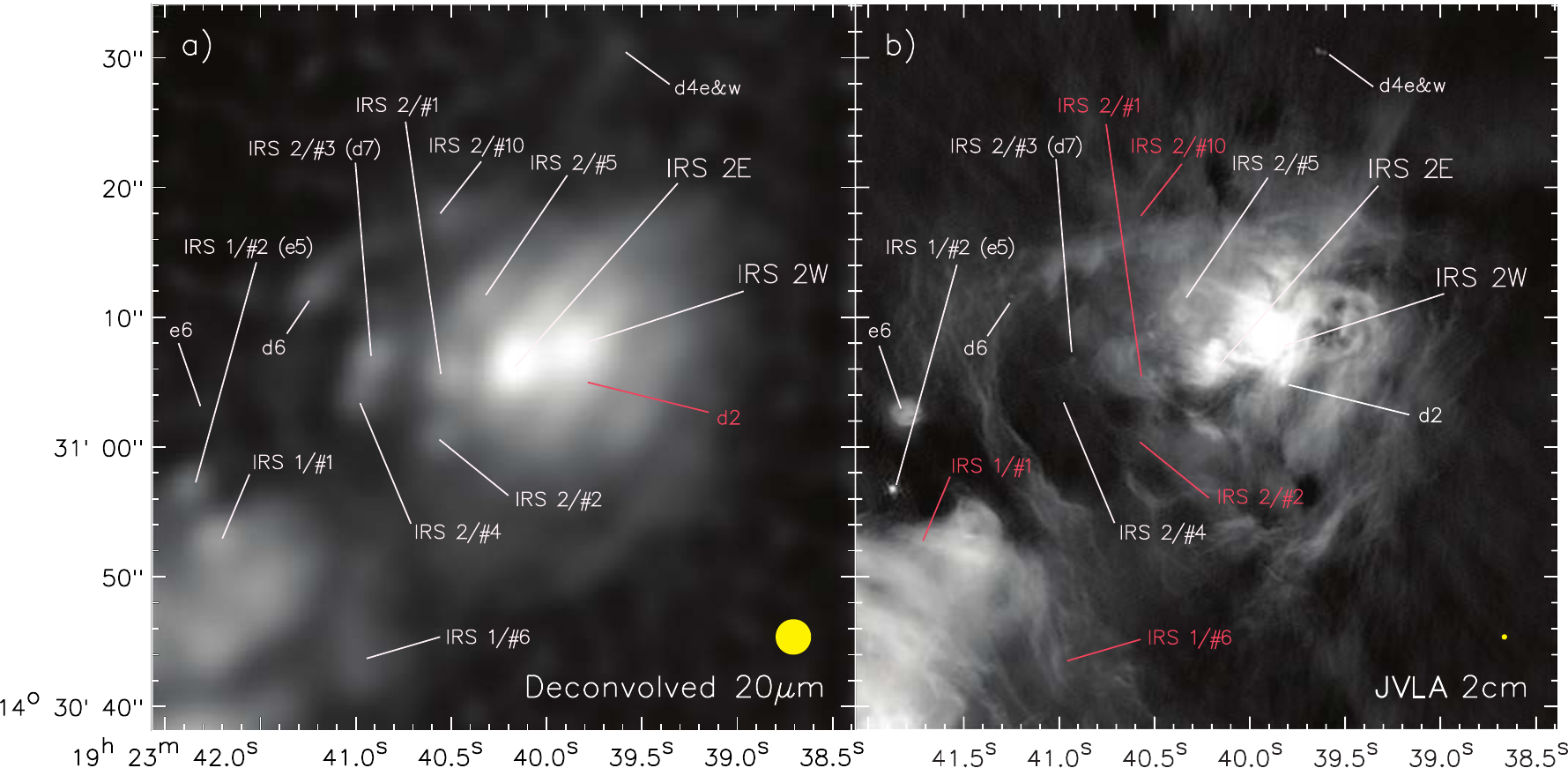}
\caption{The IRS2 (a.k.a. d) region with sources marked and labeled. If an object resembling a defined source (i.e. not simple extended emission) was detected at the source location the label is white, if not, it is red. a) The \textit{SOFIA}-FORCAST 20\,$\mu$m image deconvolved to a resolution shown by the yellow filled circle ($\sim$2.2$\arcsec$). b) \textit{JVLA} 2\, cm radio image from \citet{2016AnAp...595A..27G} of the same area at the resolution given by the yellow filled circle ($\sim$0.3$\arcsec$). 
\label{fig:f4}}
\end{figure*}

The brightest radio continuum peak in the e arc is coincident with a peak seen at both 20 and 37\,$\mu$m, as well in all \textit{Spitzer}-IRAC bands (except 8\,$\mu$m, which is saturated) which we label as IRS\,1/\#9. Interestingly, there does not seem to be a peak at this location in the \textit{Herschel} data.

\textit{The W\,51 e1/e2 cluster} --- There is a heavily studied massive star proto-cluster in the area $\sim$30$\arcsec$ interior to (east of) the arc, with radio sources designated e1, e2, e3, e4, e8n, e8s, e9, and e10 (Figure \ref{fig:f3}). This area is rich in maser emission, and is often given the moniker W\,51~MAIN (or just W\,51\,M) in maser studies of the region.   

\citet{1978MNRAS.183..435S} first found the two UC\ion{H}{2} regions in this area, and named them e1 and e2, due to their proximity to the main e feature. Later, \citet{1993ApJ...417..645G} discovered two more HC\ion{H}{2} regions near e1 and e2 at 3.6\,cm, which were named e3 and e4. \citet{1997ApJ...488..241Z} discovered an additional source at 1.3\,cm that lies between e4 and e1, which they named e8. This was later split into two sources, e8n and e8s, which were found to also be separate HC\ion{H}{2} regions \citep{2016AnAp...595A..27G}. Recently, \citet{2016AnAp...595A..27G} discovered two more HC\ion{H}{2} regions designated e9 and e10. Furthermore, there are two hot molecular cores in this area, first seen as ammonia clumps by \citet{1983ApJ...266..596H}, one very close to e2, and the other coincident with e8 (Figure \ref{fig:f3}). These hot cores have a rich line chemistry \citep{2017ApJ...842...92G} and are surrounded by multiple species of masers, which are the signposts of early massive star formation. 

\citet{2005ApJS..156..179D} observed the W\,51 e region from the ground at arcsecond resolution at both 11.7\,$\mu$m and 20.8\,$\mu$m with the \textit{IRTF}. At both wavelengths only a single point source was detected in the region near e1, but not coincident with it. The new observations made here with \textit{SOFIA} at 20\,$\mu$m with better astrometric accuracy confirm that this mid-infrared emission is not coming from e1 (Figure \ref{fig:f3}). Instead it appears that the mid-infrared point-source is coincident with a newly detected radio continuum source, e9 \citep{2016AnAp...595A..27G}, seen at 6.5\,cm which is characterized as being a HC\ion{H}{2} region.

Our image of the e9 source looks much different at 37\,$\mu$m. The morphology looks more like an arc, starting at the location of the 20\,$\mu$m point-source and stretching for $\sim$10$\arcsec$ to the east, wrapping around, but avoiding the radio sources e1 and e8. To see this emission in a little more detail, we deconvolved the 37\,$\mu$m data, which yielded an image with $\sim$2$\arcsec$ resolution (Figure \ref{fig:f3}). We see from this image a ``peanut'' of emission with two peaks at 37\,$\mu$m, a fainter one coincident with the e9 source, and the brighter one peaking just to the east of the e4/e8 sources. There is also a finger of fainter emission extending north of the brightest 37\,$\mu$m peak, which reaches the location of e2.  

While it seems clear from the deconvolved image that the peak at e9 seen at all wavelengths is clearly coming from the HC\ion{H}{2} region at this location, there are a couple of possible interpretations of why we see the brightest peak at 37\,$\mu$m just east of the e4/e8 area. First is that the combined emission from the multiple UC\ion{H}{2} and HC\ion{H}{2} regions is simply escaping from an area of lower extinction, which is located east of the e4/e8 region. Evidence for this comes from the fact that the NH$_3$ (3,3) peak \citep{1983ApJ...266..596H}, which is a dense gas tracer, peaks around the same location of the e4/e8 area, and the mm continuum emission appear to lie in a linear structure running more or less north-south just west of the ammonia peak and coincident with the ``waist'' of the peanut in 37\,$\mu$m emission (Figure \ref{fig:f3}c). This all points to a possible gradient in density in this area, with the density falling off to the east from e4/e8.

A second possible scenario is that the brightest peak at 37\,$\mu$m, and the finger of emission that connects it to the e2 area, are due to a cavity carved out of the surrounding medium by the CO outflow from e2 \citep{2010ApJ...718L.181S,2017ApJ...842...92G}. Mid-infrared emission is often seen coming from the outflow cavities carved out by the blue-shifted side of the outflows in heavily obscured MYSO regions \citep{2006ApJ...642L..57D,2017ApJ...843...33D}. The CO outflow from the e2 area does indeed have a blue-shifted outflow lobe pointing to the southeast of e2 at a position angle of 145$^{\circ}$ (see cyan arrow in Figure \ref{fig:f3}c).   

Interestingly, \citet{2016ApJ...825...54B} claim that the \textit{Spitzer} IRAC-GLIMPSE data detect infrared emission from both the e1 and e2 sources. Further scrutiny of the data show that the infrared peak mis-identified as coming from e1, is actually the same peak seen at other mid-infrared wavelengths presented here and coming from e9. The second source seen in the IRAC data is actually equidistant between e2 and e4, and not coming from e2 (see Figure \ref{fig:f3}; blue stars). It does not correspond to any known point source seen at any other infrared wavelength, but does appear to come from within the confines of the extended 37\,$\mu$m emission. The source also does not appear in \textit{2MASS} \textit{J}, \textit{H}, or \textit{K} data of this region, meaning it is likely not a foreground source. This source is apparently below the detection limits of the mid-infrared facilities that previously observed this region, but has a steep enough SED that we are beginning to pick it up at 37\,$\mu$m with \textit{SOFIA}.

\begin{figure*}[htb!]
\figurenum{5}
\epsscale{1.16}
\plotone{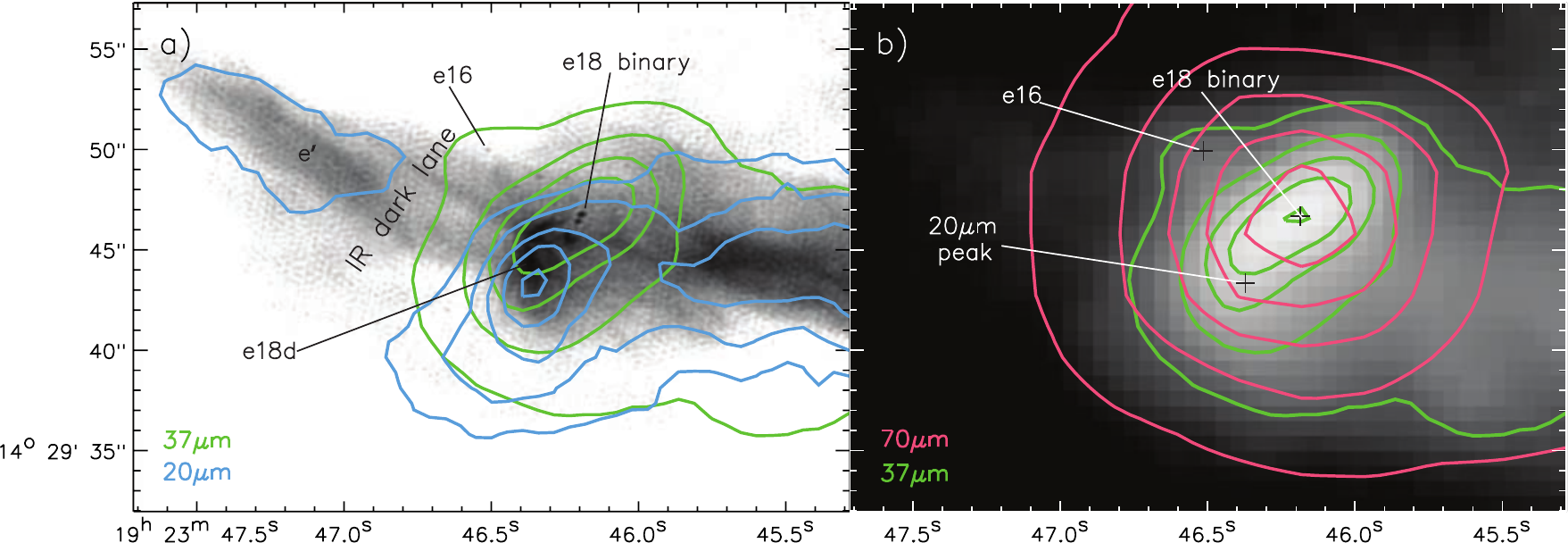}
\caption{The IRS\,4 (a.k.a. e16, e18, e18d) region. a) The inverse gray-scale image shows the \textit{JVLA} 2\,cm data from \citet{2016AnAp...595A..27G}, overlaid with contours from the \textit{SOFIA} 37\,$\mu$m (green) and 20\,$\mu$m (blue) images. The 37\,$\mu$m peak is coincident with the radio binary point sources labeled e18 by \citet{2016AnAp...595A..27G}, while the 20\,$\mu$m peak near the diffuse radio continuum emission of the \ion{H}{2} region e18d. b) Gray-scale image and green contours of the \textit{SOFIA} 37\,$\mu$m image and Herschel 70\,$\mu$m image (red contours) are displayed, and demonstrate how the peak if the infrared source IRS\,4 is co-spatial with the e18 radio binary at wavelengths $>$20\,$\mu$m. \label{fig:f5}}
\end{figure*}

\textit{Other detections in the IRS\,1/e region} --- Within the extended emission of the northern stretch of the e arc, there is a infrared point source that was detected at 20 and 37\,$\mu$m, which was first identified by \citet{2016ApJ...825...54B} as IRS\,1/\#1 (Figure \ref{fig:f4}). Several other compact radio sources (e4, e5, and e11-e23) have been identified in other areas within and around the e arc (see Figure \ref{fig:f2}). We detect compact or point-like sources in the mid-infrared at the locations of e7, e15, and e5 (also called IRS\,1/\#2 by \citet{2016ApJ...825...54B}; see Figure \ref{fig:f4}) in the \textit{SOFIA} data. While we do not resolve a point source at the location of the radio point source e11, there is an unlabeled, resolved, circular (r$\sim$3$\arcsec$) radio continuum source situated $\sim$4$\arcsec$ to the northeast of e11, where we do detect a diffuse infrared emission about the same extent at 20\,$\mu$m, however it appears as an arc-shaped structure at 37\,$\mu$m. We name this source IRS\,1/\#8 (Figure \ref{fig:f2}). We do not resolve mid-infrared point sources at the locations of the cm radio continuum point sources labeled e12-e14 and e17-23, though there is extended mid-infrared emission throughout the areas where they are situated (Figure \ref{fig:f2}). Faint infrared emission is also detected with \textit{SOFIA} at the location of e6 only at 37\,$\mu$m, though it can be seen in the \textit{Spitzer} 8\,$\mu$m data \citep{2016ApJ...825...54B}. Radio point source e16 was also detected in our mid-infrared data, but only at 37\,$\mu$m (Figure \ref{fig:f2} and \ref{fig:f5}).

We detect several sources not seen in radio continuum emission. There is a bright resolved source at both 20 and 37\,$\mu$m that was first detected at 2\,$\mu$m by \citet{1994ApJ...433..164G} and labeled IRS\,3 (Figure \ref{fig:f2}). It is also seen at 11.7\,$\mu$m in the \textit{IRTF} data from \citet{2005ApJS..156..179D}, at 8\,$\mu$m in the \textit{Spitzer} IRAC data, and at 2\,$\mu$m in the \textit{2MASS} data of the area. There are three point sources detected for the first time and only seen at 37\,$\mu$m in the vicinity of e15 (Figure \ref{fig:f2}). Following the nomenclature of \citet{2016ApJ...825...54B}, we dub these sources IRS\,1/\#3 , IRS\,1/\#4, and IRS\,1/\#5. There are also two resolved regions of mid-infrared emission in the northern part of the e region where there is no significant radio continuum emission peaks, both of which are seen in the near-infrared with \textit{2MASS}, and in the mid-infrared with \textit{SOFIA} and \textit{Spitzer}, which we will call IRS\,1/\#6 (see Figure \ref{fig:f4}) and IRS\,1/\#7(see Figure \ref{fig:f2}), respectively.

The radio point source e18 of \citet{2016AnAp...595A..27G} is actually a double separated by only $\sim$0$\farcs$5 and is coincident with the peak emission from a bar-shaped 37\,$\mu$m source on the bottom of the e arc, just west of the dark lane (Figure \ref{fig:f5}). At 20\,$\mu$m, the peak in emission is shifted to the peak of a $\sim$5$\arcsec$ in diamter radio continuum clump located $\sim$3$\farcs$5 southeast of the e18 binary named e18d, which was identified by \citet{2016AnAp...595A..27G} as a \ion{H}{2} region ( Figure \ref{fig:f5}b). At 37\,$\mu$m the source is very bright, and it appears to get brighter with increasing wavelength; at 70\,$\mu$m the emission peaks at the same location as the 37\,$\mu$m peak and this source appears as the 5th brightest source in all of W51\,A (Figure \ref{fig:f5}b). It is the 4th brightest source in all of W\,51\,A at 160\,$\mu$m after IRS\,2, IRS\,1 (peaked at IRS\,1/\#1), and the e1/e2 cluster region. We will call this infrared region IRS\,4, in keeping with the major IR emitting source nomenclature. IRS\,4 is the most-steeply rising sub-component from 20 to 37\,$\mu$m in this study which, along with the high FIR intensities, indicates the source is highly embedded and/or young. As we will see in a later section, the best fit SED model for this source yields a bolometric luminosity of 6.48$\times$10$^5\,L_{\sun}$, which is the single star equivalent spectral type of O4.5, but the SED can be fit with MYSO models with masses in the range of 24 to 96\,$M_{\sun}$. However, due presence of multiple cm radio continuum sources (e16, the e18 binary, e18d), this location is likely to be an embedded core or clump that is in the process of forming a young massive proto-cluster. 

\subsubsection{The W\,51\,A IRS\,2 region (a.k.a. G49.5-0.4\,d)}\label{sec:irs2}

In both the radio continuum and infrared imaging data, IRS\,2 breaks up into several sub-components surrounded by a $\sim$15$\arcsec\times$15$\arcsec$ cloud of emission at high spatial resolution. This extended emission was first found to be peanut-shaped in the 2\,$\mu$m images of \citet{1994ApJ...433..164G}, and they named the two peaks IRS\,2E and IRS\,2W. They also argued that the IRS\,2 region is a small cluster of ongoing star formation, identifying at least a dozen near-infrared sources. This area is also rich in masers (which are typically signposts of massive star formation), and maser studies typically refer to this region as W\,51\,A~NORTH \citep{1981ApJ...249..124S}.

High spatial resolution mid-infrared imaging by \citet{2001ApJ...553..254O} and \citet{2016ApJ...825...54B} show that the IRS\,2W component is an extended region of emission with no discernible point sources. This source is coincident with the brightest cm radio continuum feature, a cometary UC\ion{H}{2} region, with similar appearance in the radio \citep[i.e.][]{1989ApJS...69..831W,1993ApJ...417..645G} and mid-infrared. On the other hand, the IRS\,2E component is found to contain a cluster of four point-source components with $\sim$1$\arcsec$ separations. Our \textit{SOFIA} 20\,$\mu$m image of the area is shown in Figure \ref{fig:f4}. We deconvolved the image to try to resolve out as many components in the IRS\,2 region as we possible, though at the limits of our deconvolution we still cannot resolve the individual components within the IRS\,2E cluster.

\begin{figure*}[htb!]
\figurenum{6}
\epsscale{1.10}
\plotone{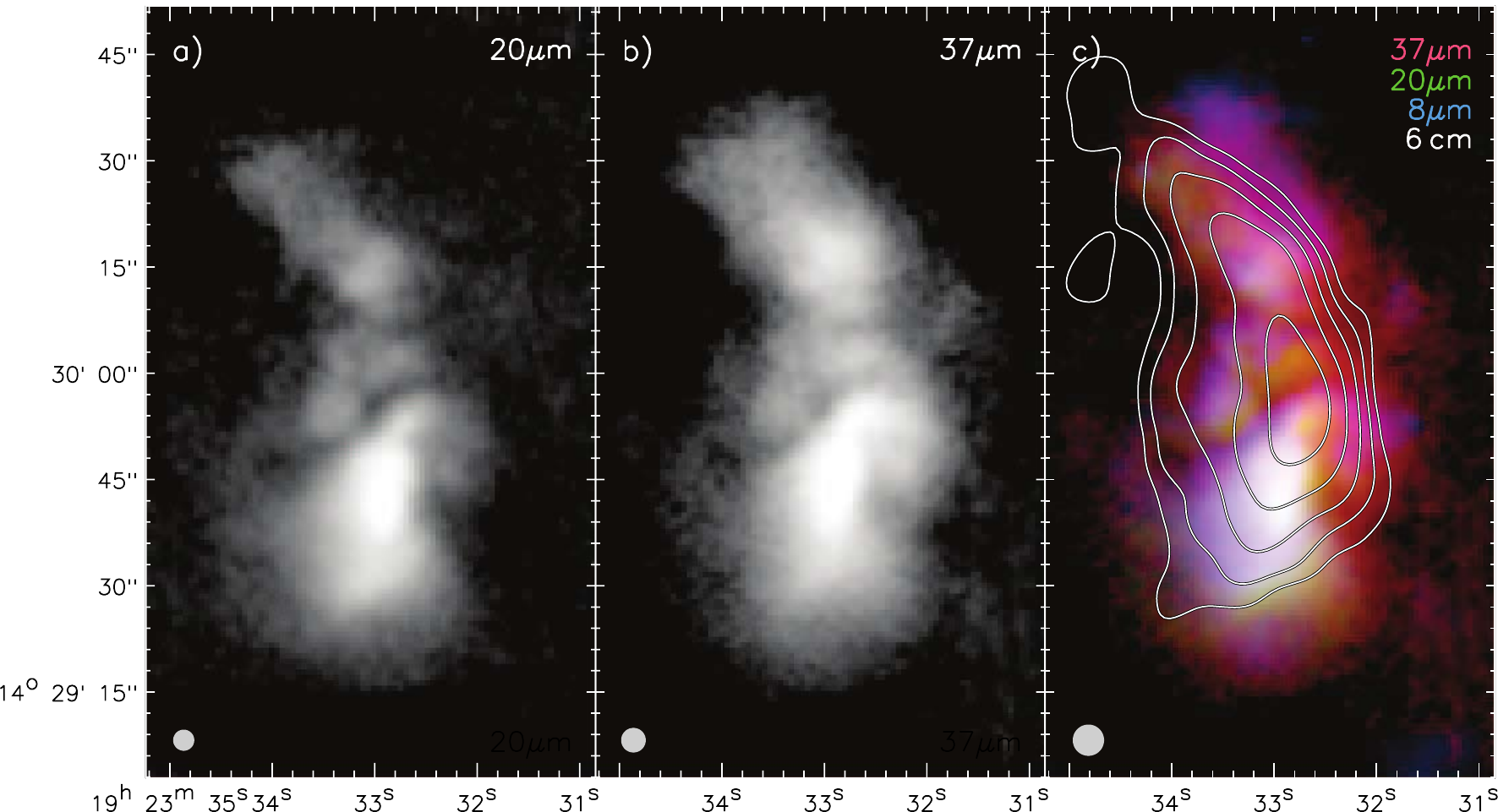}
\caption{The G49.5-0.4 b region. Panels a) and b) show the \textit{SOFIA}-FORCAST 20 and 37\,$\mu$m images, respectively, with the resolution given by the gray circles in the lower left ($\sim$3.0$\arcsec$ for 20\,$\mu$m, and $\sim$3.5$\arcsec$ for 37\,$\mu$m). c) An RGB image composed of the \textit{SOFIA} 37\,$\mu$m image (red), the \textit{SOFIA} 20\,$\mu$m image (green), and the \textit{Spitzer}-IRAC 8\,$\mu$m image (blue). Overlaid are radio continuum contours from the \textit{VLA} at 6\,cm from \citet{1994ApJS...91..713M}. The $\sim$4.4$\arcsec$ resolution of the \textit{VLA} image is shown by the circle in the lower left. \label{fig:f6}}
\end{figure*}

In the outskirts of the extended IRS~2 region there are several point-sources nested within the diffuse halo of infrared emission. Several of these sources have radio counterparts \citep[as seen by][]{2016AnAp...595A..27G}, some have been identified before in infrared observations, and others we will identify here for the first time. Radio sources d4e\&w, d6, and d7 all have infrared counterparts seen with \textit{SOFIA} (Figure \ref{fig:f4}). \citet{2016ApJ...825...54B} previously have identified the infrared emission from d7 (which they call IRS~2/\#3) and this was also seen in the mid-infrared images of \citet{2001ApJ...561..282K} and named KJD~9. It can be seen in the \textit{Spitzer} IRAC 8\,$\mu$m GLIMPSE image as well. Sources d4e\&w and d6 do not seem to have counterparts in the \textit{Spitzer} 8\,$\mu$m image, however d6 was detected in the mid-infrared by \citet{2001ApJ...561..282K} and named KJD~11. 

\citet{2016AnAp...595A..27G} identify a diffuse radio source that they label d3, however this is the previously identified radio source b2 \citep{1994ApJS...91..713M}. The b2 source does have a mid-infrared counterpart, but we will discuss it in a later section.

In addition to IRS~\,2/\#3 (KJD 9), \citet{2016ApJ...825...54B} also identify three more point-like infrared sources which they label IRS\,2/\#1, IRS\,2/\#2, and IRS\,2/\#4 on the eastern outskirts of IRS\,2. These were also seen by \citet{2001ApJ...561..282K} and labeled KJD~7, KJD~8, and KJD~10, respectively. We see all four of these sources in the \textit{SOFIA} 20 and 37\,$\mu$m images (see Figures \ref{fig:f2} and \ref{fig:f4}). IRS\,2/\#4 can also be seen as a point-source in the radio continuum images of \citet{2016AnAp...595A..27G}, though it was not labeled.   

We also detect five more infrared sources as of yet not identified in the IRS\,2/d region. Continuing the nomenclature of \citet{2016ApJ...825...54B} we will call these IRS\,2/\#6--\#10. IRS\,2/\#7 appears to not be a point source, with a slight extension from SE to NW (Figure \ref{fig:f2}). IRS\,2/\#6 appears in the \textit{Spitzer} 8\,$\mu$m image, is weakly detected in the \textit{SOFIA} 20\,$\mu$m image, and is not present in the 37\,$\mu$m \textit{SOFIA} image (Figure \ref{fig:f2}). 

We do not detect a source in the \textit{SOFIA} data at the location of KDJ~6. Though \citet{2001ApJ...561..282K} claims a source is present at this location, there is no information on the flux density or, more importantly, the significance of the detection in their paper. The source is not present in the shorter \textit{Spitzer}-IRAC bands, and the 5.8 and 8.0\,$\mu$m data are not helpful because the presumed source location resides in a region of the image that is saturated. 

\subsubsection{The G49.5-0.4\,b region}

The extended source b appears as a cometary \ion{H}{2} region or arc in the cm radio continuum images of \citet{1994ApJS...91..713M}, who also finds there is a velocity gradient from the SW to the NE as seen in H92$\alpha$. Apart from this, little else is known about this region. In the infrared, the source is bisected by a dark lane that is clearly visible in the \textit{Spitzer} IRAC data and all the way out to 37\,$\mu$m (Figure \ref{fig:f6}). The dark lane appears to be almost perpendicular to the velocity gradient seen in the radio line emission. There is a sub-mm core here, as seen in the  Herschel 160\,$\mu$m data and in the 450\,$\mu$m data of \citet{2006MNRAS.368.1223H}, with a peak close to the location of the dark lane. This may be the case of a outflowing source (or sources) buried within the dark lane, however the morphology of the radio continuum does not resemble a (partially) ionized jet or wind. The mid-infrared appearance is knotty (Figure \ref{fig:f6}). However, our source-finding algorithm found peaks at slightly different locations for sources in the 20 and 37\,$\mu$m data. This indicates that these are not likely to be individual centrally-heated sources. These sources are likely externally heated knots of dust, or optically-thin holes in the dust clump surrounding the central protostar(s) (as traced by the sub-mm and radio peak) where MIR emission is escaping. As we will discuss in Section 4, this region appears to be the least evolved (i.e. youngest) region in all of W\,51\,A, and therefore it may yet be too embedded for us to detect the YSOs within it even at wavelengths as long as 37\,$\mu$m.

\subsubsection{The G49.5-0.4\,j region}

\begin{figure*}[htb!]
\figurenum{7}
\epsscale{0.95}
\plotone{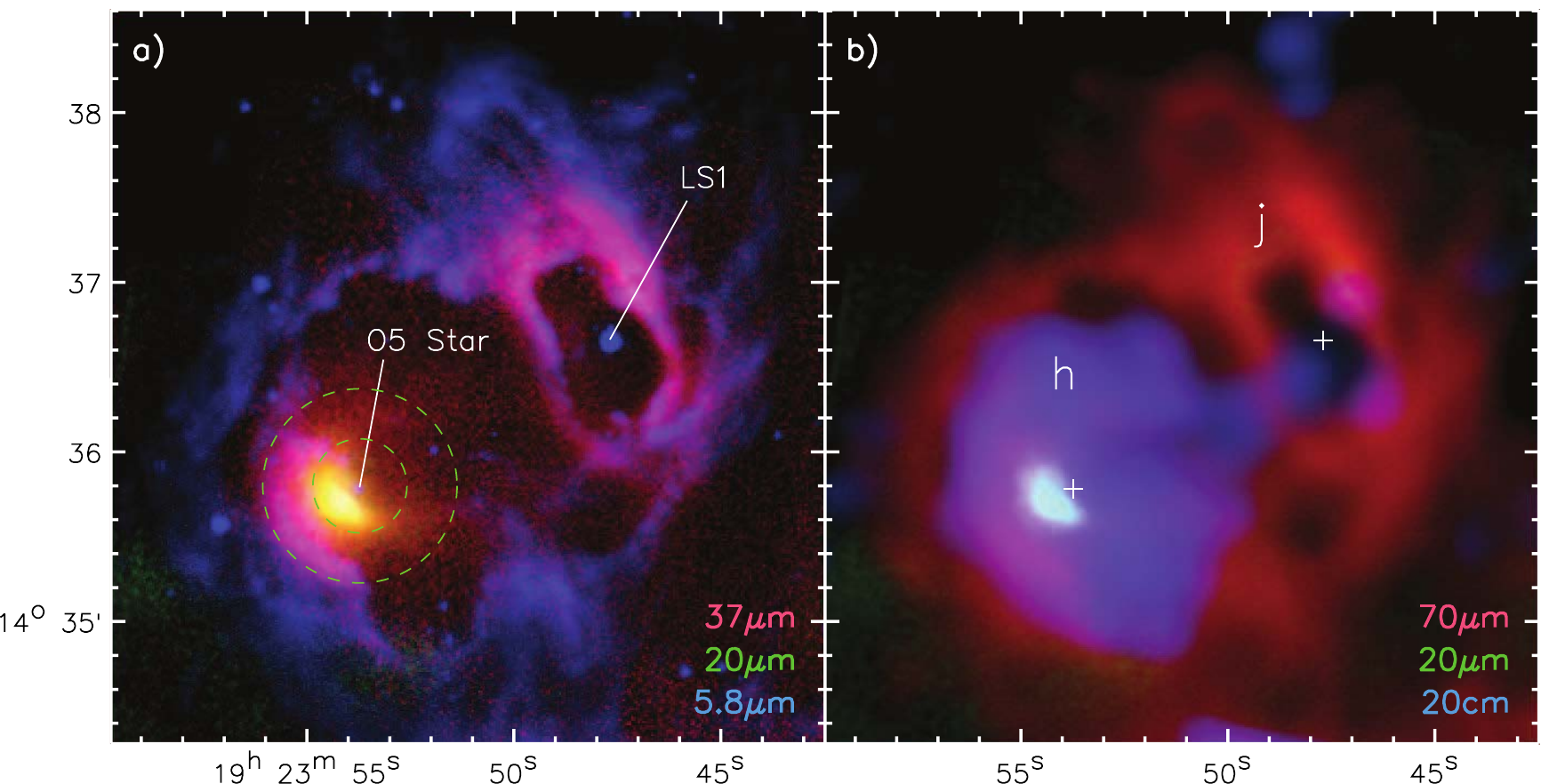}
\caption{The G49.5-0.4 h and j regions. a) Region j is the elongated ring of infrared emission which has a LBV star, LS1, at its center. It is abutted to the southeast by the h bubble, which internally has two arc structures to the southeast of a revealed O5 star. Fitting these arcs with circles (dashed green) shows them to be concentric about the O5 star. b) The rim of the h and j bubbles are traced well by the \textit{Herschel} 70\,$\mu$m emission (red), while the 20\,cm radio continuum \citep{1994ApJS...91..713M} fills in the interior of the h bubble. The locations of the O5 star and LS1 source from panel a are marked by the crosses for reference.\label{fig:f7}}
\end{figure*}

The j radio region appears as an elliptical shell in radio continuum maps \citep[e.g.][]{1994ApJS...91..713M}. In the infrared, the dust emission is fully contained within this shell tracing the ring-like structure (Figure \ref{fig:f7}). The 8\,$\mu$m \textit{Spitzer} image also shows a bright point source at the center of this shell. It is faint but detected in the \textit{SOFIA} 20\,$\mu$m images (but not at 37\,$\mu$m), and is very prominent at shorter wavelengths like the near-infrared. 

\citet{2000ApJ...543..799O} was the first to suggest the ring structure to be a wind-blown bubble driven by the star seen at its center in the near-infrared, claiming that it is a ``P Cygni-type supergiant''. This is a class of Luminous Blue Variable (LBV) star, which is thought to be a short-lived (10$^4$-10$^5$ yr) stage of massive stellar evolution between the main sequence O phase and the Wolf-Rayet (WR) phase \citep{1996ApJ...470..597M}. This short-lived phase is a time of great instability, leading to high mass loss resulting and the shedding of material that eventually forms circumstellar shells which can be seen readily in the infrared \citep{2010AJ....139.2330W}. Given the observations of \citet{2009A&A...504..429C} and the latest derivation of the distance to W51 of 5.4 kpc from \citet{2010ApJ...720.1055S}, it can be concluded that this  LBV candidate, dubbed [OMN2000]~LS1 (hereafter LS1), has a luminosity of $\sim$5$\times$10$^5$ L$_{\odot}$.

Given their fleeting nature, LBVs are rare and only a couple dozen verified LBVs have been found in our Galaxy, with about another 50 candidates awaiting confirmation \citep{2017MNRAS.466..213A}. Though naively one might think that an evolved star such should not be found in a region of active star formation, this is one of several known LBVs coincident with massive star-forming complexes (e.g. Orion, G305, W43, Westerlund 1, and the Galactic Center), which represents a significant portion of the known population of LBVs. This means that these presently active star-forming regions have had a long history of sustained star formation, and \citet{2009A&A...504..429C} claim that the data on LS1 point to an age of 3-6 Myr for the oldest observed epoch of star formation in W\,51\,A. 

\subsubsection{The other G49.5-0.4 regions}
There is very little study of the remaining G49.5-0.4 regions, which we will discuss all together in this section.

\begin{figure*}[htb!]
\figurenum{8}
\epsscale{1.16}
\plotone{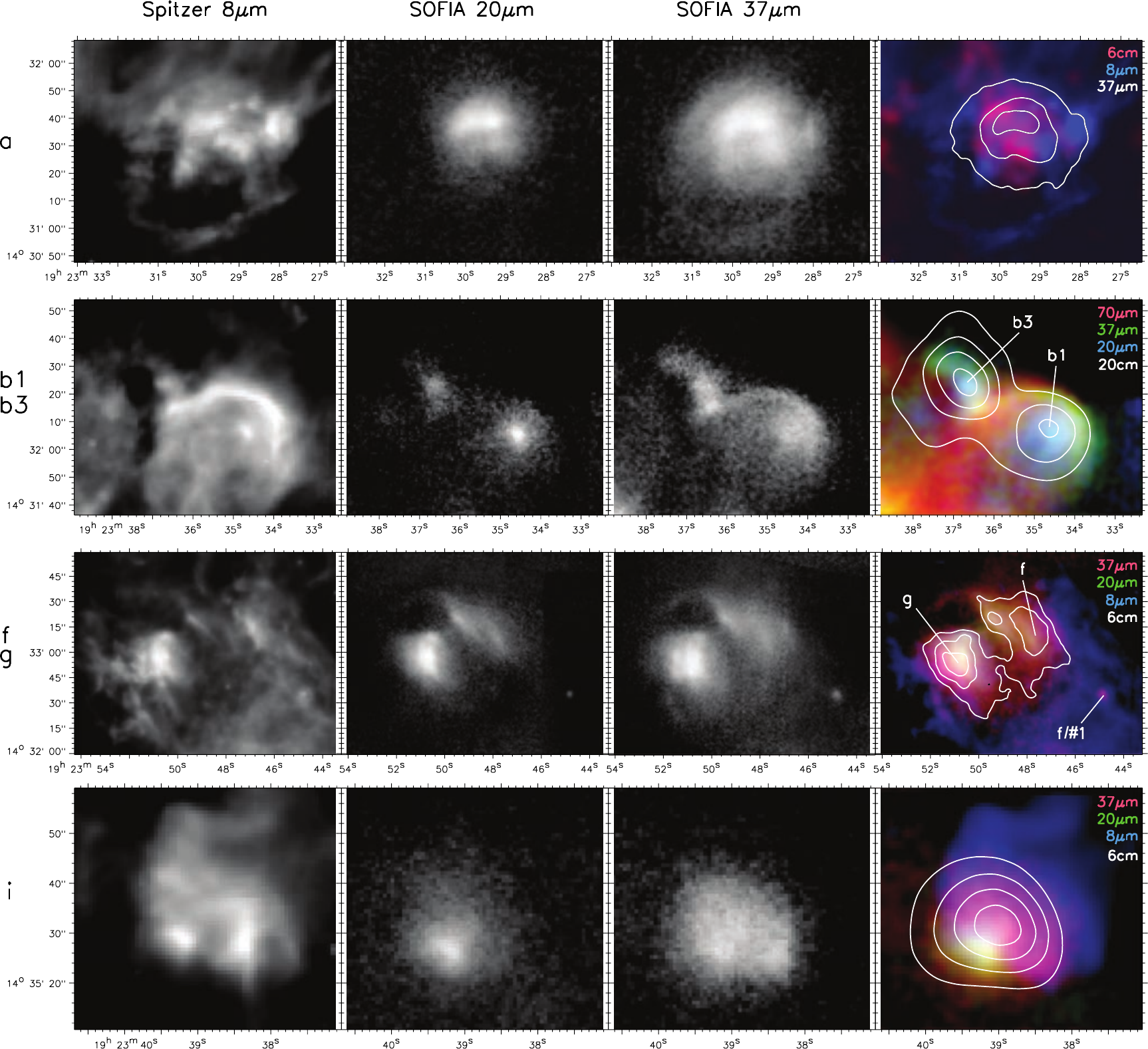}
\caption{The G49.5-0.4 a, b1/b3, f/g, and i radio continuum regions. To the left of each row of images is the radio regioin name. From left to right the images are \textit{Spitzer} 8\,$\mu$m, \textit{SOFIA} 20\,$\mu$m, \textit{SOFIA} 37\,$\mu$m, and an RGB image with the wavelengths representing each color given in the upper right corner. Contours are given by the wavelength noted in white. The 6 and 20\,cm data are \textit{VLA} data from \citet{1994ApJS...91..713M}, and 70\,$\mu$m data are from \textit{Herschel}.}
\label{fig:f8}
\end{figure*}

\textit{a} --- This region is an extended, round region of radio continuum emission with a diameter of about 20$\arcsec$, with a brighter area of emission on the northern side \citep{1994ApJS...91..713M}. In the infrared, this source takes on a different morphology at almost every wavelength (Figure \ref{fig:f8}). At 20\,$\mu$m it appears to be a `hamburger' with a brighter top than bottom, with a darker lane bisecting it through the middle. At 37\,$\mu$m it looks similar, but with more extended structures than at 20\,$\mu$m. As with the radio cm continuum images, both SOFIA wavelengths show no embedded point sources or peaks that resemble point-like sources. At 8\,$\mu$m, the dust emission is more clumpy and wispy. Comparing the 8\,$\mu$m emission to the 6\,cm radio continuum emission shows the peaks to be anti-correlated (see color image in Figure \ref{fig:f8}), and therefore the radio maybe tracing the more extinguished regions and the 8\,$\mu$m may be clumpy and wispy in appearance because to is escaping through holes that are less optically thick.  Both the \textit{Spitzer}\,8\,$\mu$m and SOFIA 37\,$\mu$m images show an arc or bubble to the south. This arc is also seen in the \textit{Spitzer}-MIPS 24\,$\mu$m image, but not in our SOFIA 20\,$\mu$m image, so is  likely fainter than our detection limit at that wavelength.

\textit{b1 and b3} --- Radio source b1 appears as a large, circularly-symmetric source in the low resolution radio images of the region \citep{1994ApJS...91..713M}. In the \textit{Spitzer} 8\,$\mu$m image (Figure \ref{fig:f8}), it consists of a sub-component surrounded to the north and west by a narrow arc structure ($\sim$20$\arcsec$ in diameter). In the \textit{SOFIA} data, the 20\,$\mu$m image shows a slightly extended source with a peak at the location of the 8\,$\mu$m compact source peak. There is very little emission at 20\,$\mu$m from the arc. At 37\,$\mu$m, emission tracing the arc seen at 8\,$\mu$m is detected, with emission also filling in the arc interior, looking more like a cometary UC\ion{H}{2} region perhaps caused by a bow shock, with a broad peak near the sub-component location.   

Source b3 looks like a slightly extended emission region on the northeast border of the b1 arc at 37\,$\mu$m, and has a similar appearance in the \textit{Spitzer} 8\,$\mu$m, and \textit{SOFIA} 20 and 37\,$\mu$m images (Figure \ref{fig:f8}). The peak at all three wavelengths is coincident with the radio peak. Like b1, this source has a bow-shock appearance.

Interestingly, the brightest 70\,$\mu$m emission is located in between b1 and b3 (see the color image for this source in Figure \ref{fig:f8}).

\textit{b2} --- Radio source b2 is a symmetric and compact source in \textit{SOFIA} 20 and 37\,$\mu$m images (Figure \ref{fig:f2}), but has a peak offset to the west in the \textit{Spitzer} 8\,$\mu$m image. 

We also detect one more sub-component in the \textit{SOFIA} images near b2 which does not have a radio continuum component. G49.5-0.4\,b2/\#1 (see Figure \ref{fig:f2}) is located $\sim$20$\arcsec$ southwest of b2, which appears as a unresolved point-source at 20\,$\mu$m, but is resolved and slightly extended at 37\,$\mu$m (and in the \textit{Spitzer} 8\,$\mu$m image). 

\textit{c1 and c} --- The naming convention for the radio emission in W\,51\,A has been to name the large regions of emission with letters, while individual peaks and sub-components within or near these regions are indexed with numbers. It is puzzling that there does not appear to be an extended radio region labeled c, but only the individual source peak c1 has been identified. Though there is a large and diffuse radio continuum region surrounding c1 and extending east towards the b region, it has never been labeled in radio studies, and is simply referred to as ``the arc-like area between regions b and c1'' by \citep{1994ApJS...91..713M}. The peak of c1, lies in an arc-shaped structure in the southeastern edge of a larger (r$\sim$40$\arcsec$), diffuse region of extended mid-infrared and radio continuum emission (Figure \ref{fig:f2}). This region appears to be separated from c1 and b by gaps in radio continuum emission, however the 20 and 37\,$\mu$m maps look very different, with diffuse infrared emission from this region forming a continuous region of dust emission all the way east to c1. In keeping with previous nomenclature we will refer to this entire  extended radio and infrared continuum region as region c. 

\citet{2008AJ....136..221F} identified two revealed O9 stars (sources \#62 and \#64 in their list) near the peak of radio source c1. The radio continuum source identified as e15 by \citet{2016AnAp...595A..27G} and the three newly discovered MIR sources (IRS\,1/\#3 , IRS\,1/\#4, and IRS\,1/\#5) all lie in the northern edge of the extended c region (Figure \ref{fig:f2}).  

\textit{f and g} --- Observations in the near-infrared by \citet{2000ApJ...543..799O} find 5 revealed O stars and 23 early B stars in the combined f and g regions. \citet{1997ApJS..108..489K} used HI absorption studies to determine that f and g are located either near the front or northern the edge of the molecular cloud containing W\,51\,A, while components a, b, and e are likely to be embedded in or behind it. 

With \textit{SOFIA} we see the same morphology and extent as what is seen in the low spatial resolution radio continuum images of this region (see color image for this source in Figure \ref{fig:f8}). \textit{Spitzer} images at 3--8\,$\mu$m show extended emission from the e region continuing north and surrounding the f and g radio regions to the south, east, and west. While there seems to be some emission and near-infrared point sources near the peak of radio source g, near-infrared emission is conspicuously absent from the areas of most of the extended radio and mid-infrared continuum emission of the f and g region.  This suggests that this region is being carved out by the O stars present here, heating and ionizing the areas we see in the \textit{SOFIA} mid-infrared and radio continuum images, consistent with the hypothesis by \citet{1997ApJS..108..489K} that the f and g region is likely in front of the W\,51\,A molecular cloud.

\citet{2005MNRAS.363..405H} find a 1.2\,mm dust clump coincident with the peak of the g source, and estimate it has 180~M$_{\odot}$ of dust. We detect a sub-component $\sim$1$\arcmin$ southwest of the center of radio source f in the \textit{SOFIA} data at both 20 and 37\,$\mu$m (Figure \ref{fig:f8}), which we label as f/\#1. There are some peaks at 20\,$\mu$m not present at 37\,$\mu$m, and vice-versa, with in the extended MIR emission of f and g, but no discernable embedded or point-like sources.

\begin{figure*}[htb!]
\figurenum{9}
\epsscale{0.95}
\plotone{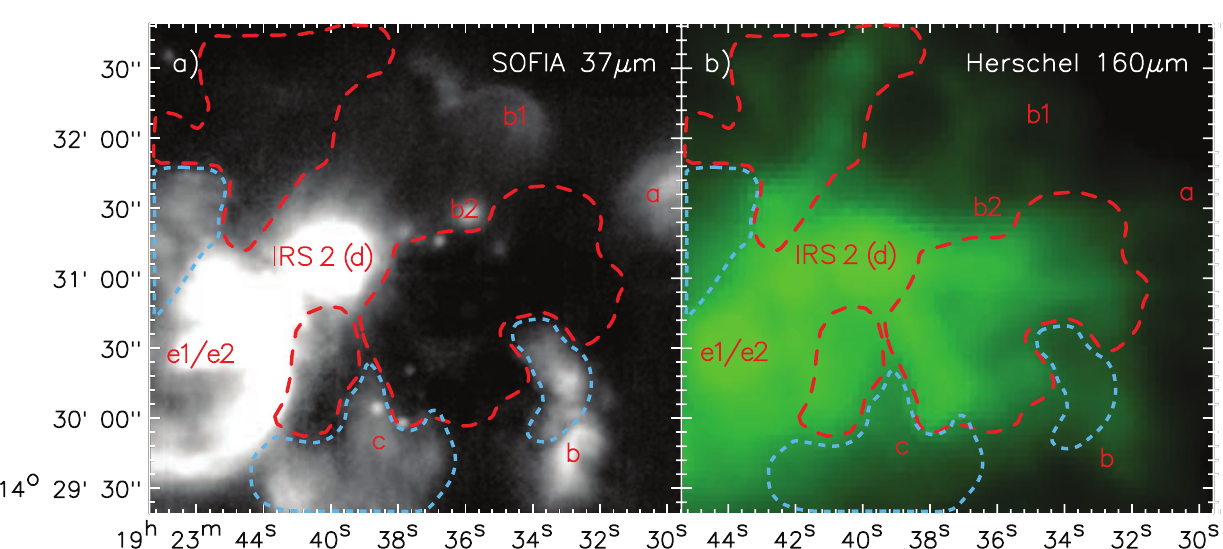}
\caption{Areas near IRS\,2 where the mid-infrared and far-infrared emission are anti-correlated. a) The \textit{SOFIA} 37\,$\mu$m image in gray-scale with source locations labeled. Interior to the three red dashed regions there is faint or no extended mid-infrared emission, while interior to the three blue dashed regions there is extended mid-infrared emission. b) The \textit{Herschel} 160\,$\mu$m image is shown in green-scale.  Interior to the three red dashed regions there is extended far-infrared emission, while interior to the three blue dashed regions there faint or no extended far-infrared emission. \label{fig:f9}}
\end{figure*}

\textit{h} --- This region was found to contain class II methanol masers \citep{2000A&AS..143..269S,2010RAA....10...67L}, which are a tracer of the earliest stages of massive star formation.  In the near-infrared, \citet{2000ApJ...543..799O} find dozens of revealed B stars around the h radio region, with an evolved O5 star near its center. This star can be seen in the \textit{Spitzer} images of the region (Figure \ref{fig:f7}). It is bordered to the southeast by two concentric arcs, the nearest bright at both 20 and 37\,$\mu$m, but the outer arc is only bright at wavelengths 37\,$\mu$m and longer. Encircling the O5 star and the two arcs is an outer bright-rimmed bubble that can be most easily seen in the 70\,$\mu$m Herschel image, which is filled in by radio continuum emission (Figure \ref{fig:f7}b). The radio continuum peak is close to the 20 and 37\,$\mu$m peak, indicating that the whole h region may be ionized and heated by the O5 star located near there. This is unlike the region j, which abuts the rim of h to the west, which is devoid of emission inside its wind-blown shell at infrared and cm radio wavelengths. \citet{2000ApJ...543..799O}  state that the h and j region has the lowest extinction in the whole of G49.5-0.4, which is likely due to the evolved state of these two regions.

\begin{figure*}[htb!]
\figurenum{10}
\epsscale{1.15}
\plotone{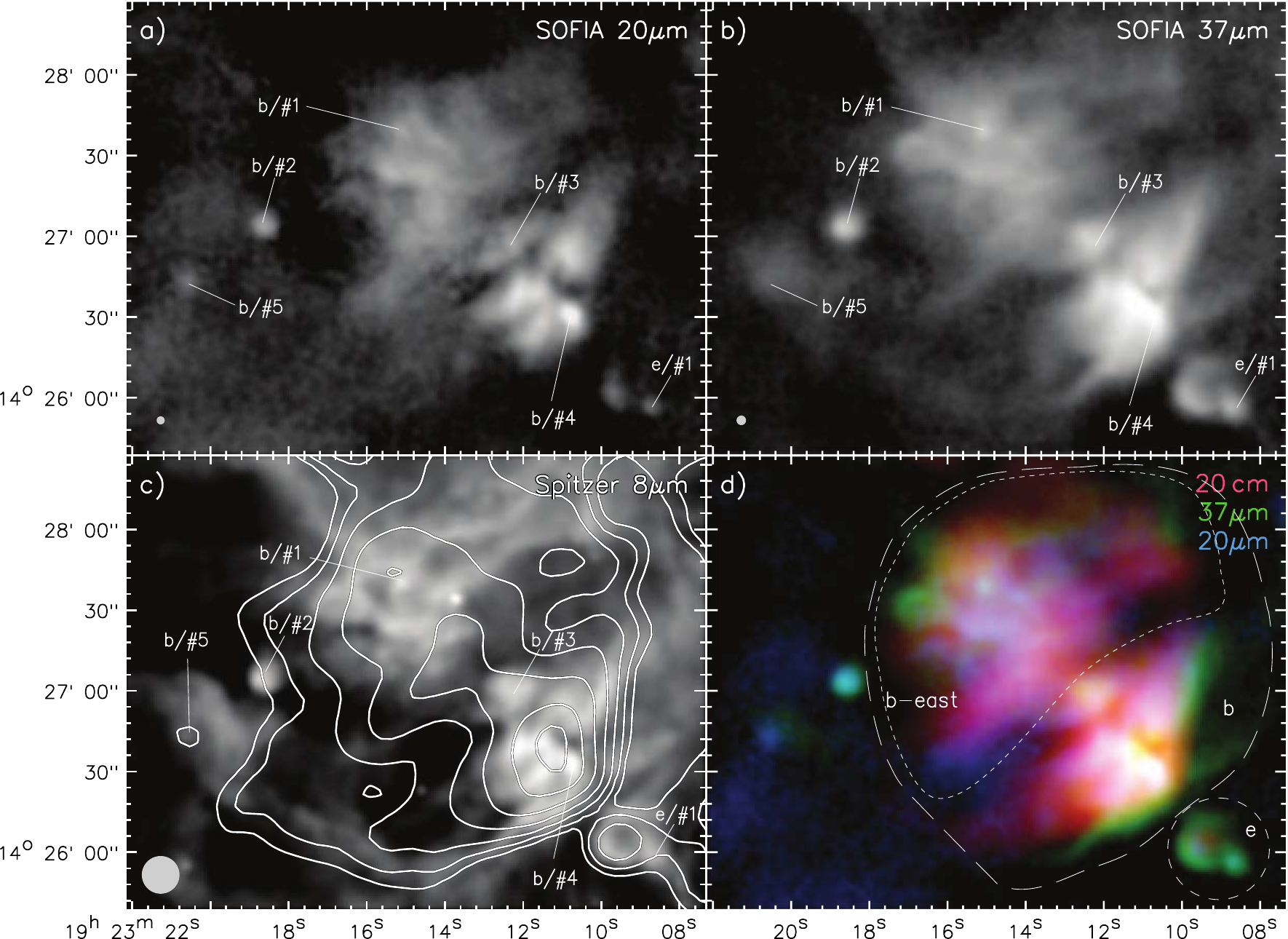}
\caption{G49.4-0.3 regions b and e. a) The \textit{SOFIA} 20\,$\mu$m image with infrared sources and peaks labeled. b) The \textit{SOFIA} 37\,$\mu$m image with infrared sources and peaks labeled. c) The \textit{Spitzer} 8\,$\mu$m image with infrared sources and peaks labeled. Overlaid are contours from the \textit{VLA} at 20\,cm \citep{1994ApJS...91..713M}. d) An RGB image with the \textit{VLA} 20\,cm emission in red, the \textit{SOFIA} 37\,$\mu$m emission in green, and the \textit{SOFIA} 20\,$\mu$m in blue. Encompassed in dashed lines and labeled are the major regions b, b-east, and e. \label{fig:f10}}
\end{figure*}

\textit{i} --- One O9 star and one B1 star is seen in the near-infrared in this region by \citet{2000ApJ...543..799O}  The region appears to be a multi-peaked, extended region with a radius of $\sim$14$\arcsec$ in the \textit{Spitzer} 8\,$\mu$m image (Figure \ref{fig:f8}). Interestingly, the 20\,$\mu$m SOFIA image shows a much less extended emission with a peak coincident with the southwestern peak seen at 8\,$\mu$m. The color image for this source in Figure \ref{fig:f8} shows that the combined emission across all mid-infrared wavelengths is fan-shaped, with the 20\,$\mu$m emission being most compact, the 37\,$\mu$m emission extending out to the north and west beyond that, with the 8\,$\mu$m emission extending yet farther beyond both the 20 and 37\,$\mu$m emission to the north and west. Given the morphology as a function of wavelength in the infrared could be a ``blister''-type \ion{H}{2} where the source lies on the edge of a dense region and the emission is breaking out on one side (where the density is lowest).

\subsubsection{Mid-infrared ``dark'' areas of G49.5-0.4}
In addition to the infrared-dark lanes discussed above in the previous sections, the \textit{Herschel} 160\,$\mu$m image show that the infrared-dark area south of b2, west of d and e, north of c, and east of b and a is ``filled in'' by 160\,$\mu$m dust emission. This signifies that this area is infrared-dark due to the presence of wide-spread cold dust (Figure \ref{fig:f9}). 

The 160\,$\mu$m emission is strongest around the d and e1/e2 regions and mimics the shape seen by \textit{SOFIA} of those regions to first order. However, the brightest 160\,$\mu$m emission actually wraps around and ``avoids'' the hot infrared emission seen by \textit{SOFIA} of the b and c sources. Further to the north, the outskirts of the 160\,$\mu$m emission also look like they wrap around and avoid source g and f. This appears to indicate that the much of the appearance of G49.5-0.4 in the mid-infrared is dominated by us only seeing emission on the surfaces of the sub-cloud structure and/or leaking out through less dense areas devoid of large dust grains carved out by ionization fronts and outflows within this region of the W\,51\,A molecular cloud. 

\subsection{G49.4-0.3}

There is very little study of this region, even though it is only $\sim$2$\farcm$5 west of the well-studied G49.5-0.4 region. Though \citet{1972MNRAS.157...31M} was the first to resolve the radio continuum emission of G49.4-0.3 into three regions (labeled a through c), it was the observations of \citet{1994ApJS...91..713M} that resolve and identified further radio continuum sources (labeled d through f). Source b was identified as the brightest radio continuum component, and it is also the brightest far-infrared \citep{1986ApJ...300..737H} source. Since most of the studies of this region have focused on the areas around source b, we will discuss this source first before discussing the remaining sources.

\subsubsection{The G49.4-0.3\,b region}

\begin{figure*}[htb!]
\figurenum{11}
\epsscale{1.16}
\plotone{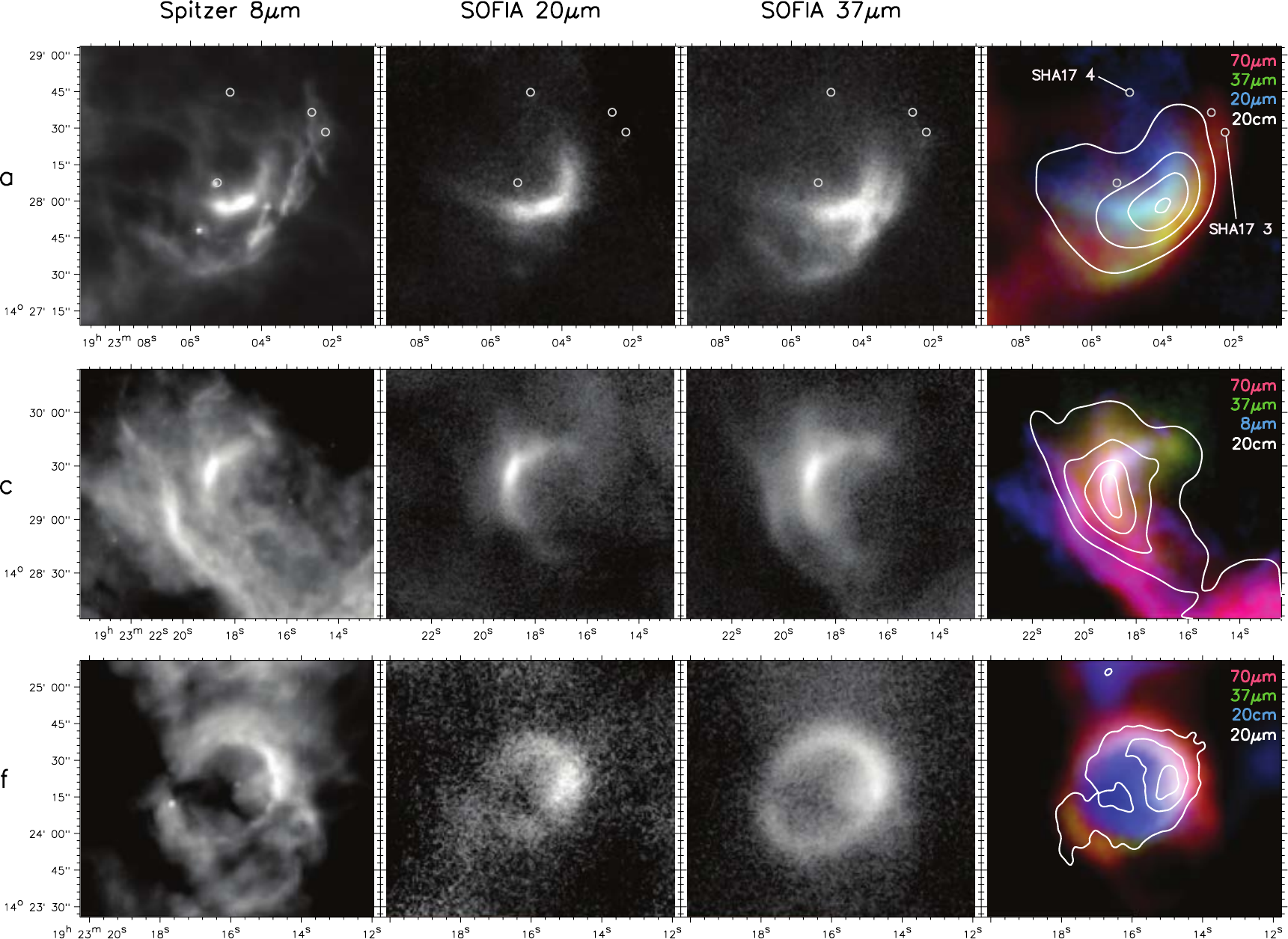}
\caption{The G49.4-0.3 a, c, and f regions. To the left of each row of images is the source name. From left to right the images are \textit{Spitzer} 8\,$\mu$m, \textit{SOFIA} 20\,$\mu$m, \textit{SOFIA} 37\,$\mu$m, and an RGB image with the wavelengths representing each color given in the upper right corner. Contours are given by the wavelength noted in white. The 20\,cm data are \textit{VLA} data from \citet{1994ApJS...91..713M}, and 70\,$\mu$m data are from \textit{Herschel}. For source a the white circles mark the locations of the MYSO candidates of \citet{2017ApJ...839..108S}.\label{fig:f11}}
\end{figure*}

\citet{1986ApJ...300..737H} resolved this region into two components in the far-infrared with the \textit{Kuiper Airborne Observatory}. The brightest peak in the far-infrared is near the cm radio continuum peak b. But there is a secondary peak $\sim$1$\arcmin$ to the northeast in the far-infrared which they named b-east (Figure \ref{fig:f10}). This peak is seen in the 20\,cm images of \citet{1994ApJS...91..713M}, but was not labeled. At all wavelengths from the near-infrared to the radio there is a dark gap or decrease in emission running NW to SE and separating the southwestern part of source b from b-east, and is therefore likely due to a decrease or absence of gas and dust at that location.  

Source b has a peak in the cm radio continuum that is close to, but not exactly coincident with, the far-infrared peak seen at 70\,$\mu$m ($\sim$5$\arcsec$ offset). Both components appear to reside in a infrared-dark area (as seen in \textit{Spitzer} IRAC and \textit{SOFIA} data) that bisects the b source and runs NE to SW (Figure \ref{fig:f10}). The 160\,$\mu$m \textit{Herschel} peak seems to be exactly centered and the same shape as the ``lane'' in the near- and mid-infrared emission. Given the fact that this infrared-dark area has radio continuum emission and water maser emission \citep{1988A&AS...76..445C}, and is surrounded by YSOs \citep{2017ApJ...839..108S}, it is likely the site of very embedded massive star formation that is infrared-dark at wavelengths $<$40\,$\mu$m due to very high extinction. Most of the peaks within this region shift as a function of wavelength in the MIR, indicating they are externally heated knots or holes in the otherwise optically thick emission in the region where MIR light is escaping. However we find two sources where the peaks does not change with wavelength, and are therefore likely to be MYSO candidates, which we label b/{\#}3 and b/{\#}4 (Figure \ref{fig:f10}).

Source b-east is a much more diffuse area of mid-infrared emission, but it does appear to have one embedded point source which we name b/\#1 (Figure \ref{fig:f10}). It is likely a MYSO, since it is apparent in the \textit{Spitzer} IRAC data, is seen at both 20 and 37\,$\mu$m, and is coincident with a radio continuum peak at 1.5\,cm \citep{2016AnAp...595A..27G}.

We have found in the \textit{SOFIA} data a resolved but compact source detected at 10$\arcsec$ east of the extended b-east region, that is also seen in \textit{Spitzer} 8\,$\mu$m that we have named b/\#2 (Figure \ref{fig:f10}). This source is also a very bright object in the \textit{Herschel} 70 and 160\,$\mu$m images of this region, but has no associated cm radio continuum emission. Given its high flux in the mid- and far-infrared and lack of radio continuum emission, we believe that it is likely an MYSO in a very young evolutionary state prior to the onset of a UC\ion{H}{2} region (as we will see in a later section, this source does indeed appear to be a MYSO from SED model fitting). We also see another isolated source $\sim$15$\arcsec$ southeast of b/\#2, which also has a radio component which we label b/\#5. 

\subsubsection{The G49.4-0.3\,a, c, d, e and f regions}
These four sources encircle the main intensity peak near source b and all have cometary or shell-like structure in the radio and in the infrared  (Figures \ref{fig:f1} and \ref{fig:f11}). It appears from the morphologies of the sources in the \textit{Spitzer} NIR data alone that these sources (and indeed most of this region's structure) are due to wind-blown bubbles and/or are bright-rimmed clouds. The outer layer of the shells is generally demarcated by the dust emission as seen in the \textit{Spitzer} IRAC and \textit{SOFIA} images, and generally the interiors of the shells/arcs are filled with cm radio continuum emission. 

\textit{a} --- This source has an interesting double arc structure, with the 8\,$\mu$m emission, 37\,$\mu$m emission, and cm radio continuum tracing both arcs.  Interestingly, the \textit{SOFIA}\,20\,$\mu$m emission dominantly traces the interior arc. The peak fluxes of the inner and outer arcs differ only by $\sim$20\,\% in 8\,$\mu$m and 37\,$\mu$m, while at 20\,$\mu$m there is almost an order of magnitude differences in flux at the same positions. As we will discuss later in \S\,\ref{sec:cps}, we can use a color-color diagram to determine if a source has flux in the IRAC bands that is dominated by PAH emission. Using that method we have found that this source falls well within the definition of PAH-dominated. Therefore, a plausible explanation for the behavior of the flux of this source as a function of wavelength is that the continuum emission of the outer arc may be low at wavelengths $\leq$20\,$\mu$m, and that the IRAC 8\,$\mu$m flux is high because of strong PAH emission. 

It appears that there is a cluster of YSOs identified by \citet{2017ApJ...839..108S} located interior to (or east of) this double arc (Figure \ref{fig:f11}), which we assume is most likely responsible for the shaping, heating, and ionizing source a. The four massive YSO candidates from \citet{2017ApJ...839..108S} are shown in Figure \ref{fig:f11}, though we only detect sources in the SOFIA data at the locations of the sources labeled SHA17\,3 and SHA17\,4. (We will show in the section on SED model fitting that these sources are unlikely to be MYSOs).

\textit{c} --- This source also has a double rimmed structure, with the eastern arc traced by \textit{Spitzer} 8\,$\mu$m emission, and the western arc traced by the \textit{SOFIA} 20 and 37\,$\mu$m emission (Figure \ref{fig:f11}). The cm radio continuum emission fills in the area interior to the eastern arc, with a peak near the inner, western arc. Interestingly, \citet{2017ApJ...839..108S} finds a cluster of $\sim$15 YSOs to the east and south of source c. Feedback from these YSOs may be responsible for shaping the arc-shaped dust structure seen here in the MIR, however there is no evidence of any truly energetic YSOs in the cluster given that none of the cluster members display radio continuum emission, and the region appears to be devoid of continuum sources from the near-infrared out to 160\,$\mu$m (i.e. no indication of massive and/or young and active cluster members). 

\textit{d} --- This source has a ring-shape with a radius of $\sim$1$\arcmin$, which is brightest to the southeast and faintest to the northwest. This southeastern rim appears as a bright arc in the \textit{Spitzer} NIR data and \textit{Herschel} far-IR data (see the large arc of 70\,$\mu$m emission to the west of source e in Figure \ref{fig:f1}), but we barely detect it at 37\,$\mu$m and do not see any evidence of it at 20\,$\mu$m.

\textit{e} --- This source is a very tiny bright-rimmed source located on the eastern rim of source d (Figure \ref{fig:f10}). This rim can be seen in the \textit{Spitzer} NIR and \textit{SOFIA} 37\,$\mu$m data, and the center is filled by unresolved radio continuum emission at cm. Only the brighter eastern part of the shell is detected at 20\,$\mu$m. There is what appears to be a point source in the mid-infrared, just to the southwest of e. We name this source e/\#1 (Figure \ref{fig:f10}). There is no cm radio emission detected from this source.

\textit{f} --- This source is a smaller ring-shaped region (r$\sim$25$\arcsec$), with the outer rim radiating brightly in the \textit{Spitzer} NIR images as well as the \textit{SOFIA} 37\,$\mu$m image (Figure \ref{fig:f11}). Interior to this is a ring seen in radio continuum emission as well as 20\,$\mu$m, so is likely an ionized bubble (i.e., Stromgren sphere). At all wavelengths the ring is brightest to the west, giving it a cometary UC\ion{H}{2} appearance. 

\section{Data Analysis}\label{sec:data}

\subsection{Physical Properties of Sub-Components and Point Sources: SED Model Fitting and Determining MYSO Candidates}\label{sec:cps}

In order to identify what sources may be MYSOs within our W\,51\,A field, we compile the list of subcomponents and point-sources already identified and discussed in Section 3.  We add to this list two sources in G49.4-0.3 (a/\#1 and b/\#6) and three in G49.5-0.4 (i/\#1, IRS\,1/\#10, and  IRS\,1/\#11), which are sources detected in the field covered by \textit{SOFIA} but outside the main areas of infrared emission discussed in Section 3.  Table\,\ref{tb:cps1} contains the information regarding the position, radius employed for aperture photometry, and 20 and 37\,$\mu$m flux densities (before and after background subtraction) of all these sources.

\begin{deluxetable*}{rrrrrrrrrr}
\centering
\tabletypesize{\scriptsize}
\tablecolumns{8}
\tablewidth{0pt}
\tablecaption{Observational Parameters of Sub-components and Point Sources in W51A}
\tablehead{\colhead{  }&
           \colhead{  }&
           \colhead{  }&
           \multicolumn{3}{c}{${\rm 20\mu{m}}$}&
           \multicolumn{3}{c}{${\rm 37\mu{m}}$}&
           \colhead{  }\\
           \cmidrule(lr){4-6} \cmidrule(lr){7-9} \\
           \colhead{ Source }&
           \colhead{ R.A. } &
           \colhead{ Dec. } &
           \colhead{ $R_{\rm int}$ } &
           \colhead{ $F_{\rm int}$ } &
           \colhead{ $F_{\rm int-bg}$ } &
           \colhead{ $R_{\rm int}$ } &
           \colhead{ $F_{\rm int}$ } &
           \colhead{ $F_{\rm int-bg}$ } &
           \colhead{  Notes  }\\
	   \colhead{  } &
	   \colhead{  } &
	   \colhead{  } &
	   \colhead{ ($\arcsec$) } &
	   \colhead{ (Jy) } &
	   \colhead{ (Jy) } &
	   \colhead{ ($\arcsec$) } &
	   \colhead{ (Jy) } &
	   \colhead{ (Jy) } &
	   \colhead{} 
}
\startdata
{\bf G49.4-0.3} & & & & & & & & & \\
    a/{\#}1 & 19 22 59.2 &+14 29 39.7 &  3.84 &     1.90 &     1.46 &  4.61 &     5.39 &     7.20 &                \\
    b/{\#}1 & 19 23 15.1 &+14 27 39.0 &  4.61 &     7.02 &     1.60 &  4.61 &     43.6 &     11.4 &                \\
    b/{\#}2 & 19 23 18.7 &+14 27 03.7 &  9.98 &     11.6 &     6.33 &  9.98 &     61.5 &     57.4 &                \\
    b/{\#}3 & 19 23 12.3 &+14 26 57.4 &  9.98 &     24.7 &     6.51 &  9.98 &      180 &      116 &                \\
    b/{\#}4 & 19 23 10.7 &+14 26 30.0 &  6.14 &     24.5 &     21.8 &  6.14 &      187 &      111 &                \\
    b/{\#}5 & 19 23 20.5 &+14 26 42.4 &  9.22 &     9.13 &     3.07 &  15.4 &     43.6 &     35.1 &                \\
    b/{\#}6 & 19 23 32.1 &+14 26 55.4 &  4.61 &     2.96 &     0.67 &  6.91 &     17.2 &     9.19 &                \\
    e/{\#}1 & 19 23 08.7 &+14 25 56.1 &  6.14 &     3.10 &     1.63 &  6.14 &     33.6 &     36.2 &                \\
    SHA17 3 & 19 23 02.2 &+14 28 24.6 &  3.84 &     0.77 &     0.75 &  6.91 &     10.7 &     4.50 &                \\
    SHA17 4 & 19 23 04.8 &+14 28 43.3 &  3.84 &     1.83 &     0.53 &  5.38 &     5.32 &     0.88 &                \\
{\bf G49.5-0.4} & & & & & & & & & \\
         b1 & 19 23 34.5 &+14 32 05.5 &  18.4 &     41.1 &     25.7 &  25.3 &      258 &      156 &                \\
         b2 & 19 23 35.8 &+14 31 27.8 &  7.68 &     24.7 &     20.5 &  9.98 &      105 &     76.5 &                \\
   b2/{\#}1 & 19 23 34.9 &+14 31 11.9 &  6.91 &     3.69 &     1.65 &  7.68 &     14.6 &     10.9 &  SHA17\,2              \\
         b3 & 19 23 36.7 &+14 32 23.4 &  12.3 &     15.4 &     6.29 &  10.8 &     59.3 &     33.5 &                \\
    d4e+d4w & 19 23 39.7 &+14 31 29.4 &  4.61 &     7.50 &     3.49 &  4.61 &     $<$55.9\tablenotemark{u} &     \nodata &                \\
         d6 & 19 23 41.2 &+14 31 11.1 &  3.07 &     8.58 &     5.36 &  3.84 &      137 &     74.5 &       KJD\,11  \\
         e7 & 19 23 44.8 &+14 29 10.3 &  6.91 &     31.7 &     21.8 &  9.98 &      124 &     84.5 &                \\
         e9 & 19 23 43.6 &+14 30 26.7 &  4.61 &     16.8 &     6.65 &  10.8 &     1300 &      614 &                \\
        e15 & 19 23 38.6 &+14 30 04.9 &  4.61 &     9.19 &     6.06 &  4.61 &     43.2 &     22.6 &                \\
    f/{\#}1 & 19 23 44.8 &+14 32 35.0 &  5.38 &     1.09 &     2.01 &  6.91 &     11.0 &     10.1 &                \\
          i & 19 23 39.2 &+14 35 26.8 &  13.8 &     46.6 &     23.0 &  19.2 &      155 &      138 &                \\
    i/{\#}1 & 19 23 37.6 &+14 33 59.1 &  6.14 &     1.07 &     1.48 &  15.4 &     0.49 &     18.3 &                \\
 IRS1/{\#}1 & 19 23 41.7 &+14 30 51.9 &  3.84 &     78.7 &    54.6\tablenotemark{n} &  4.61 &   613 &   499\tablenotemark{n} &                \\
 IRS1/{\#}2 & 19 23 41.9 &+14 30 56.2 &  3.07 &     16.3 &     12.8 &  4.61 &      403 &   275\tablenotemark{n} &            e5  \\
 IRS1/{\#}3 & 19 23 37.9 &+14 29 59.4 &  3.84 &    $<$0.14 &    \nodata &  3.84 &    21.3 &    10.1 &                \\
 IRS1/{\#}4 & 19 23 37.6 &+14 30 21.1 &  3.84 &    $<$0.14 &    \nodata &  3.84 &     5.08 &     3.62 &                \\
 IRS1/{\#}5 & 19 23 37.3 &+14 30 10.8 &  3.84 &    $<$0.14 &    \nodata &  3.84 &     2.09 &     0.83 &                \\
 IRS1/{\#}6 & 19 23 41.0 &+14 30 43.6 &  3.84 &     18.1 &     9.17 &  3.84 &      121 &     49.1 &                \\
 IRS1/{\#}7 & 19 23 45.2 &+14 31 14.2 &  8.45 &     21.1 &     9.12 &  8.45 &      144 &     35.7 &      $\sim$20$\arcsec\times$14$\arcsec$          \\
 IRS1/{\#}8 & 19 23 45.9 &+14 30 30.3 &  6.91 &     21.6 &     11.8 &  9.98 &      195 &      135 &   e11d, bubble \\
 IRS1/{\#}9 & 19 23 41.8 &+14 30 35.6 &  5.38 &      308 &      234 &  5.38 &     1030 &      574 &                \\
IRS1/{\#}10 & 19 23 44.5 &+14 31 28.1 &  9.98 &     23.8 &     6.11 &  10.8 &      204 &     38.0 &                \\
IRS1/{\#}11 & 19 23 54.0 &+14 28 25.1 &  3.07 &     1.50 &     0.36 &  3.84 &     0.71 &     3.60 &                \\
 IRS2/{\#}1 & 19 23 40.5 &+14 31 05.0 &  3.07 &      120 &    90.6\tablenotemark{n} &  3.84 &  1280 &  1060\tablenotemark{n} &        KJD\,7  \\
 IRS2/{\#}2 & 19 23 40.6 &+14 30 59.9 &  3.07 &     22.5 &     13.9 &  3.84 &      394 &   298\tablenotemark{n} &        KJD\,8  \\
 IRS2/{\#}3 & 19 23 40.9 &+14 31 06.0 &  3.07 &     24.8 &     13.8 &  3.84 &      260 &   146\tablenotemark{n} &     d7, KJD\,9 \\
 IRS2/{\#}4 & 19 23 41.0 &+14 31 03.0 &  3.07 &     15.1 &     10.3 &  3.84 &      185 &    89.3\tablenotemark{n} &       KJD\,10  \\
 IRS2/{\#}5 & 19 23 40.3 &+14 31 10.7 &  3.07 &      141 &   125\tablenotemark{n} &  3.84 &  1340 &  1080\tablenotemark{n} &                \\
 IRS2/{\#}6 & 19 23 38.3 &+14 31 11.5 &  3.07 &     2.31 &     0.31 &  3.84 &     $<$20.4\tablenotemark{u} &     \nodata &   SHA17\,17    \\
 IRS2/{\#}7 & 19 23 37.8 &+14 31 20.1 &  4.61 &     3.96 &     2.54 &  4.61 &     29.1 &     16.2 &                \\
 IRS2/{\#}8 & 19 23 37.3 &+14 31 16.5 &  3.07 &     1.79 &     1.04 &  3.07 &     7.61 &     2.12 &                \\
 IRS2/{\#}9 & 19 23 36.7 &+14 31 15.9 &  4.61 &     5.44 &     4.45 &  4.61 &     17.2 &     9.48 &                \\
IRS2/{\#}10 & 19 23 40.5 &+14 31 16.9 &  3.84 &     21.2 &     13.6 &  3.84 &      235 &   151\tablenotemark{n} &                \\
      IRS2E & 19 23 40.2 &+14 31 05.9 &  3.84 &      817 &   806\tablenotemark{n} &  4.61 &  4350 &  4220\tablenotemark{n} &                \\
      IRS2W & 19 23 39.9 &+14 31 06.6 &  3.84 &      824 &   811\tablenotemark{n} &  4.61 &  3880 &  3800\tablenotemark{n} &                \\
       IRS3 & 19 23 43.2 &+14 30 50.2 &  3.84 &     52.5 &     14.7 &  4.61 &      322 &     59.5 &                \\
       IRS4 & 19 23 46.3 &+14 29 43.3 &  6.91 &     48.5 &     35.4 &  9.22 &      551 &      454 &  e16,e18,e18d  \\
        LS1 & 19 23 47.8 &+14 36 38.4 &  3.84 &     0.14 &     0.15 &  3.84 &   $<$0.33 &   \nodata &   LBV candidate  \\
\enddata
\tablecomments{\scriptsize R.A. and Dec. are for the center of apertures. $F_{\rm int}$ indicates total flux inside the aperture. Values preceded by a ``$<$'' denote a 3-$\sigma$ upper limit.}
\tablenotetext{n}{The peak at this wavelength is not well-resolved from nearby sources or extended emission which likely affects the accuracy of the background subtracted photometry.}
\tablenotetext{u}{The $F_{\rm int}$ value is used as the upper limit since the source is highly contaminated by extended source G49.5-0.4\,d that makes difficult to determine the source flux.}
\label{tb:cps1}
\end{deluxetable*}

In addition to using the photometry from the SOFIA data, we performed multi-band aperture photometry on the \textit{Spitzer} IRAC 3.6, 4.5, 5.8, 8.0\,$\mu$m and \textit{Herschel} PACS 70 and 160\,$\mu$m image data for W\,51\,A to create the NIR to FIR spectral energy distributions (SEDs) of the identified sub-components and point sources ( see Appendix \ref{appendix} for \textit{Spitzer} and \textit{Herschel} photometry). Since the \textit{Spitzer} IRAC images at all four wavelengths are completely saturated at the location of IRS\,2\,E and partially saturated at IRS\,2\,W, for these two sources we added to our SED data Two Micron All Sky Survey \citep[2MASS][]{2006AJ....131.1163S} $K_{\rm s}$ band ($\lambda$=2.159\,$\mu$m) measured photometry values (using a 4$\farcs$0 aperture size). While the $K_{\rm s}$ band is not ideal in general for these fits because of the potential of contamination due to scattered emission and bright line emission, particularly from H$_2$ and CO ($\nu$=2--0) transitions, having an unsaturated data point at wavelengths shorter than those from \textit{SOFIA} will at least provide some constraint to the SED fits at NIR wavelengths for IRS\,2\,E and IRS\,2\,W. 

 The infrared positions and aperture sizes that were used for photometry of the sub-components and point sources were determined using the FORCAST 20 and 37\,$\mu$m images and employing an optimal extraction method \citep{1998MNRAS.296..339N} that measures the radial intensity profile of each sub-component and determines the radial angular distance at which the intensity profile starts to be flat. For each source, we chose the angular distance between the center of the source and the `turn over' point as the radius of the aperture. We then determined the background value from an annulus outside the aperture radius that shows relatively flat profile and is as close as possible to the inner aperture. However, in order to minimize contamination from extended emission and/or nearby sources, the location and sizes of the chosen background annuli differ for each source. 

While the flux error in the flux calibration factor (Jy/ADU) of the FORCAST data is quite small ($<$15\%), the backgrounds around sources can be quite large and variable (i.e. not flat under the source), the fluxes obtained through background subtraction can carry a larger uncertainty.  Since the upper limit uncertainty on the flux cannot be significantly larger than the background amount we subtracted, we set the upper error bar as the background flux value. The lower error bar values for all sources come from the average total photometric error at each wavelength (as discussed in Section 2) which are set to be the estimated photometric errors of 20, 15 and 10\,\% for 4.5, 20 and 37\,$\mu$m bands, respectively.  In the few cases where the background around a source is negative (see discussion of data issues in \S\,\ref{sec:obs}), the errors in photometry are handled in the opposite manner as above, i.e. the background value is used as the lower error bar, and the average total photometric error is used as the upper error bar.

\begin{figure}[htb!]
\figurenum{12}
\epsscale{1.13}
\plotone{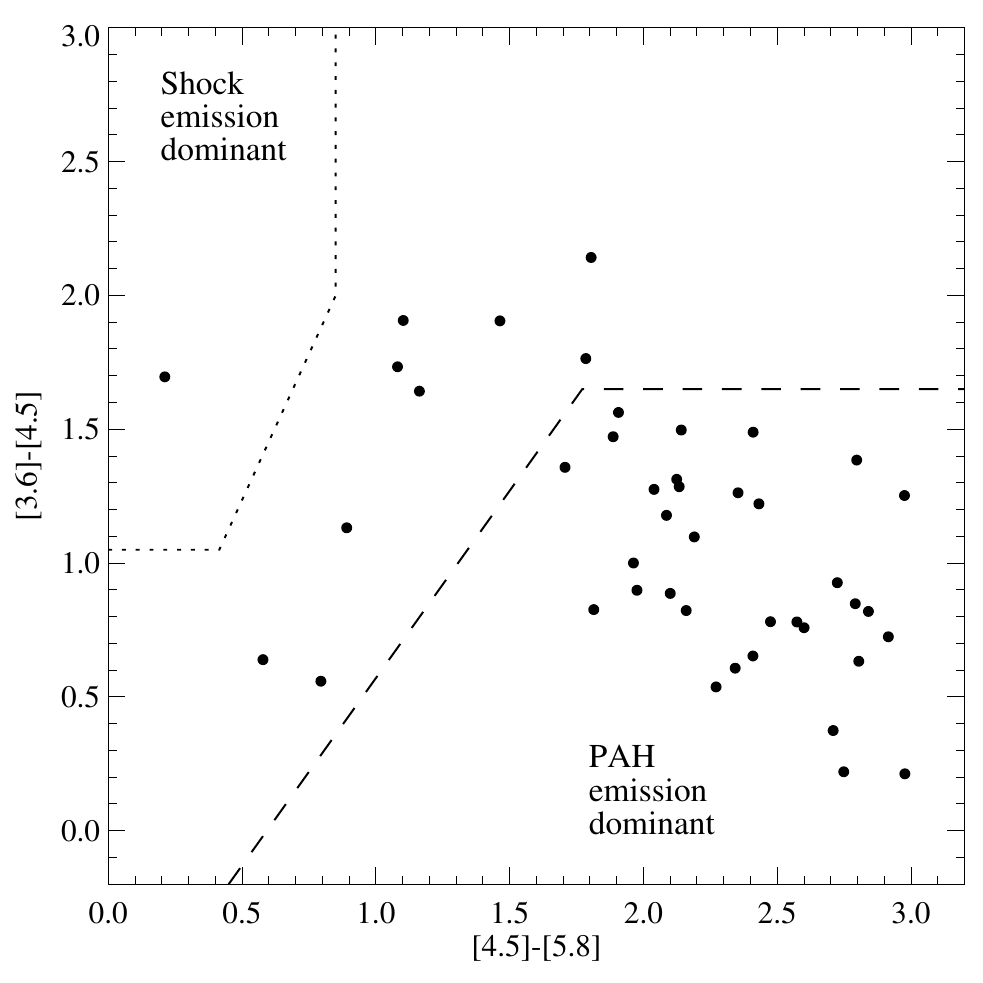}
\caption{A color-color diagram utilizing our background-subtracted \textit{Spitzer} IRAC  3.6, 4.5, 5.8 and 8\,$\mu$m source photometry to distinguish ``shocked emission dominant'' and ``PAH emission dominant'' YSO candidates from our list of sub-components and point sources. Above (up-left) of dotted line indicates shock emission dominant regime. Below (bottom-right) dashed line indicates PAH dominant regime. We adopt this metric from \citet{2009ApJS..184...18G}.}
\label{fig:ccd}
\end{figure}

 One problem with using \textit{Spitzer} IRAC data for MYSO SED model fitting is that the 3.6, 5.8 and 8\,$\mu$m fluxes can be contaminated by PAH emission \citep{2001ApJ...548L..73H,2007ApJ...657..810D}, and the 4.5\,$\mu$m fluxes can be contaminated by shock-excited H$_2$ emission \citep{2006AJ....131.1479R}. Figure\,\ref{fig:ccd} shows a simple color-color diagram ([3.6]-[4.5] vs. [4.5]-[5.8]) method which can be used to determine if sources are highly contaminated by shock emission and/or PAH emission \citep{2009ApJS..184...18G} based on analytic estimation of the emission line contribution to the \textit{Spitzer}-IRAC bands \citep{2006AJ....131.1479R}. This analysis used the measured background subtracted IRAC band fluxes for each source (see Table\,\ref{tb:cps2}), so that we could determine which \textit{Spitzer} IRAC data would be the least contaminated in order to create accurate SEDs for our sources. 

We found that, out of the 43 sub-components and point sources plotted in Figure\,\ref{fig:ccd}, only one source, IRS1/{\#}3, can be categorized as a ``shock emission dominant'' source. Note that IRS1/{\#}3 shares a location with OH masers \citep{2000ApJS..129..159A}, which are shock-excited, supporting the idea that IRS1/{\#}3 is a massive YSO generating shocks. Therefore, in our SED for IRS1/{\#}3 we set IRAC 4.5\,$\mu$m data point as and upper limits due to shock emission. We also set IRAC 3.6, 5.8 and 8\,$\mu$m data points as upper limits since we do not know how the PAH emission affects shock emission dominant sources \citep{2008AJ....136.2391C}. We further find that the vast majority of our sources plotted in Figure\,\ref{fig:ccd}, 33 out of the 43, can be identified as ``PAH emission dominant'' sources, so we set the IRAC 3.6, 5.8 and 8\,$\mu$m fluxes in the SEDs of these sources as upper limits. Hence, only the IRAC flux values trusted in the SED fits for these sources is the uncontaminated 4.5\,$\mu$m values. There are 9 sources in Figure\,\ref{fig:ccd} that appear to not be contaminated by shock and/or PAH emission. Thus, we use the fluxes from all IRAC bands for these 9 sources as nominal data points in their SEDs, assigning them a total photometric error of 20\,\%.

There are some sources missing from the analysis in Figure\,\ref{fig:ccd}. Two sources, IRS1/{\#}4 and IRS1/{\#}5, could not be included in the color-color diagram due to non-detections at 5.8\,$\mu$m. For these sources, we simply treat them as average sources, i.e. ``PAH emission dominant'' with IRAC 3.6, 5.8 and 8\,$\mu$m fluxes as upper limits. Furthermore, IRS\,2\,E and IRS\,2\,W  are saturated in all four IRAC bands, and thus could not be included in the color-color diagram. Therefore, in the SEDs we set all four IRAC band fluxes for IRS\,2\,E and IRS\,2\,W as lower limits.

In the SEDs for all sources, the \textit{Herschel} 70 and 160\,$\mu$m band fluxes are also set as upper limits since their poorer angular resolution ($\sim$10$\arcsec$) would include high levels of contamination from extended nearby sources. We also set \textit{Spitzer} 8\,$\mu$m band fluxes of IRS2/{\#}1 and IRS2/{\#}5 as lower limits due to partial saturation. Both lower and upper limits utilize the band flux before background subtraction, $F_{\rm int}$. Additionally, IRS1/{\#}3, IRS1/{\#}4 and IRS1/{\#}5 are not detected in the FORCAST 20\,$\mu$m image, so we set a 3-$\sigma$ upper limit for these three sources at 20\,$\mu$m. The \textit{SOFIA} 37\,$\mu$m fluxes of d4e+d4w and IRS2/{\#}6 are set as upper limits since the strong 37\,$\mu$m extended emission from IRS2 make difficult to distinguish the relatively weak emission from  d4e+d4w and IRS2/{\#}6.

The next step in determining whether the infrared sources are MYSOs is to use the photometry data for each source and investigate whether they could be fit with theoretical MYSO SED models. We consider the Turbulent Core Accretion model of massive star formation \citep{2003ApJ...585..850M} as the fiducial models for this study since, 1) W\,51\,A is an active massive star forming region, and 2) MIR-revealed sources that were not detected in optical and NIR regimes are likely deeply embedded objects, i.e. presumed to be in the early stages of massive star formation development. \citet{2011ApJ...733...55Z} developed an IDL SED fitter program based on the Core Accretion model. In a series of papers \citep{2013ApJ...766...86Z,2014ApJ...788..166Z,2018ApJ...853...18Z}, the detailed physical mechanisms of the Core Accretion models and effects of different conditions (e.g. foreground extinction, inclination of rotational axis, and outflow opening angles) toward observed MYSO SEDs was investigated (hereafter, we call these ZT models). This SED fitter estimates the intrinsic SEDs of YSOs by correcting foreground extinction and inclination angle. It then finds the best model fits that match those SEDs employing a $\chi^2$-minimization method  that is normalized by the number of nominal data points (i.e. neither upper nor lower limits). The $\chi^2$ values derived from fits to only nominal data points are called $\chi^2_{\rm nonlimit}$ in the ZT model fitter. \citet{2018ApJ...853...18Z} describe that for the same observed SED, the number of nominal data points is dependent on the model SED being fit. If a data point is being used as an upper limit and the model SED is higher than that data point, it is counted in number of nominal data points. If the model SED is lower than that data point, it is not counted in in number of nominal data points, since it is not constraining the fit. Consequently, ``$\chi^2_{\rm nonlimit}$ is a measurement of the average deviation of the model SED from the constraining data points'' \citep{2018ApJ...853...18Z}.

By plotting a histogram of the $\chi^2_{\rm nonlimit}$ values of the model fits for each source, we determine a group of best fit models that all have values similar to the lowest value and are distinguishable as a group from next group of models showing consistent yet significantly larger $\chi^2_{\rm nonlimit}$ values. The number of the best fit models found via this $\chi^2_{\rm nonlimit}$ method varies from source to source and are given in Table\,\ref{tb:sedp}. Note that the $\chi^2_{\rm nonlimit}$ values can only be utilized for relative comparison of the goodness of fit. The usage of absolute $\chi^2_{\rm nonlimit}$ values to determine good fits (e.g. $\chi^2_{\rm nonlimit}$-$\chi^2_{\rm nonlimit,min}\lesssim$3) is not recommended by various authors of SED model fitters \citep{2006ApJS..167..256R,2018ApJ...853...18Z}.

\begin{figure*}[htb!]
\figurenum{13}
\epsscale{1.13}
\plotone{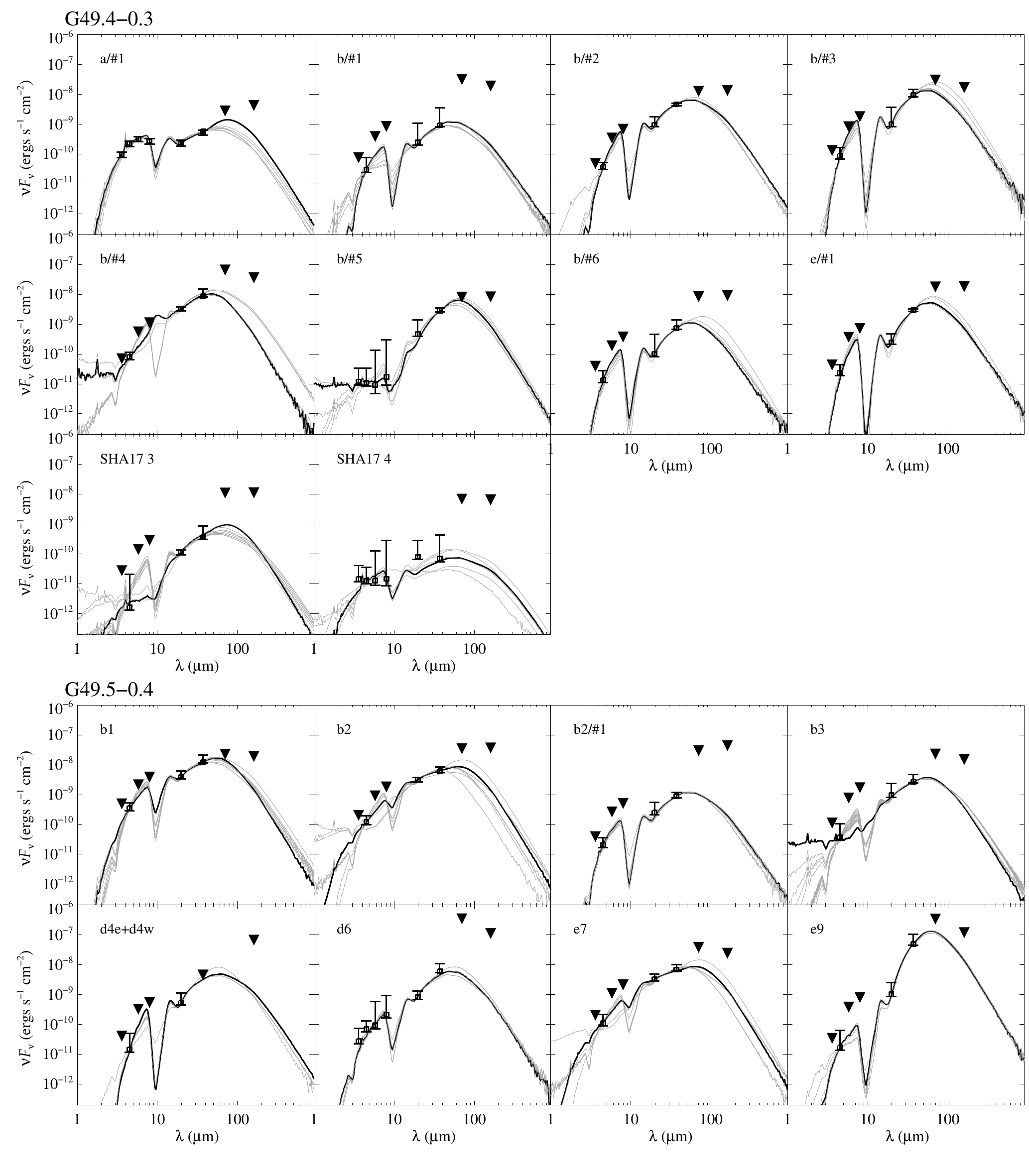}
\caption{SED fitting with ZT model for sub-components and point sources in W\,51\,A. For each source, the absolute best fit model (i.e. lowest $\chi^2_{\rm nonlimit}$ value) is shown in black, and the rest of the best fit models are shown in gray.}
\label{fig:fd1}
\end{figure*}

\begin{figure*}[htb!]
\figurenum{13}
\epsscale{1.13}
\plotone{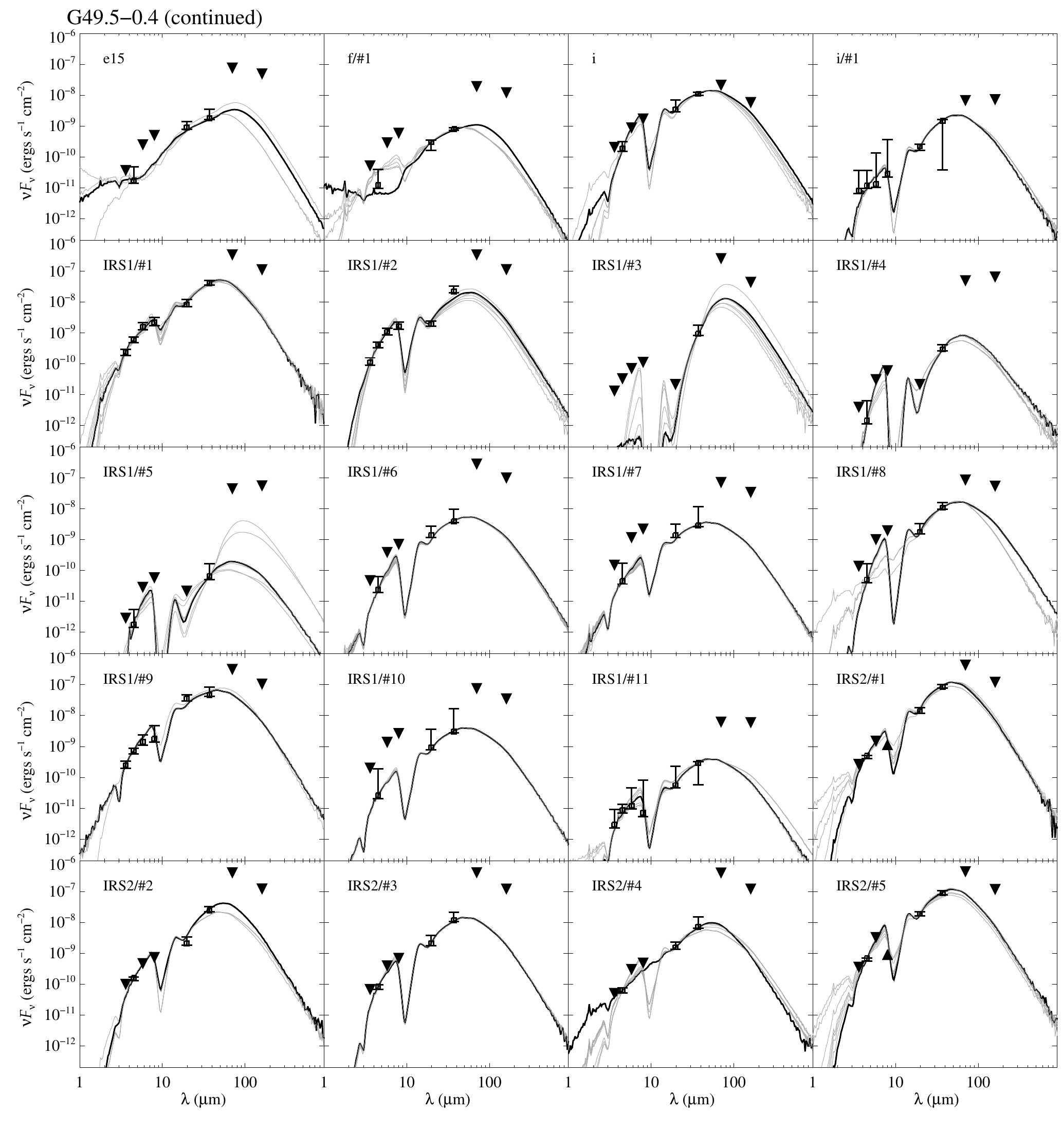}
\caption{\textit{Continued.}}
\label{fig:fd2}
\end{figure*}

\begin{figure*}[htb!]
\figurenum{13}
\epsscale{1.13}
\plotone{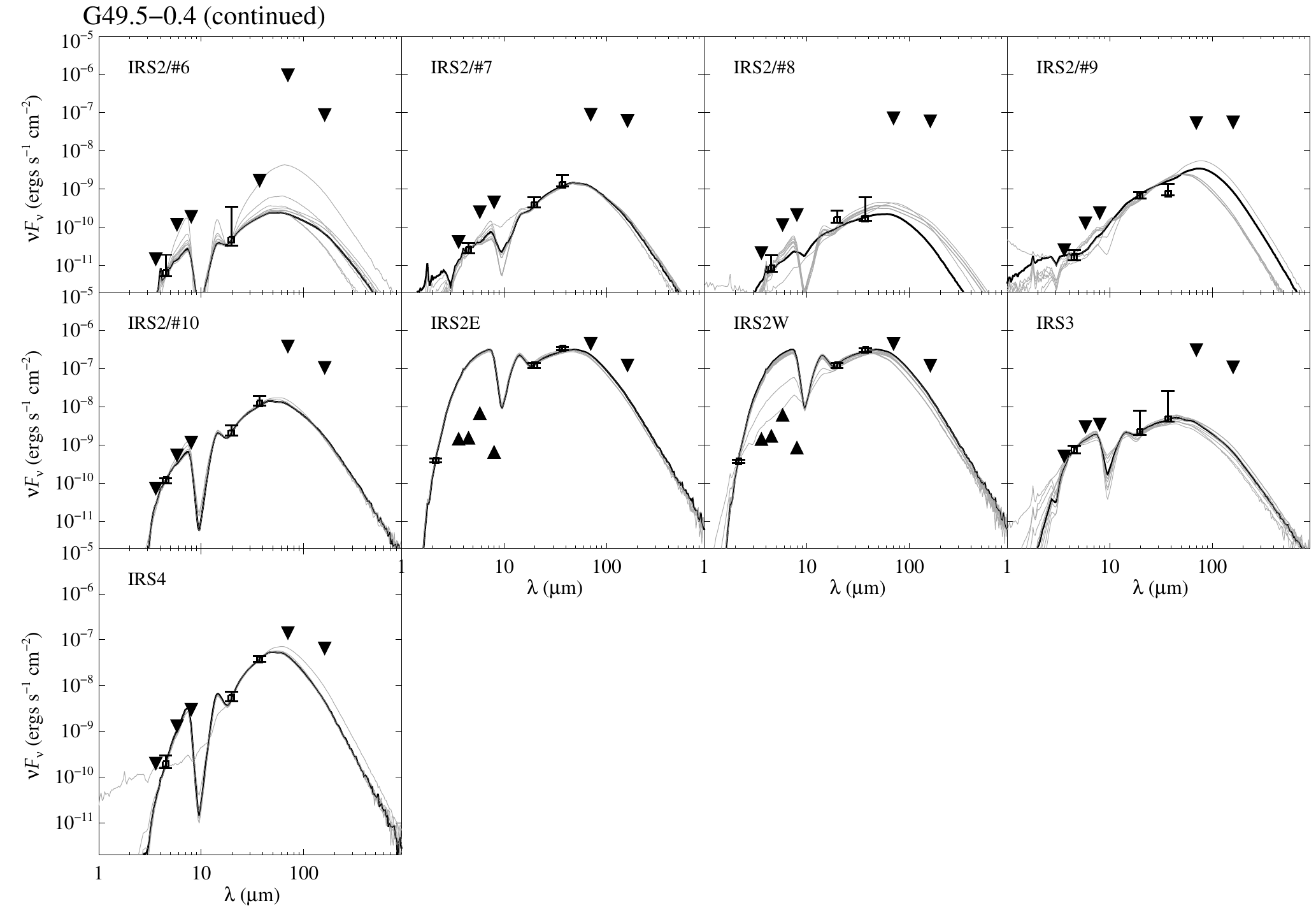}
\caption{\textit{Continued.}}
\label{fig:fd2b}
\end{figure*}

Figure\,\ref{fig:fd1} shows the photometry data as a function of wavelength and the ZT model fits to those SED data for the sources in G49.4-0.3 and G49.5-0.4. Table\,\ref{tb:sedp} lists physical parameters of the model fits for all sources. The column\,2 of Table\,\ref{tb:sedp} is the observed bolometric luminosity of the absolute best fit model (i.e. the model with the lowest $\chi^2_{\rm nonlimit}$ value), $L_{\rm obs}$, and column\,3 is the true total bolometric luminosity, $L_{\rm tot}$, which corrects for foreground extinction and disk inclination angle.  The absolute best model fit foreground extinction and stellar mass are shown in column\,4, and column\,5, respectively. The number of best fit models for each source is given in column\,7 (these models are plotted as gray lines for each source in Figure\,\ref{fig:fd1}), and the range of extinction values and the range of stellar mass values derived from that group of best fit models is given in column\,8 and column\,9, respectively. Column\,6 shows the spectral types of the YSOs derived from the best fit stellar masses,  comparing them to the masses of Zero Age Main Sequence (ZAMS) stars \citep{2000AJ....119.1860B}. It is important to point out that the ZT models assume a single YSO within a core.  Given the distance to W\,51\,A \citep[5.4\,kpc;][]{2010ApJ...720.1055S} and the angular resolution limits of FORCAST ($\sim$3$\arcsec$), we are only able to resolve structures as small as $\sim$0.08\,pc. It is likely, therefore, given the high multiplicity fraction of massive stars \citep[e.g.][]{2009AJ....137.3358M} that in many cases the IR sources discussed here contain proto-binaries or even proto-clusters. Though the assumption of a single YSO can be reasonable when the core contains a dominant primary YSO while other companions are relatively low-mass, we cannot be certain that this would be the case in general. Consequently, even though the ZT model fits provide more output parameters than luminosity and mass, those derived parameters are likely not meaningful given their assumption of a singe central star. The luminosity and mass parameters, at a minimum, inform us as to the likelihood of an IR source in our sample as being a massive YSO or not, which is our main interest for performing the fitting.  The multiplicity can also affect the derived extinction but one can find the range of extinction we obtain among nearby sources agree reasonably well (Table\,\ref{tb:sedp}).

From the SEDs shown in Figure\,\ref{fig:fd1} one can see that the Herschel 70 and 160\,$\mu$m flux points, which we use as upper limits, are in most cases much higher than the fitted SEDs curves at those wavelengths. In these cases, if the Herschel photometry values were instead used in the fit (i.e. not as upper limits) the SED fitter would not be able to fit both the Herschel and \textit{SOFIA}-FORCAST data points due to such a large, discontinuous jump in flux from 37 to 70\,$\mu$m. The Herschel 70 and 160\,$\mu$m data (and certainly the SPIRE 250, 300, and 500\,$\mu$m data) are too coarse in resolution, and combined with the likelihood of contamination from cold dust from other nearby sources, the Herschel photometry is only useful as upper limits. This shows the importance of the SOFIA data (especially 37\,$\mu$m) in helping to define accurate SEDs for these sources, which in turn allows us to get a more accurate understanding of their true nature.

Looking at the results in Table\,\ref{tb:sedp}, the absolute best model fits for the mid-infrared detected YSO candidates in the all of W\,51\,A yield protostellar masses in the range  $m_*$\,=\,1--96\,$M_{\sun}$, which is approximately equivalent to a range of ZAMS spectral type G5--O3 stars. Note that ZT models have sampled protostellar mass at at 0.5, 1, 2, 4, 8, 12, 16, 24, 32, 48, 64, 96, 128, 160\,$M_{\sun}$ thus there is a minimum mass granularity that can be explored with the models \citep{2018ApJ...853...18Z}. The most massive sources in W\,51\,A are, perhaps unsurprisingly, in the IRS\,1 and IRS\,2 regions; they are G49.5-0.4\,IRS\,1/\#1, G49.5-0.4\,IRS\,2/\#1, and G49.5-0.4\,IRS\,5/\#5. All three are best fit with a stellar mass of  96$M_{\sun}$, or the equivalent of spectral type O3 ZAMS star. 

\begin{deluxetable*}{rccccccrclrcll}
\tabletypesize{\scriptsize}
\tablecolumns{8}
\tablewidth{0pt}
\tablecaption{SED Fitting Parameters of All Sub-components and Point Sources in W51A}
\tablehead{\colhead{   Source   }                                              &
           \colhead{  $L_{\rm obs}$   } &
           \colhead{  $L_{\rm tot}$   } &
           \colhead{ $A_v$ } &
           \colhead{  $M_{\rm star}$  } &
           \colhead{ Spectral } &
           \colhead{ Best } &
           \multicolumn{3}{c}{$A_v$ Range}&
           \multicolumn{3}{c}{$M_{\rm star}$ Range}&
           \colhead{ Note }\\
	   \colhead{        } &
	   \colhead{ ($\times 10^3 L_{\sun}$) } &
	   \colhead{ ($\times 10^3 L_{\sun}$) } &
	   \colhead{ (mag.) } &
	   \colhead{ ($M_{\sun}$) } &
	   \colhead{ Type\tablenotemark{a} } &
       \colhead{  Models\tablenotemark{b}   } &
       \multicolumn{3}{c}{(mag.)}&
       \multicolumn{3}{c}{($M_{\sun}$)}&
       \colhead{     }\\
}
\startdata
{\bf G49.4-0.3} & & & & & & & & & & & & & \\
   a/{\#}1 &      2.05 &     1.99 &     50.3 &      4 &       B5 &    7 &  45.0 & - &  75.5 &   4 & - &   4 & \\
   b/{\#}1 &      1.58 &     8.76 &     66.2 &      8 &       B1 &   10 &  13.2 & - &  83.8 &   8 & - &  24 & MYSO, 1.5cm \\
   b/{\#}2 &      8.30 &     18.7 &     55.3 &     12 &       B1 &    5 &  15.9 & - &  71.5 &   8 & - &  24 & MYSO\\
   b/{\#}3 &      15.0 &      617 &      138 &     64 &     O4.5 &    9 &   101 & - &   225 &  12 & - &  64 & MYSO, 1.5cm\\
   b/{\#}4 &      18.1 &     22.4 &     1.70 &     12 &       B1 &    7 &   0.8 & - &  26.5 &  12 & - &  48 & MYSO\\
   b/{\#}5 &      6.34 &     12.5 &     25.2 &      8 &       B1 &    6 &   2.7 & - &  67.1 &   8 & - &   8 & MYSO, 1.5cm\\
   b/{\#}6 &      1.27 &     92.9 &      117 &     24 &     O8.5 &    7 &  74.2 & - &   201 &   4 & - &  24 & pMYSO\\
   e/{\#}1 &      5.62 &     88.4 &      143 &     24 &     O8.5 &    5 &   114 & - &   218 &  12 & - &  32 & MYSO\\
   SHA17 3 &      1.14 &     1.99 &     26.5 &      1 &       G5 &   11 &   1.7 & - &  75.5 &   1 & - &  32 & \\
   SHA17 4 &      0.14 &     0.15 &     38.6 &      1 &       G5 &    6 &   5.3 & - &  47.7 & 0.5 & - &  12 & \\
{\bf G49.5-0.4} & & & & & & & & & & & & & \\
        b1 &      21.0 &     47.3 &     50.3 &     12 &       B1 &   10 &  23.8 & - &  63.7 &  12 & - &  32 & MYSO, 6cm\\
        b2 &      12.8 &     16.6 &     26.5 &      8 &       B1 &    9 &   1.7 & - &  53.0 &   8 & - &  48 & MYSO, 6cm\\
  b2/{\#}1 &      1.49 &     8.76 &     71.3 &      8 &       B1 &   10 &  50.3 & - &   101 &   8 & - &  24 & MYSO\\
        b3 &      4.55 &     9.67 &     27.7 &      8 &       B1 &   18 &   2.7 & - &  41.9 &   8 & - &  12 & MYSO, 6cm\\
   d4e+d4w &      6.21 &     9.45 &     47.0 &      8 &       B1 &    5 &   7.9 & - &  49.5 &   8 & - &  24 & MYSO, 2cm (cCWB)\\
        d6 &      6.79 &      158 &     45.0 &     32 &       O7 &    8 &  42.4 & - &  75.5 &  24 & - &  32 & MYSO, 2cm (cCWB)\\
        e7 &      12.8 &     16.6 &     26.5 &      8 &       B1 &    7 &   1.7 & - &  53.0 &   8 & - &  16 & MYSO, 2cm (UC\ion{H}{2})\\
        e9 &       118 &      528 &      101 &     48 &     O5.5 &    6 &  25.2 & - &   151 &  24 & - &  96 & MYSO, 2cm (HC\ion{H}{2})\\
       e15 &      3.01 &     13.3 &     26.5 &      8 &       B1 &    4 &   8.4 & - &  28.5 &   8 & - &  16 & MYSO, 2cm (UC\ion{H}{2})\\
   f/{\#}1 &      1.22 &     13.6 &     14.3 &     12 &       B1 &    6 &   5.3 & - &  25.2 &  12 & - &  16 & MYSO\\
         i &      20.0 &     22.9 &     10.9 &     12 &       B1 &    6 &   3.3 & - &  78.0 &  12 & - &  32 & MYSO, 2cm
\\
   i/{\#}1 &      2.57 &     36.0 &     45.0 &     16 &       B1 &    8 &  45.0 & - &  75.5 &   8 & - &  16 & MYSO\\
IRS1/{\#}1 &      58.3 &     1410 &     33.5 &     96 &       O3 &   10 &  13.2 & - &  58.7 &  16 & - & 128 & MYSO\\
IRS1/{\#}2 &      24.1 &     50.3 &     71.5 &     16 &       B1 &    7 &  71.5 & - &   126 &  12 & - &  32 & MYSO, 2cm (HC\ion{H}{2})\\
IRS1/{\#}3 &      6.07 &     9.17 &      196 &      8 &       B1 &    7 &   184 & - &   212 &   4 & - &   8 & pMYSO\\
IRS1/{\#}4 &      0.84 &     92.9 &      352 &     24 &     O8.5 &    9 &   127 & - &   361 &   8 & - &  24 & MYSO\\
IRS1/{\#}5 &      0.23 &     1.06 &      233 &      4 &       B5 &    7 &  54.5 & - &   260 &   2 & - &   4 & \\
IRS1/{\#}6 &      7.30 &     9.45 &     8.40 &      8 &       B1 &   18 &   0.8 & - &  14.3 &   8 & - &   8 & MYSO\\
IRS1/{\#}7 &      4.95 &     9.67 &     10.9 &      8 &       B1 &   11 &   3.3 & - &  22.6 &   8 & - &   8 & MYSO\\
IRS1/{\#}8 &      20.6 &     37.7 &     58.7 &     16 &       B1 &    7 &   2.7 & - &  61.2 &  12 & - &  48 & MYSO, 2cm (\ion{H}{2})\\
IRS1/{\#}9 &      85.2 &      161 &     3.40 &     32 &       O7 &    6 &   3.3 & - &  53.0 &  24 & - &  32 & MYSO, 2cm\\
IRS1/{\#}10 &      5.17 &     9.95 &     21.0 &      8 &       B1 &    8 &  20.1 & - &  27.7 &   8 & - &   8 & MYSO\\
IRS1/{\#}11 &      0.49 &     6.29 &     67.1 &      8 &       B1 &    8 &   3.3 & - &  72.9 &   4 & - &   8 & pMYSO\\
IRS2/{\#}1 &      127 &      1314 &     39.7 &     96 &       O3 &    6 &  2.7& - &  67.1 &  48 & - &  96 & MYSO, 3.5cm\\
IRS2/{\#}2 &      43.4 &      732 &     75.5 &     64 &     O4.5 &    5 &  29.1 & - &  75.5 &  32 & - &  64 & MYSO\\
IRS2/{\#}3 &      17.1 &     80.6 &     65.4 &     24 &     O8.5 &    7 &  65.4 & - &  76.3 &  24 & - &  24 & MYSO, 2cm (cCWB)\\
IRS2/{\#}4 &      10.8 &      196 &     8.40 &     32 &       O7 &   10 &   8.4 & - &  49.5 &  12 & - &  32 & MYSO\\
IRS2/{\#}5 &       133 &     1310 &     60.9 &     96 &       O3 &    8 &   7.9 & - &  60.9 &  48 & - &  96 & MYSO, 3.5cm\\
IRS2/{\#}6 &      0.35 &     0.77 &     21.8 &      4 &       B5 &   11 &  16.8 & - &   246 &   4 & - &  32 & \\
IRS2/{\#}7 &      1.83 &     19.6 &     5.30 &     12 &       B1 &   10 &   5.3 & - &  41.9 &   8 & - &  24 & MYSO\\
IRS2/{\#}8 &      0.35 &     1.87 &     26.5 &      4 &       B5 &   10 &   3.3 & - &  39.4 &   4 & - &   4 & \\
IRS2/{\#}9 &      3.01 &     13.3 &     26.5 &      8 &       B1 &    8 &   8.4 & - &  79.5 &   4 & - &  24 & pMYSO\\
IRS2/{\#}10 &      16.6 &      151 &     78.8 &     32 &       O7 &    7 &  71.5 & - &   101 &  32 & - &  64 & MYSO\\
     IRS2E &       598 &      841 &     75.5 &     64 &     O4.5 &    6 &  75.5 & - &  75.5 &  64 & - & 128 & MYSO, 3.5cm\\
     IRS2W &       598 &      841 &     75.5 &     64 &     O4.5 &   13 &  25.2 & - &  75.5 &  64 & - & 128 & MYSO, 3.5cm\\
      IRS3 &      8.16 &     30.4 &     40.2 &     16 &       B1 &   11 &  11.7 & - &   132 &   8 & - &  48 & MYSO\\
      IRS4 &      57.7 &      648 &     92.7 &     64 &     O4.5 &    6 &  63.6 & - &   103 &  24 & - &  96 & MYSO, 2cm (\ion{H}{2})\\
\enddata
\tablenotetext{a}{\scriptsize Determined by using the absolute best model fitted YSO mass in column 5 and finding the ZAMS equivalent spectral type from \citet{2000AJ....119.1860B}.}
\tablenotetext{b}{\scriptsize The number of models in the group of best fit models (see section 4.1). These models were used to determine the ranges of $M_{\rm star}$ and $A_v$.}
\tablecomments{\scriptsize A ``MYSO'' in the right column denotes a MYSO candidate. A ``pMYSO'' indicates that their is greater uncertainty in the derived physical parameters and that these sources are possible MYSO candidates. If the source is a point source in cm radio continuum, or at the location of a prominent radio continuum peak, the wavelength of this is given in the right column, along with any previous identification of the nature of the source by \citet{2016AnAp...595A..27G} [HCHII: hypercompact HII region; UCHII: ultracompact HII region; HII: extended HII region; cCWB candidate colliding-wind binary]. }
\label{tb:sedp}
\end{deluxetable*}

IRS\,2E and IRS\,2W, which are the two brightest infrared sources in the IRS\,2 region, along with IRS\,4 are all best fit with models with stellar masses of 64\,$M_{\sun}$, equivalent to O4.5 stars. Again, this is under the assumption of a single central heating source. \citet{2016ApJ...825...54B} distinguished four infrared sources at the position of IRS\,2E which cannot be resolved by our \textit{SOFIA}-FORCAST observations (see \S\,\ref{sec:irs2}). The total mass of the four sources was derived to be 80$M_{\sun}$ in \citet{2016ApJ...825...54B} based on the stellar evolutionary tracks of \citet{1996A&A...307..829B}. This is in agreement with our result for IRS\,2E under the assumption of a single protostar (the best fit models range from 64 to 128\,$M_{\sun}$, with the the absolute best fit being 64\,$M_{\sun}$).

With the physical parameters from the SED fits given in Table\,\ref{tb:sedp}, we can deduce the likelihood of each YSO being massive. If a source has an absolute best fit stellar mass equal to or greater than 8\,$M_{\sun}$, and a minimum mass range value equal to or greater than 8\,$M_{\sun}$, we identify it as a MYSO candidate and label it as a `MYSO' in Table\,\ref{tb:sedp}. If the mid-infrared source is also an isolated cm radio continuum source or coincident with a radio peak, this adds further evidence that the source may be massive and this is also given in Table\,\ref{tb:sedp}. If the absolute best fit stellar mass is equal to or greater than 8\,$M_{\sun}$, but the minimum mass range value is lower than 8\,$M_{\sun}$, we identify the source as a potential MYSO candidate (labeled `pMYSO' in Table\,\ref{tb:sedp}). Overall, we find 41 MYSO and potential MYSO candidates, many identified as such here for the first time.

For sources SHA17\,3 and SHA17\,4, which were previously identified as potential MYSOs \citep{2017ApJ...839..108S}, we find that with the added photometry at longer infrared wavelengths\footnote{Comparing our IRAC photometry (see Table 5) using the optimal extraction technique discussed in \S\,\ref{sec:cps} to that of \citet{2017ApJ...839..108S}, our measurements yield much higher integrated fluxes. This is because \citet{2017ApJ...839..108S} used a fixed radius at 2$\farcs$4 for all point sources, which in all cases smaller than the apertures we employed.  \citet{2017ApJ...839..108S} also estimate the background intensity of each source using an annulus abutting the point source aperture, i.e. the inner and outer radii of the annuli are always 2$\farcs$4 and 7$\farcs$2, respectively, which provides overestimated background intensities if the PSF of the source is bigger than the aperture size and/or if there is contamination from nearby sources of emission.}, the absolute best fits yield masses of only 1\,$M_{\sun}$, however we do have fits in the group of best fits that yield a stellar masses for these sources greater than 8\,$M_{\sun}$.

Roughly half of the MYSO candidates that we have identified (20 of 41) have no detected radio continuum emission. This means that in W\,51\,A half of the population of the presently forming massive stars are likely in a very young state prior to the onset of a hypercompact \ion{H}{2} region \citep{2010ApJ...721..478H} and not observable via radio continuum emission. This demonstrates that the mid-infrared is vital in completing the inventory of the entire population of massive YSOs within W\,51\,A.   

\subsection{Physical Properties of Extended Sources: Kinematic Status and Global History}\label{sec:es}

In this section, we investigate the global evolutionary state of star cluster formation in W\,51\,A by utilizing two different molecular clump evolutionary tracers, the luminosity-to-mass ratio, $L/M$, and virial parameter, $\alpha_{\rm vir}$, toward the radio-defined extended sources. We assume the extended radio sources are star-forming molecular clumps that contain embedded massive young star clusters that are ionizing the extended \ion{H}{2} regions seen in radio continuum. Using the \textit{SOFIA} FORCAST 20 and 37\,$\mu$m mosaics, the central positions and mid-infrared extent of the sources associated with the major radio continuum regions of W\,51\,A were measured  where the central positions agree to within $\sim$10\,$\arcsec$ and the extents typically vary by factor of 2. These regions are listed in Table\,\ref{tb:es} with their total integrated fluxes. These values have been background subtracted, with the background levels determined from regions nearby ($\le2\arcmin$) each source. 

\subsubsection{The Luminosity-to-mass Ratio}\label{sec:lm}

The luminosity-to-mass ratio, $L/M$ is considered as a good tracer of stellar cluster formation and molecular clump evolution where $L$ is the bolometric luminosity of young stellar clusters (or molecular clump) and $M$ is the mass of the cluster. The theoretical study of \citet{2007ApJ...654..304K} showed that the $L/M$ of a massive stellar cluster (1,000\,$M_{\sun}$) had a positive correlation with the age of the cluster, i.e. $L/M$ increases with the evolutionary stage of stellar cluster.  \citet{2013ApJ...779...79M} studied 303 massive molecular clumps defined by HCO$^+$(1-0) emission \citep{2011ApJS..196...12B} in order to constrain physical properties along the complete span of protocluster evolution. They analyzed mid-infrared to sub-mm dust continuum images to derive bolometric luminosity and cold component dust temperature ($T_{\rm c}$) and adopted a mass estimate from HCO$^+$(1-0). They found that $L/M$ of all molecular clumps were in the range from $\sim$0.1 to $\sim$1000 $L_\odot$/$M_\odot$. The $L/M$ of the molecular clumps showed positive correlation with $T_{\rm c}$ as well as MIR surface brightness (as measured in {\it Spitzer}-IRAC data). These results imply that $L/M$ traces star cluster formation and evolution.

We estimated the mass of each extended source by producing a mass surface density ($\Sigma$) map and using the estimated distance of W\,51\,A \citep[5.4\,kpc,][]{2010ApJ...720.1055S}. The pixel-by-pixel $\Sigma$ values were derived via the method investigated in \citet{2016ApJ...829L..19L}. In this method, the optically thin assumption of dust continuum emission is adopted to perform the graybody fit (i.e. modified blackbody fit). We used \textit{Herschel}-PACS 160\,$\mu$m, -SPIRE 250, 350 and 500\,$\mu$m images for the fitting while the PACS 70\,$\mu$m data were excluded since that wavelength could still be optically thick under the conditions encountered in these regions \citep{2015A&A...573A.106G,2014ApJ...780L..29L}. The convolution of 160, 250 and 350\,$\mu$m data to match to the angular resolution of SPIRE 500\,$\mu$m images ($\sim{36}\arcsec$) was performed with the methods introduced by Gordon et al. (2008). We then estimated the diffuse Galactic background emission via Galactic Gaussian (GG) method \citep{2015A&A...573A.106G,2016ApJ...829L..19L} that assumes the Galactic background follows Gaussian profiles along latitude. Each pixel of the background-subtracted flux density maps (160 to 500\,$\mu$m) were treated to derive $\Sigma$ by using the standard graybody equation, 
\begin{equation}
	I_{\rm \nu} \simeq B_{\nu}(T) (1-e^{-\tau_{\nu}}) = B_{\nu}(T) (1-e^{-\Sigma \kappa_{\nu}})
\label{eq:gb}
\end{equation} 
where $I_{\rm \nu}$ is observed intensity of the corresponding band, $B_{\rm \nu}(T)$ is the temperature-based filter-weighted blackbody radiation, $\tau_{\nu}$ is the optical depth, $\Sigma$ is mass surface density, and $\kappa_{\nu}$ is the filter weighted opacity. We adopted the thin ice mantle dust opacity model of \citet{1994A&A...291..943O} and a dust-to-gas mass ratio of 1/142 \citep{2011ApJ...732..100D} to estimate dust opacity, $\kappa_{\nu}$.

The bolometric luminosity of each clump was then derived from the integrated intensities inside the given apertures (Table\,\ref{tb:es}) through the following method. We found that the measured radius of any given extended source is similar at all wavelength bands $\leq$\,160\,$\mu$m, and thus for each source we use the same aperture size for photometry at these wavelengths. We also found that we could use radii of approximately 2.0, 2.5, and 3.0 times the aperture size used at shorter wavelengths to perform the aperture photometry at 250, 350 and 500\,$\mu$m, while we use the 20\,$\mu$m aperture size for all \textit{Spitzer}-IRAC bands. In order to reproduce the intrinsic fluxes of MIR extended sources, which we assume are young stellar clusters embedded in the middle of dense molecular structures, we had to de-redden the observed $F_{\rm \nu,tot}$. The 1-D radiative transfer equation of absorption, $F_{\rm \nu,tot,1} \simeq F_{\rm \nu,tot,0}^{-\tau}$, was used to determine the intrinsic intensity, $F_{\rm \nu,tot,0}$, where $F_{\rm \nu,tot,1}$ is the observed fluxes (i.e., with extinction). Here we use the median mass surface density value, $\tilde{\Sigma}$, of each extended source to derive the optical depth, $\tau_{\nu}$. Checking these values against those of previous studies \citep[e.g.][]{2010ApJS..190...58K}, we find that our derived optical depth values are in agreement to within a factor of 2. We make the simple assumption that the young clusters are embedded at the center of the molecular clumps so that the dust structures in front and back of the cluster are symmetric along the line of sight. Therefore, since we assumed the material to be optically thin, and since the $\Sigma$ values are derived for the total column density along the line of sight, we divided $\tilde{\Sigma}$ by 2 to de-redden only the foreground extinction of the cluster so that $\tau_{\nu}$\,=\,1/2\,$\kappa_{\nu}$\,$\tilde{\Sigma}$. In addition to the photometric uncertainty levels of each band (\S\,\ref{sec:cps}), one needs to also consider the de-reddening effect, the contribution of different temperature components, and nearby source contamination as additional errors. With all these aspects, we assume $\sim$\,30$\%$ total uncertainty for 4.5\,$\mu$m, $\sim$\,40$\%$ for 20 and 37\,$\mu$m, and $\sim$\,50$\%$ for 70 and 160\,$\mu$m. The 3.5,  5.8 and 8\,$\mu$m bands are treated as upper limits due the the expectation of high PAH contributions. The 250, 350 and 500\,$\mu$m bands are also assumed to be upper limits due to the coarse angular resolution and possible contamination from extended emission of nearby sources. 

\begin{deluxetable}{rrrrrrrr}
\centering
\tabletypesize{\scriptsize}
\tablecolumns{8}
\tablewidth{0pt}
\tablecaption{Observational Parameters of Extended Sources in W\,51\,A}
\tablehead{\colhead{   }&
           \colhead{   }&
           \colhead{   }&
           \multicolumn{2}{c}{${\rm 20\mu{m}}$}&
           \multicolumn{2}{c}{${\rm 37\mu{m}}$}\\
           \cline{4-5} \cline{6-7} \\
           \colhead{  Source  }&
           \colhead{   R.A.  } &
           \colhead{   Dec.  } &
           \colhead{ $R_{\rm tot}$ } &
           \colhead{      $F_{\rm tot}$      } &
           \colhead{ $R_{\rm tot}$ } &
           \colhead{      $F_{\rm tot}$      } \\
	   \colhead{  } &
	   \colhead{  } &
	   \colhead{  } &
	   \colhead{ ($\arcsec$) } &
	   \colhead{ (Jy) } &
	   \colhead{ ($\arcsec$) } &
	   \colhead{ (Jy) } &
}
\startdata
{\bf G49.4-0.3} & & & & & & \\
         a & 19 23 05.5 &+14 28 09.6 & 44.0      &    384         & 53.3      &   2040       \\
         b & 19 23 13.0 &+14 27 09.4 & 72.2      &    1020        & 72.2      &   5680       \\
         c & 19 23 17.5 &+14 29 15.8 & 48.0      &    465         & 56.5      &   2300       \\
         e & 19 23 09.2 &+14 26 02.0 & 13.8      &     22.9       & 16.8      &    246       \\
         f & 19 23 16.2 &+14 24 16.9 & 31.7      &    185         & 35.9      &    1040       \\
{\bf G49.5-0.4} & & & & & & \\
         a & 19 23 29.5 &+14 31 35.6 & 30.6      &    367         & 30.63      &   1340       \\
         b & 19 23 33.3 &+14 29 59.6 & 42.8      &    607         & 42.8      &   2120      \\
         c & 19 23 39.2 &+14 29 35.7 & 44.1      &    660         & 44.1      &   3110       \\
         d & 19 23 40.1 &+14 31 05.8 & 22.6      &   2240         & 40.5      &  17700       \\
         e & 19 23 44.8 &+14 30 26.8 & 59.5      &   3540         & 59.5      &  18200       \\
         f & 19 23 48.5 &+14 33 18.3 & 29.5      &    333         & 33.3      &    1130       \\
         g & 19 23 50.8 &+14 32 52.5 & 25.8      &    411         & 28.6      &    1150       \\
         h & 19 23 54.2 &+14 35 42.8 & 35.0      &    338         & 35.0      &    1110      \\
         i & 19 23 39.2 &+14 35 29.5 & 16.6      &     88.9       & 19.7      &    336       \\
         j & 19 23 47.7 &+14 36 44.0 & 63.9      &    394         & 63.9      &   2200       \\
\enddata
\tablecomments{\footnotesize R.A. and Dec. are for the center of the apertures which have radii defined by \textit{$R_{\rm int}$}.}
\label{tb:es}
\end{deluxetable}

\begin{figure}[t]
\figurenum{14}
\epsscale{1.1}
\plotone{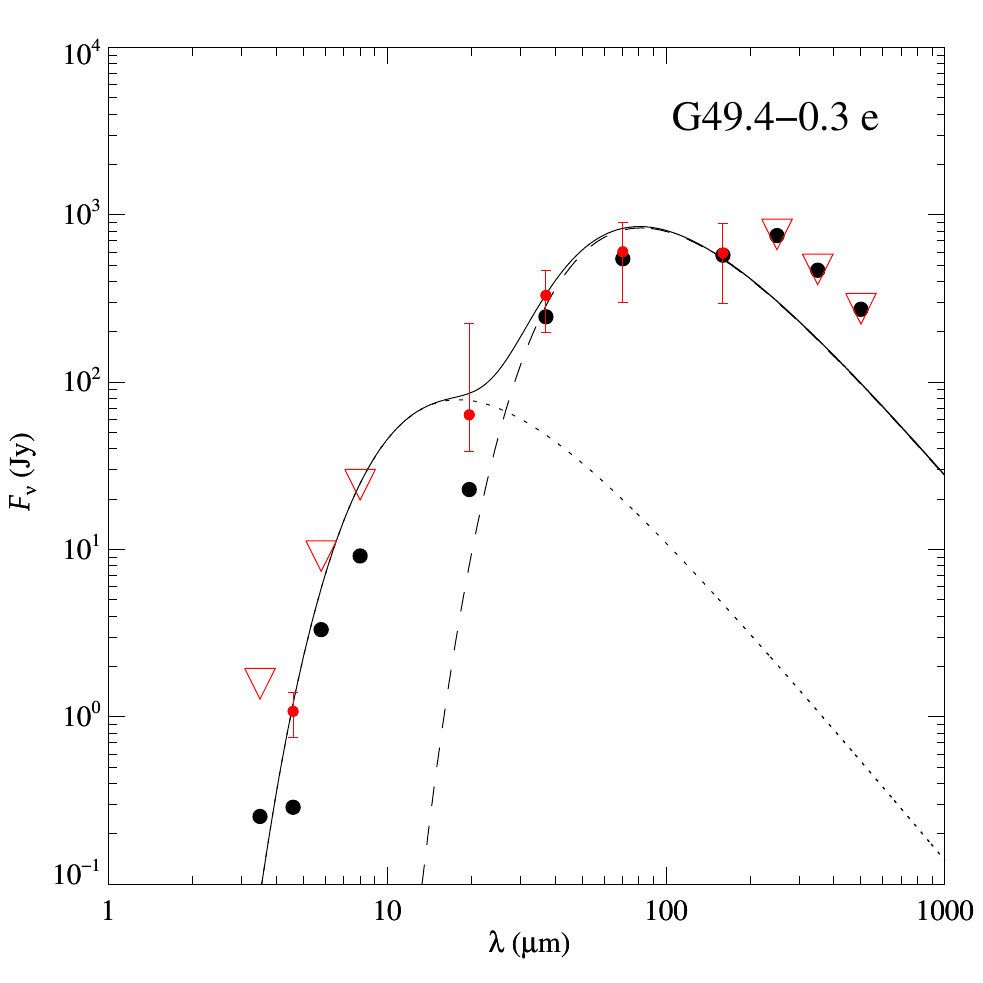}
	\caption{An example of the observed SED of extended source `e' in G49.4-0.3 with two temperature graybody fitting (the best fit model). The black dots are uncorrected observed data points. The red symbols show the extinction corrected data points, inverse triangles for the upper limits, and data points with error bars for the nominal values. The black dotted line shows the warm temperature component and the black dashed line shows the cold dust temperature component of graybody fitting. The black solid line shows the combination of warm and cold components. This shows that the \textit{SOFIA}-FORCAST 20 and 37\,$\mu$m data are crucial to determine $\lambda$ ranges of cold and warm components. We fit the warm graybody component at $\lambda\leq$\,20\,$\mu$m and the cold component at $\lambda\geq$\,37\,$\mu$m.}
\label{fig:essed}
\end{figure}

When trying to fit the 3--500\,$\mu$m photometry data of the extended sources with graybody fits, it was found that a single graybody was not sufficient, but that a two-component fit worked nicely for all sources (Figure\,\ref{fig:essed}). Therefore, following the work of \citet{2013ApJ...779...79M}, we derived the bolometric luminosity via a best-fit graybody model with two temperature components, i.e. cold and warm dust components. Based on these SEDs, we discovered that the \textit{SOFIA}-FORCAST 20 and 37\,$\mu$m photometry points are crucial in distinguishing the presence of the different temperature components.  Figure\,\ref{fig:essed} shows an example that represents well the two-component nature of the SEDs of all of the sources in Table\,\ref{tb:vir}. Integrating under these SEDs allows us to derive the bolometric luminosity of each source. 

Table\,\ref{tb:vir} shows the $M$, $L$, and $L/M$ values of the extended sources in column\,3, 4, and 7, respectively. We did not retrieved $M$ and $L$ of extended source d or e since the \textit{Herschel} images are mostly saturated in those regions. From the 13 remaining extended sources, we see large variation of $L/M$:  25\,$\lesssim$\,$(L/M)$/$(L_{\sun}/M_{\sun})$\,$\lesssim$\,790. The G49.4-0.3 sources show typically smaller $L/M$ than sources in G49.5-0.4, i.e. 40\,$\lesssim$\,$(L/M)$/$(L_{\sun}/M_{\sun})$\,$\lesssim$\,110. The extended sources in G49.5-0.4\,f--j show high $L/M$ values ($\sim$270\,--\,790\,$L_{\sun}/M_{\sun}$) and the sources in the G49.5-0.4\,a--c area have 25\,$\lesssim$\,$(L/M)$/$(L_{\sun}/M_{\sun})$\,$\lesssim$\,100. One might assume that the relative ages of stellar clusters in W\,51\,A are high at G49.5-0.4\,f--j while the sources in G49.5-0.4\,a--c and G49.4-0.3 regions are possibly in similar evolutionary stages. This may also be seen from the derived cold temperature components (column\,5 of Table\,\ref{tb:vir}). We find the $T_{\rm cold}$ of high $L/M$ sources (i.e. G49.5-0.4\,f--j) are $\sim$\,20--30\,K higher than the other extended sources which can indicate that the dust grains are internally heated by embedded young stellar clusters so that both $T_{\rm cold}$ and $L/M$ should increase simultaneously.

\subsubsection{Virial Analysis}\label{sec:vir}

\begin{deluxetable*}{rrrcrrrrrrrr}
\centering
\tabletypesize{\scriptsize}
\tablecolumns{12}
\tablewidth{0pt}
\tablecaption{Virial Parameters of Extended Sources in W51A}
\tablehead{\colhead{  Source  }&
	   \colhead{   $M_{\rm vir}$  } &
           \colhead{   $M$  } &
           \colhead{   $L$  } &
           \colhead{   $T_{\rm cold}$  } &
           \colhead{   $T_{\rm warm}$  } &
           \colhead{   $L/M$  } &
           \colhead{ $\alpha_{\rm vir}$ } \\
	   \colhead{            } &
	   \colhead{   ($M_{\sun}$)   } &
	   \colhead{   ($M_{\sun}$)   } &
	   \colhead{   ($\times10^{5}L_{\sun}$)   } &
	   \colhead{   ($K$)   } &
	   \colhead{   ($K$)   } &
	   \colhead{   ($L_{\sun}/M_{\sun}$)   } &
	   \colhead{             }  }
\startdata
	{\bf G49.4-0.3} &    &    &    &  &  &  &  \\
	a & 580 & 2560 & 3.66 & 76.2 & 264.6 & 71.4 & 0.23 \\
	b & 2490 & 9510 & 12.0 & 71.0 & 261.6 & 63.0 & 0.26 \\
	c & 1370 & 1580 & 3.61 & 91.0 & 261.2 & 113.8 & 0.87 \\
	e & 373 & 819 & 0.65 & 62.6 & 283.2 & 39.6 & 0.45 \\
	f & 125 & 916 & 1.64 & 91.0 & 273.0 & 89.4 & 1.36 \\
	{\bf G49.5-0.4} & & & & & & & \\
	a & 1900  &  1430  & 2.77 & 88.6 & 253.2 & 96.8 &  1.21 \\
	b & 1810  & 9930 & 5.14 & 69.9 & 247.6 & 25.9 & 0.18 \\
	c  & 3950  &  6870   & 10.3 & 66.1 & 270.7 & 81.0 & 0.58 \\
	d  &  947  & \nodata  & \nodata & \nodata & \nodata & \nodata & \nodata \\
	e &   5420   & \nodata & \nodata & \nodata & \nodata & \nodata & \nodata \\
	f  &  1800  &   432 & 1.98 & 93.0 & 256.7 & 228.5 & 4.16 \\
	g  &  1950  &   320 & 2.18 & 100.5 & 255.7 & 340.8 &  6.11 \\
	h  &  1520  &   122 & 1.93 & 116.6 & 250.7 & 790.1 &  12.50 \\
	i  &  511  &    107  & 0.57 & 116.4 & 254.7 & 266.4 & 4.77 \\
	j   & 1570  &   386 & 3.26 & 104.8 & 270.7 & 421.6 &  4.06 \\
\enddata
	\tablecomments{\footnotesize Source d and e of G49.5-0.4 are saturated in \textit{Herschel}-SPIRE observations.}
\label{tb:vir}
\end{deluxetable*}

The virial analysis is an effective tool to determine the importance of kinematic and gravitational energies of ISM structures, especially for molecular clumps and cores \citep{1992ApJ...395..140B}. As molecular clumps evolve, kinematic energy from internal sources (e.g., radiative pressure, outflow, and shock) are assumed to increase their influences on the system. This can be traced by comparing the estimated mass at virial equilibrium (i.e. virial mass), $M_{\rm vir}$, and the mass of each clump. \citet{1992ApJ...395..140B} defined the virial mass of molecular clumps as $M_{\rm vir} = 5 \sigma^2 R / G$,  where $M_{\rm vir}$ is the mass of the structure if it were in virial equilibrium, $\sigma$ is velocity dispersion ($\sigma$ = $\Delta$v / (8 ln2)$^{1/2}$) for the Gaussian line profile of corresponding clump with $\Delta$v as FWHM of molecular line profiles), $R$ is the radius of the clump, and G is the gravitation constant. The virial parameter, $\alpha_{\rm vir}$, is defined as $\alpha_{\rm vir}$ = $M_{\rm vir}$ / $M$, where $M$ is the intrinsic mass of the clump. For simplicity, the effect of magnetic fields is ignored even though the magnetic field is important in regulating the dynamics of molecular clumps \citep{1992ApJ...395..140B,2013ApJ...779...96T}. In this assumption, the virial status of a clump can be explained as self-gravitationally collapsing ($\alpha_{\rm vir}$ $<$ 1), virial equilibrium ($\alpha_{\rm vir}$ = 1, the clump is gravitationally stable, i.e. virialized), quasi-virial equilibrium (1 $<$ $\alpha_{\rm vir}$ $\leq$ 2, the clump is slightly expanding but still gravitationally bound), or gravitationally unbound ($\alpha_{\rm vir}$ $>$ 2). 
In order to inspect the virial state of the extended sources in W\,51\,A (i.e., the regions in Table\,\ref{tb:es}), we utilized public $^{13}$CO(2-1) data cube from 10m Heinrich Hertz Telescope \citep{2010ApJS..190...58K}. Note that we measure the ratio of internal kinetic energy to gravitational binding energy in the extended sources, ignoring surface pressure terms and effects of magnetic fields. We used the integrated $^{13}$CO line profile of each clump to fit a Gaussian. In order to determine the central gas $^{13}$CO velocity of each extended source, we used the literature values of velocity ranges defined in \citet{2010ApJS..190...58K} and \citet{2015A&A...573A.106G}.

We derive the virial parameter, $\alpha_{\rm vir}$, assuming constant density for the extended sources so that 
\begin{equation}\label{eqn:alpha}
	\alpha_{\rm vir} = \frac{M_{\rm vir}}{M} \sim 210\,\times \bigg( \frac{\sigma}{\rm km\,s^{-1}} \bigg)^2 \times \bigg( \frac{R}{\rm pc} \bigg) \times \bigg( \frac{M_\sun}{M} \bigg)
\end{equation}
where R is radius of the clump in parsec scale, $\sigma$ is the FWHM  of the $^{13}$CO(2-1) line in km/s, and $M$ is derived from sub-mm dust emission-based $\Sigma$ in units of $M_{\sun}$. If we assume the density profile falls off as 1/r, the $\alpha$ values will be $\sim$10\% smaller than constant density case \citep{1988ApJ...333..821M}. As we can see from Eqn.\,\ref{eqn:alpha}, the uncertainty of $\alpha_{\rm vir}$ is derived from the errors of gas velocity width, derived clump mass, and distance estimation so that conservative total uncertainty of $\alpha_{\rm vir}$ is about factor of 2 \citep[e.g.][]{2013ApJ...779..185K}. 

We present these parameters in Table~\ref{tb:vir}. The extended sources in G49.4-0.3 region show a mean $\alpha_{\rm vir}\sim$0.63, while the median $\alpha_{\rm vir}\sim$0.45. The sources in G49.5-0.4\,a--c show mean and median $\alpha_{\rm vir}$ values of 0.66 and 0.58, respectively. The G49.5-0.4\,f--j sources show a mean $\alpha_{\rm vir}$\,$\sim$\,6.32 and median $\alpha_{\rm vir}$\,$\sim$\,4.77. These derived values of $\alpha_{\rm vir}$ show that the sources in G49.4-0.3 are mostly sub-virial which indicates these sources are probably undergoing self-gravitational collapse. The extended sources G49.5-0.4\,f--j are all super-virial with $\alpha_{\rm vir}\gtrsim4$, indicating the sources are gravitationally unbound and expanding. The source G49.5-0.4\,a is close to the virial equilibrium status (i.e., gravitationally stable). Source G49.5-0.4\,b shows the lowest virial parameter, $\alpha_{\rm vir}$\,$\sim$\,0.18, which is unique from any other extended source in this study. We do not detect any individual MYSO sources within G49.5-0.4\,b. From the lowest $\alpha_{\rm vir}$ value and absence of significant YSOs, we assume G49.5-0.4\,b is the youngest molecular clump in W\,51\,A area.

\subsubsection{The History of Stellar Cluster Formation in W\,51\,A}\label{sec:hist}

\begin{figure}[t]
\figurenum{15}
\epsscale{1.15}
\plotone{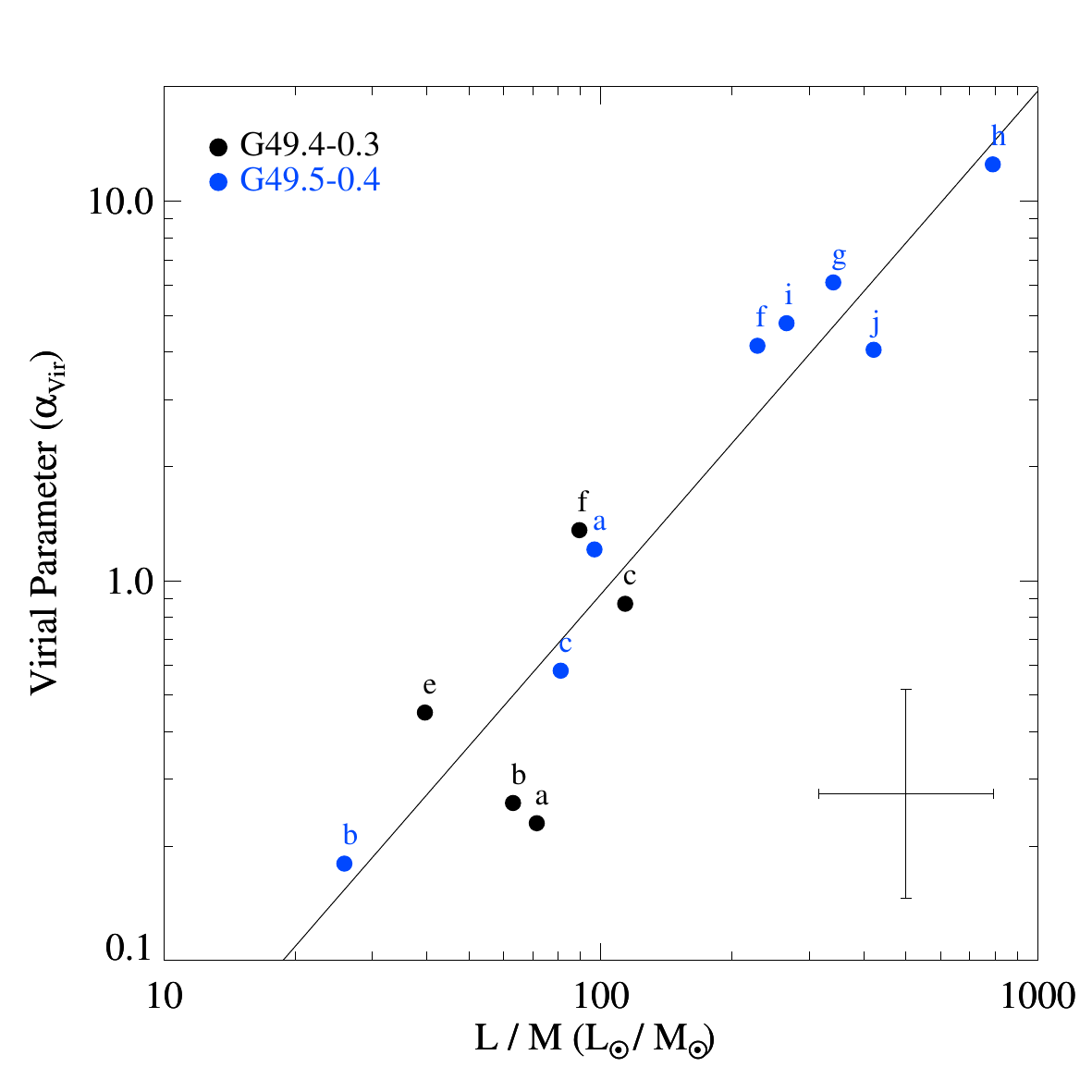}
	\caption{Virial parameter ($\alpha_{\rm vir}$) vs. $L/M$ of all extended sources in W\,51\,A definded by ${\it SOFIA}$-FORCAST 20 \& 37\,$\mu$m images. The blue and black dots are the extended sources in G49.5-0.4 and G49.4-0.3 regions, respectively. The name of each source is shown at the top of each dot. The solid line indicates the best line fit ($\alpha\sim$1.28 in log-space). The error bar at the bottom left shows the typical uncertainty (a factor of $\sim$2) on both $L/M$ and $\alpha_{\rm vir}$ directions.}
\label{fig:virlm}
\end{figure}

Figure\,\ref{fig:virlm} shows $L/M$ versus $\alpha_{\rm vir}$ of all extended sources in W\,51\,A. The correlation shows that both evolutionary tracers of stellar cluster formation are under good agreement, indicating G49.5-0.4\,f--j are relatively older than the G49.5-0.4\,a--c sources, while the entire G49.04-0.3 region is likely younger on average than the entire G49.5-0.4 region. The result of $L/M$ versus $\alpha_{\rm vir}$ is not only consistent with previous studies, but as we will now discuss, the result helps to further clarify our understanding of the star formation history of the W\,51 area.

\citet{elmegreen1992} suggested that several Myrs of age difference in between two nearby ($\sim$10--50\,pc away) sources may be evidence for triggered star(-cluster) formation. Using this logic, \citet{2000ApJ...543..799O} hypothesized that G49.5-0.4 had undergone triggered sequential star formation. They suggest the stellar clusters in the region around G49.5-0.4\,h \& j (which they call `Region\,1') are the oldest, while the region around G49.5-0.4\,a--e (their `Region\,3') are the youngest, and that the star cluster formation in Region\,3 is triggered by the stellar wind from the evolved stars in Region\,1 and the expansion of the G49.5-0.4\,f \& g \ion{H}{2} regions (Region\,2').   

We derived the total mass ratio between NIR revealed stars \citep[from][]{2000ApJ...543..799O} to MIR revealed stars \citep[from][and this study]{2009ApJ...706...83K}, $\Sigma_{\rm M*,MIR}$ / $\Sigma_{\rm M*,NIR}$, in Regions\,1, 2, and 3. We assume the NIR-detected stars are less embedded and therefore relatively older than the deeply embedded MIR-detected stars, meaning that the ratio $\Sigma _{\rm M*,MIR}$ / $\Sigma _{\rm M*,NIR}$ should get smaller with cluster age. We find that $\Sigma _{\rm M*,MIR}$ / $\Sigma _{\rm M*,NIR}$\,$\sim$ 0, 0.01, and 0.10 for Regions\,1, 2, and 3, respectively. This relative decrease in evolutionary state from Regions 1 to 2 to 3 is basically consistent with the relative ages that claimed by \citet{2000ApJ...543..799O}, and consistent with the trends we see from L/M and the virial parameter (Figure\,\ref{fig:virlm}). 

In contrast, \citet{2009A&A...504..429C} claimed that they could not find any evidence of sequential triggered star formation in W\,51\,A area but found the indication of independent star formation activities in multiple positions that could be generated by external triggering effects as \citet{2004MNRAS.353.1025K} suggested. While \citet{2009A&A...504..429C} and \citet{2004MNRAS.353.1025K} investigated the star formation history of W\,51 region via NIR observations, \citet{2015A&A...573A.106G} pointed out the absence of signs of triggered star formation from expanding \ion{H}{2} regions toward G49.5-0.4\,a--e based on Karl G. Jansky Very Large Array (JVLA) centimeter observations. \citet{2010ApJS..190...58K} studied the overall structure of W\,51 area by observing CO isotopologues. They suggested the W\,51\,A region underwent cloud-cloud collision to produce the stellar clusters possessing high-mass stars and the extended source G49.5-0.4\,b is possibly at the colliding location between two molecular clouds that could be distinguished by one at $\sim$58\,km/s (encompassing G49.5-0.4\,a-e) and one at 68\,km \citep[a.k.a. the High Velocity Stream;][]{1998AJ....116.1856C}. They found a `bridge' in the position-velocity diagram which connect two different CO velocity components ($<$61\,km/s and $\sim$68\,km/s) as well as the self-absorption line at the location of G49.5-0.4\,b region. They insisted the self-absorption line was caused since the colliding interface (G49.5-0.4\,b) had been heated while surroundings were still cold while the `bridge' showed the interaction between two different clouds.  

The comparison of two relative evolutionary tracers, $L/M$ and $\alpha_{\rm vir}$, could be a more accurate way to determine the evolutionary states of star cluster formation in the W\,51\,A area than the absolute age calculations that were performed in previous NIR studies \citep[e.g.][]{2000ApJ...543..799O}. In general, these previous studies focused on analyzing the morphologies of IR and/or mm bubbles around YSOs and the estimated the ages of YSOs as $\sim$\,0.5\,--\,3\,Myrs. The ages were typically determined from the isochrones \citep[e.g.][]{1992A&AS...96..269S,2000A&A...361..101M} sometimes comparing them to the expansion ages of \ion{H}{2} regions \citep{1989ApJS...69..831W}. However, these calculations can have relatively large errors \citep[up to 100\,\%,][]{2011ApJ...732....8V}, and thus assuming star-forming history based on these values and the morphologies of molecular bubble structures is likely to be highly uncertain. An example of this is that \citet{2009A&A...504..429C} estimated the age of LS1 was at least 3\,Myr (possibly 6\,Myr or older), while \citet{2000ApJ...543..799O} estimated the age as $\sim$\,2.3\,Myr. The difference in derived ages between \citet{2000ApJ...543..799O} and \citet{2009A&A...504..429C} led to the different interpretations of star forming history in W\,51\,A. 

From Figure\,\ref{fig:virlm}, we can see the evolutionary states of G49.5-0.4\,f--j sources are clearly separated from sources on G49.5-0.4\,a--c and G49.4-0.3 regions. If the internal feedback of G49.5-0.4\,f--j regions could affect the star formation in G49.5-0.4\,a--c, one would expect to find a smoothly continuous trend of $L/M$ versus $\alpha_{\rm vir}$ along all G49.5-0.4 sources. This might support the scenario sketched by \citet{2015A&A...573A.106G} that the location of the GMC possessing G49.5-0.4\,f--j regions is different from either G49.5-0.4\,a--e or G49.4-0.3. In this case, the independent formation of stellar clusters, as suggested in \citet{2009A&A...504..429C}, would mean that the non-interacting (and thus separated) clouds induce their own star formation history. 

Given its very low $\alpha_{\rm vir}$ and $L/M$, the source G49.5-0.4\,b might be the youngest clump in the W\,51\,A area. The low $\alpha_{\rm vir}$ and $L/M$ can be due to the lack of internal heating sources (i.e. young stars) so that the internal gas motion is not strong enough to overcome the gravitational pressure of the molecular clump. The evolutionary state of G49.5-0.4\,b as the uniquely young molecular clump in W\,51\,A region can be explained by the recent cloud-cloud collision that is suggested by \citet{2010ApJS..190...58K}. Comparing CO observational results of \citet{2010ApJS..190...58K} to a synthetic CO emission lines from theoretical simulation of cloud-cloud collision scenario \citep[e.g.][]{2017ApJ...835..137W} can be helpful to address the effect and evidence of cloud-cloud collision on the molecular clump formation and evolution.

\section{Summary} \label{sec:conclusion}

We obtained \textit{SOFIA}-FORCAST images at 20 and 37\,$\mu$m of the central 10$\arcmin\times$20$\arcmin$ region of W\,51\,A. The 37\,$\mu$m images are the highest spatial resolution observations of W\,51\,A yet obtained at wavelengths beyond 25\,$\mu$m. We compared these images to data at multiple other wavelengths to get a clearer picture of the nature of this giant \ion{H}{2} region and star-forming complex. We discussed the observations of all of the individual sources and sub-components within W\,51\,A, and based on our new imaging data and previous multi-wavelength observations, we conjecture (for the first time for several sources) on their nature. In summary, we itemize our most significant results:

1) The most-studied area of W\,51\,A is the e1/e2 cluster area. The \textit{SOFIA}-FORCAST images show that the only thermal infrared source present at wavelengths less than 20\,$\mu$m is coincident with the hypercompact \ion{H}{2} region e9. Though this source appears point-like at these shorter infrared wavelengths, the \textit{SOFIA} 37\,$\mu$m image reveals a source with a double peak surrounded by an extended fainter structure. The secondary 37\,$\mu$m peak is coincident with the 20\,$\mu$m peak (and thus coming from radio source e9), however the primary peak is located $\sim$5$\arcsec$ to the northeast and closer to (but not coincident with) the peak of the hot core seen at mm wavelengths. We suggest that the primary peak of emission at 37\,$\mu$m is either due to IR emission leaking from gaps on the eastern edge of the otherwise optically thick hot core, and/or from emission from the blue-shifted outflow cavity of the MYSO at the location of the radio source e2.

2) We detect an extended infrared source at 20 and 37\,$\mu$m that becomes the 5th brightest source in W\,51\,A at 70\,$\mu$m and the 4th brightest source behind IRS\,1, IRS\,2, and the e1/e2 cluster at 160\,$\mu$m. It is the most-steeply rising source from 20 to 37\,$\mu$m in this study, indicating the source is highly embedded and/or young. The best fit SED model for this source yields a bolometric luminosity of  6.48$\times$10$^5\,L_{\sun}$. We dub this new infrared region as IRS\,4. At its location lies a resolved radio continuum emission point source, e16, as well as a resolved radio binary, e18, along with an extended HII region, e18d. Given the high-luminosity, steeply rising IR SED, presence of multiple radio continuum sources, and prominence in the FIR, this location is likely to be an embedded core or clump hosting a young massive proto-cluster.

3) Some individual regions and much of the G49.5-0.4 area seem to owe their observed mid-infrared morphology to extinction effects. Individual extended sources like G49.5-0.4\,a and b, and G49.4-0.3\,b appear as a collection of peaks that shift with wavelength in the infrared. This indicates that the stellar sources forming within them are not being directly viewed in our infrared images, but we instead are likely seeing the mid-infrared light escaping from gaps in the less-dense regions of the surrounding clumpy material. Likewise, extinction appears to be affecting larger-scale mid-infrared morphology, especially around radio sources in G49.5-0.4. Sources a, b, c, d, and e encircle large MIR-dark areas that are ``filled in'' by cold dust emission seen at far-infrared wavelengths by \textit{Herschel}. 

4) Most sources in G49.4-0.3 and many in G49.5-0.4 are ring- or arc-shaped in the infrared. These are likely to be wind-blown bubbles or Stromgren spheres from older generations of massive star formation.

5) We used \textit{SOFIA}-FORCAST photometry in conjunction with \textit{Spitzer}-IRAC and \textit{Herschel}-PACS photometry data to construct SEDs of sub-components and point sources detected in the infrared. We fit those SEDs with young stellar object models, and found 41 sources that are likely to be massive young stellar objects, many of which are identified as such in this work for the first time. Almost half of the MYSOs  (20/41) do not have radio continuum emission, implying a very young state of formation. Due to the relatively good spatial resolution of the \textit{Spitzer} and \textit{SOFIA} data, especially at 37\,$\mu$m, we are able to isolate the emission from many sources that are unresolved or confused in the \textit{Herschel} FIR data. Furthermore, we showed that the 37\,$\mu$m data point was crucial in getting good SED fits for these MYSOs.

6) In calculating the luminosity of the large sub-regions of W\,51\,A, we found that a two-temperature fit is needed, and that the \textit{SOFIA}-FORCAST photometry at 20 and 37\,$\mu$m was essential in determining these two temperatures, since they straddle and define the transition wavelengths in the SEDs between the warm and cold dust components. 

7) We used the luminosity-to-mass ratio and virial parameters of the extended sub-regions of W51A to estimate their relative ages. We are able to confirm analytically what previous authors have determined qualitatively concerning the relative ages of the different sub-regions of W51A.

8) We suggest the extended source G49.5-0.4\,b is the youngest molecular clump in W\,51\,A region because of its lowest luminosity-to-mass ratio and virial parameters. The absence of enough internal heating sources (YSOs) can explain the low $\alpha_{\rm vir}$ and $L/M$. The recent cloud-cloud collision occurring at the position of G49.5-0.4\,b could be the mechanism responsible for creating this newly-formed young stellar cluster.

\acknowledgments
Authors thank an anonymous referee for constructive comments that help to improve the manuscript significantly. Authors also thank J. M. Jackson, J. T. Radomski, W. T. Reach, J. C. Tan, W. D. Vacca and Y. Zhang for discussions. This research is based on observations made with the NASA/DLR Stratospheric Observatory for Infrared Astronomy (SOFIA). SOFIA is jointly operated by the Universities Space Research Association, Inc. (USRA), under NASA contract NAS2-97001, and the Deutsches SOFIA Institut (DSI) under DLR contract 50 OK 0901 to the University of Stuttgart. This work is also based in part on archival data obtained with the Spitzer Space Telescope, which is operated by the Jet Propulsion Laboratory, California Institute of Technology under a contract with NASA. This work is also based in part on archival data obtained with Herschel, an European Space Agency (ESA) space observatory with science instruments provided by European-led Principal Investigator consortia and with important participation from NASA. Financial support for this work was provided by NASA through awards 01\_0007, 02\_0113, 03\_0008, and 03\_0009 issued by USRA. 

\vspace{5mm}
\facility{SOFIA(FORCAST)}

\appendix

\section{Data release}

The reduced images and used in this paper are publicly available at: {\it https://dataverse.harvard.edu/dataverse/SOFIA-GHII} 

The data include the \textit{SOFIA} FORCAST 20 and 37\,$\mu$m final image mosaics and their exposure maps.

\section{{\it Spitzer} and {\it Herschel} Photometry of Sub-components and point sources in W\,51\,A}\label{appendix}

As we mentioned in \S\,\ref{sec:cps}, we performed optimal extraction photometry for the FORCAST 20 and 37\,$\mu$m images to define the location of all sub-components and point sources, and to determine the aperture radii to be used for photometry. Using these source locations, we employed the optimal extraction technique on the \textit{Spitzer}-IRAC 8\,$\mu$m data for all sources to find the optimal aperture to be used for all IRAC bands (since the source sizes are typically similar or smaller at the shorter IRAC bands). As we have done for the FORCAST images, we estimated the background emission from the annuli that showed the least contamination from nearby sources, i.e. showing relatively flat radial intensity profile (\S\,\ref{sec:cps}). Table\,\ref{tb:cps2} shows the photometry values we derive for all sources from the \textit{Spitzer}-IRAC bands. 

Table\,\ref{tb:cps3} shows the photometry result for the \textit{Herschel}-PACS bands. We use fixed aperture radii for all PACS bands ($R_{\rm int}$=16$\farcs$0 for 70\,$\mu$m and $R_{\rm int}$=22$\farcs$5 for 160\,$\mu$m, except for G49.5-4\,b1 and i due to their larger sizes) that are based on the PSFs of relatively isolated sources (e.g. G49.4-0.3 a/{\#}1, b/{\#}6 and G49.5-0.4 IRS1/{\#}11) and using a generous aperture size. In general, this aperture size cannot be determined accurately using the optimal extraction technique due to the ubiquity of extended emission from nearby sources that are overlapping the source being measured. We compared our aperture sizes to those typically used in the Hi-GAL Compact Source Catalogue \citep{2017MNRAS.471..100E,2016A&A...591A.149M}. That catalogue employs aperture sizes comparable to the ones we used in this study. Note, however, Hi-GAL catalogue sources are also hugely contaminated by nearby sources (especially in G49.5-0.4\,d and e regions). We therefore believe that the fixed aperture size we employ here is reasonable, especially since the data are only being used to provide upper limits to our SED model fits. 

\begin{deluxetable}{rccccccccc}
\tabletypesize{\scriptsize}
\tablecaption{{\it Spitzer}-IRAC bands Observational Parameters of Sub-components and Point Sources in W51A}
\tablehead{\colhead{  }&
           \colhead{  }&
           \multicolumn{2}{c}{${\rm 3.6\mu{m}}$}&
           \multicolumn{2}{c}{${\rm 4.5\mu{m}}$}&
           \multicolumn{2}{c}{${\rm 5.8\mu{m}}$}&
           \multicolumn{2}{c}{${\rm 8.0\mu{m}}$}\\
           \cmidrule(lr){3-4} \cmidrule(lr){5-6} \cmidrule(lr){7-8} \cmidrule(lr){9-10}\\
           \colhead{ Source }&
           \colhead{ $R_{\rm int}$ } &
           \colhead{ $F_{\rm int}$ } &
           \colhead{ $F_{\rm int-bg}$ } &
           \colhead{ $F_{\rm int}$ } &
           \colhead{ $F_{\rm int-bg}$ } &
           \colhead{ $F_{\rm int}$ } &
           \colhead{ $F_{\rm int-bg}$ } &
           \colhead{ $F_{\rm int}$ } &
           \colhead{ $F_{\rm int-bg}$ } \\
	   \colhead{  } &
	   \colhead{ ($\arcsec$) } &
	   \colhead{ (mJy) } &
	   \colhead{ (mJy) } &
	   \colhead{ (mJy) } &
	   \colhead{ (mJy) } &
	   \colhead{ (Jy) } &
	   \colhead{ (Jy) } &
	   \colhead{ (Jy) } &
	   \colhead{ (Jy) } \\
}
\startdata
{\bf G49.4-0.3} & & & & & & & & & \\
    a/{\#}1 &  4.80 &      123 &      116 &      347 &     337  &     0.68 &     0.63 &     0.87 &     0.72  \\
    b/{\#}1 &  6.00 &     91.0 &     43.2 &      117 &     45.4 &     0.75 &     0.24 &     2.25 &     0.40  \\
    b/{\#}2 &  7.20 &     57.9 &     37.5 &     77.5 &     56.4 &     0.67 &     0.44 &     1.82 &     1.06  \\
    b/{\#}3 &  9.60 &      158 &     95.3 &      247 &     130  &     1.59 &     1.14 &     4.80 &     2.52  \\
    b/{\#}4 &  6.00 &     83.8 &     56.7 &      175 &     122  &     1.07 &     0.55 &     2.99 &     1.57  \\
    b/{\#}5 &  6.00 &     40.0 &     14.1 &     49.6 &     16.3 &     0.26 &     0.02 &     0.80 &     0.05  \\
    b/{\#}6 &  4.80 &     46.8 &     26.2 &     41.5 &     20.5 &     0.36 &     0.17 &     1.00 &     0.39  \\
    e/{\#}1 &  7.20 &     51.9 &     25.2 &     66.6 &     35.3 &     0.70 &     0.30 &     1.92 &     0.74  \\
    SHA17 3 &  4.80 &     33.0 &     1.20 &     31.8 &     2.40 &     0.27 &     0.01 &     0.77 &     0.04  \\
    SHA17 4 &  6.00 &     48.9 &     17.5 &     54.6 &     18.7 &     0.25 &     0.02 &     0.75 &     0.04  \\
{\bf G49.5-0.4} & & & & & & & & & \\
         b1 & 21.6 &       831 &      590 &     788  &     533  &     7.23 &     4.13 &    20.5 &    10.3  \\
         b2 & 10.8 &       241 &      136 &     296  &     186  &     1.76 &     0.63 &     4.85 &     1.66  \\
   b2/{\#}1 &  7.20 &     46.9 &     21.4 &     53.7 &     31.3 &     0.51 &     0.12 &     1.35 &     0.46  \\
         b3 & 12.0 &       134 &     45.1 &     160  &     56.2 &     1.49 &     0.53 &     4.53 &     1.34  \\
    d4e+d4w &  4.80 &     48.7 &     9.70 &     74.1 &     22.1 &     0.62 &     0.19 &     1.43 &     0.41  \\
         d6 &  4.80 &     88.4 &     33.4 &     200  &     105  &     1.15 &     0.18 &     2.47 &     0.57  \\
         e7 &  9.60 &      246 &      116 &     322  &     168  &     2.12 &     0.74 &     5.64 &     1.95  \\
         e9 &  4.80 &     41.0 &     12.5 &     95.5 &     25.3 &     0.73 &     0.25 &     2.12 &     0.91  \\
        e15 &  4.80 &     43.7 &     16.2 &     72.1 &     26.1 &     0.48 &     0.10 &     1.34 &     0.30  \\
    f/{\#}1 &  6.00 &     61.9 &     16.0 &     60.9 &     18.4 &     0.56 &     0.16 &     1.57 &     0.38  \\
          i & 18.0 &       433 &      247 &     479  &     288  &     2.82 &     1.69 &     8.20 &     4.57  \\
    i/{\#}1 &  7.20 &     43.9 &     9.90 &     55.5 &     17.9 &     0.26 &     0.03 &     0.99 &     0.08  \\
 IRS1/{\#}1 &  4.80 &      352 &      284 &     1080 &     921  &     4.26 &     3.05 &     8.63 &     5.86  \\
 IRS1/{\#}2 &  4.80 &      188 &      136 &     717  &     626  &     2.71 &     2.11 &     5.99 &     4.43  \\
 IRS1/{\#}3 &  4.80 &     23.1 &     14.4 &     66.1 &     42.0 &     0.22 &     0.03 &     0.54\tablenotemark{u} &     \nodata  \\
 IRS1/{\#}4 &  3.60 &     4.70 &     0.50 &     9.60 &     2.20 &     0.06\tablenotemark{u} &     \nodata &     0.16\tablenotemark{u} &   \nodata  \\
 IRS1/{\#}5 &  3.60 &     3.40 &     0.20 &     8.10 &     2.70 &     0.06\tablenotemark{u} &    \nodata &     0.15\tablenotemark{u} &   \nodata  \\
 IRS1/{\#}6 &  3.60 &     56.6 &     27.7 &     96.1 &     36.4 &     0.74 &     0.23 &     1.87 &     0.76  \\
 IRS1/{\#}7 &  9.60 &      178 &     53.6 &     258  &     69.0 &     2.25 &     0.48 &     5.89 &     1.08  \\
 IRS1/{\#}8 &  9.60 &      165 &     37.5 &     247  &     76.7 &     1.93 &     0.43 &     5.21 &     0.86  \\
 IRS1/{\#}9 &  6.00 &      401 &      296 &     1260 &     1100 &     4.63 &     2.70 &    12.7 &     4.71  \\
IRS1/{\#}10 &  9.60 &      237 &     49.7 &     278  &     38.7 &     2.61 &     0.38 &     6.98 &     1.10  \\
IRS1/{\#}11 &  3.60 &     11.0 &     3.50 &     20.2 &     13.1 &     0.09 &     0.02 &     0.22 &     0.02  \\
 IRS2/{\#}1 &  3.07 &      312 &      302 &     765  &     751 &     2.85 &     2.73 &     3.09\tablenotemark{l} &     \nodata  \\
 IRS2/{\#}2 &  3.07 &      117 &      107 &     254  &     240  &     0.89 &     0.74 &     1.93 &     1.56  \\
 IRS2/{\#}3 &  3.07 &     79.7 &     70.9 &     142  &     125  &     0.75 &     0.60 &     1.82 &     1.50  \\
 IRS2/{\#}4 &  3.07 &     59.3 &     51.1 &     113  &     96.8 &     0.57 &     0.42 &     1.27 &     0.94  \\
 IRS2/{\#}5 &  3.07 &      418 &      410 &     1040 &     1030 &     6.21 &     6.08 &     2.43\tablenotemark{l} &     \nodata  \\
 IRS2/{\#}6 &  3.60 &     17.4 &     5.00 &     28.5 &     9.80 &     0.22 &     0.06 &     0.50 &     0.13  \\
 IRS2/{\#}7 &  4.80 &     49.0 &     29.5 &     58.9 &     38.7 &     0.48 &     0.27 &     1.19 &     0.57  \\
 IRS2/{\#}8 &  3.60 &     25.2 &     11.4 &     27.6 &     12.8 &     0.22 &     0.07 &     0.56 &     0.12  \\
 IRS2/{\#}9 &  3.60 &     30.0 &     18.7 &     37.4 &     25.5 &     0.25 &     0.12 &     0.62 &     0.27  \\
IRS2/{\#}10 &  3.84 &     86.2 &     72.6 &     201  &     184  &     1.03 &     0.85 &     3.05 &     2.60  \\
      IRS2E &  3.84 &     1770\tablenotemark{l} &  \nodata &     2370\tablenotemark{l} &  \nodata &    13.3\tablenotemark{l} &  \nodata &     1.78\tablenotemark{l} &  \nodata  \\
      IRS2W &  3.84 &     1750\tablenotemark{l} &  \nodata &     2670\tablenotemark{l} &  \nodata &    12.1\tablenotemark{l} &  \nodata &     2.31\tablenotemark{l} &  \nodata  \\
       IRS3 &  4.80 &      598 &      420 &     1460 &     1130 &     5.69 &     4.20 &     9.04 &     3.32  \\
       IRS4 &  9.60 &      234 &      138 &     442  &     289  &     2.49 &     1.32 &     7.84 &     3.97  \\
        LS1 &  4.80 &      740 &      717 &     782  &     759  &     0.93 &     0.84 &     0.87 &     0.64  \\
\enddata
\tablecomments{\footnotesize Same as Table\,\ref{tb:cps1} but for \textit{Spitzer}-IRAC bands. The center positions of the apertures are based on \textit{SOFIA} observation in Table\,\ref{tb:cps1}.}
\tablenotetext{l}{The $F_{\rm int}$ value is used as the lower limit since the source is partially/entirely saturated.}
\tablenotetext{u}{The $F_{\rm int}$ value is used as the upper limit since the source is difficult to distinguish from the background due to the relatively weak source emission.}
\label{tb:cps2}
\end{deluxetable}
 
\begin{deluxetable}{rcccc}
\tabletypesize{\scriptsize}
\tablewidth{0pt}
\tablecaption{{\it Herschel}-PACS bands Observational Parameters of Sub-Components and Point Sources in W51A}
\tablehead{\colhead{  }&
           \multicolumn{2}{c}{${\rm 70\mu{m}}$}&
           \multicolumn{2}{c}{${\rm 160\mu{m}}$}\\
           \cmidrule(lr){2-3} \cmidrule(lr){4-5} \\
           \colhead{ Source }&
           \colhead{ $R_{\rm int}$ } &
           \colhead{ $F_{\rm int}$ } &
           \colhead{ $R_{\rm int}$ } &
           \colhead{ $F_{\rm int}$ }  \\
	   \colhead{  } &
	   \colhead{ ($\arcsec$) } &
	   \colhead{ ($\times10^{6}$Jy) } &
	   \colhead{ ($\arcsec$) } &
	   \colhead{ ($\times10^{6}$Jy) } \\
}
\startdata
{\bf G49.4-0.3} & & & & \\
    a/{\#}1 & 16.0 &     0.27 & 22.5 &     0.37  \\
    b/{\#}1 & 16.0 &     3.08 & 22.5 &     2.03  \\
    b/{\#}2 & 16.0 &     1.23 & 22.5 &     1.38  \\
    b/{\#}3 & 16.0 &     2.93 & 22.5 &     1.79  \\
    b/{\#}4 & 16.0 &     6.37 & 22.5 &     3.92  \\
    b/{\#}5 & 16.0 &     0.78 & 22.5 &     0.82  \\
    b/{\#}6 & 16.0 &     0.80 & 22.5 &     0.90  \\
    e/{\#}1 & 16.0 &     1.72 & 22.5 &     1.92  \\
    SHA17 3 & 16.0 &     1.07 & 22.5 &     1.15  \\
    SHA17 4 & 16.0 &     0.67 & 22.5 &     0.62  \\
{\bf G49.5-0.4} & & & & \\
         b1 & 25.6 &     2.19 & 27.0 &     1.37  \\
         b2 & 16.0 &     3.33 & 22.5 &     3.99  \\
   b2/{\#}1 & 16.0 &     2.86 & 22.5 &     4.77  \\
         b3 & 16.0 &     2.22 & 22.5 &     1.58  \\
    d4e+d4w & 16.0 &     18.5 & 22.5 &     7.36  \\
         d6 & 16.0 &     33.5 & 22.5 &     12.4  \\
         e7 & 16.0 &     3.68 & 22.5 &     2.61  \\
         e9 & 16.0 &     32.9 & 22.5 &     13.1  \\
        e15 & 16.0 &     7.35 & 22.5 &     5.40  \\
    f/{\#}1 & 16.0 &     1.86 & 22.5 &     1.24  \\
          i & 30.4 &     0.98 & 27.0 &     0.35  \\
    i/{\#}1 & 16.0 &     0.65 & 22.5 &     0.70  \\
 IRS1/{\#}1 & 16.0 &     32.7 & 22.5 &     12.4  \\
 IRS1/{\#}2 & 16.0 &     32.7 & 22.5 &     12.5  \\
 IRS1/{\#}3 & 16.0 &     6.35 & 22.5 &     4.89  \\
 IRS1/{\#}4 & 16.0 &     6.19 & 22.5 &     7.16  \\
 IRS1/{\#}5 & 16.0 &     5.62 & 22.5 &     6.10  \\
 IRS1/{\#}6 & 16.0 &     27.1 & 22.5 &     11.1  \\
 IRS1/{\#}7 & 16.0 &     6.91 & 22.5 &     3.70  \\
 IRS1/{\#}8 & 16.0 &     8.26 & 22.5 &     5.97  \\
 IRS1/{\#}9 & 16.0 &     30.2 & 22.5 &     11.6  \\
IRS1/{\#}10 & 16.0 &     7.23 & 22.5 &     3.77  \\
IRS1/{\#}11 & 16.0 &     0.60 & 22.5 &     0.55  \\
 IRS2/{\#}1 & 16.0 &     41.6 & 22.5 &     13.5  \\
 IRS2/{\#}2 & 16.0 &     38.9 & 22.5 &     13.4  \\
 IRS2/{\#}3 & 16.0 &     39.2 & 22.5 &     13.3  \\
 IRS2/{\#}4 & 16.0 &     38.5 & 22.5 &     13.3  \\
 IRS2/{\#}5 & 16.0 &     42.2 & 22.5 &     12.9  \\
 IRS2/{\#}6 & 16.0 &     18.3 & 22.5 &     9.57  \\
 IRS2/{\#}7 & 16.0 &     8.54 & 22.5 &     6.63  \\
 IRS2/{\#}8 & 16.0 &     6.86 & 22.5 &     6.55  \\
 IRS2/{\#}9 & 16.0 &     5.19 & 22.5 &     6.08  \\
IRS2/{\#}10 & 16.0 &     36.3 & 22.5 &     11.6  \\
      IRS2E & 16.0 &     42.7 & 22.5 &     13.3  \\
      IRS2W & 16.0 &     42.5 & 22.5 &     13.2  \\
       IRS3 & 16.0 &     29.4 & 22.5 &     12.0  \\
       IRS4 & 16.0 &     13.4 & 22.5 &     7.06  \\
        LS1 & 16.0 &     0.47 & 22.5 &     0.14  \\
\enddata
\tablecomments{\footnotesize Same as Table\,\ref{tb:cps2} but for \textit{Herschel}-PACS 70 and 160\,$\mu$m observation. $R_{\rm int}$=5\,pixels are used as the fixed aperture size for each band except G49.5-0.4\,b1 and i.}
\label{tb:cps3}
\end{deluxetable}


\begin{thebibliography}{}

\bibitem[Agliozzo et al.(2017)]{2017MNRAS.466..213A} Agliozzo, C., Nikutta, R., Pignata, G., et al.\ 2017, \mnras, 466, 213 
\bibitem[Argon et al.(2000)]{2000ApJS..129..159A} Argon, A. L., Reid, M. J. \& Menten, K. M.\ 2000, \apjs, 129, 159
\bibitem[Barbosa et al.(2016)]{2016ApJ...825...54B} Barbosa, C.~L., Blum, R.~D., Damineli, A., Conti, P.~S., \& Gusm{\~a}o, D.~M.\ 2016, \apj, 825, 54 
\bibitem[Bertoldi \& McKee(1992)]{1992ApJ...395..140B} Bertoldi, F. \& McKee, C.~F.\ 1992, \apj, 395, 140
\bibitem[Barnes et al.(2011)]{2011ApJS..196...12B} Barnes, P.~J., Yonekura, Y., Fukui, Y., et al.\ 2011, \apjs, 196, 12
\bibitem[Battersby et al.(2011)]{2011A&A...535A.128B} Battersby, C., Bally, J., Ginsburg, A., et al.\ 2011, \aap, 535, A128
\bibitem[Bernasconi \& Maeder(1996)]{1996A&A...307..829B} Bernasconi, P.~A., \& Maeder, A.\ 1996, \aap, 307, 829 
\bibitem[Bisbas et al.(2018)]{2018MNRAS.478L..54B} Bisbas, T. G., Tan, J. C., Csengeri, T., et al.\ 2018, \mnras, 478, 54
\bibitem[Blum et al.(2000)]{2000AJ....119.1860B} Blum, R.~D., Conti, P.~S., \& Damineli, A.\ 2000, \aj, 119, 1860
\bibitem[Carpenter \& Sanders(1998)]{1998AJ....116.1856C} Carpenter, J.~M., \& Sanders, D.~B.\ 1998, \aj, 116, 1856
\bibitem[Cesaroni et al.(1988)]{1988A&AS...76..445C} Cesaroni, R., Palagi, F., Felli, M., et al.\ 1988, \aaps, 76, 445 
\bibitem[Churchwell (2002)]{churchwell02} Churchwell E.\ 2002, \araa, 40, 27
\bibitem[Churchwell et al.(2009)]{2009PASP..121..213C} Churchwell, E., Babler, B.~L., Meade, M.~R., et al.\ 2009, \pasp, 121, 213
\bibitem[Clark et al.(2009)]{2009A&A...504..429C} Clark, J.~S., Davies, B., Najarro, F., et al.\ 2009, \aap, 504, 429
\bibitem[Conti \& Crowther(2004)]{2004MNRAS.355..899C} Conti, P.~S., \& Crowther, P.~A.\ 2004, \mnras, 355, 899 \bibitem[Cyganowski et al.(2008)]{2008AJ....136.2391C} Cyganowski, C. J., Whitney, B. A., Holden, E., et al.\ 2008, \aj, 136, 2391
\bibitem[De Buizer et al.(2005)]{2005ApJS..156..179D} De Buizer, J.~M., Radomski, J.~T., Telesco, C.~M., \& Pi{\~n}a, R.~K.\ 2005, \apjs, 156, 179 
\bibitem[De Buizer(2006)]{2006ApJ...642L..57D} De Buizer, J.~M.\ 2006, \apjl, 642, L57
\bibitem[De Buizer et al.(2017)]{2017ApJ...843...33D} De Buizer, J.~M., Liu, M., Tan, J.~C., et al.\ 2017, \apj, 843, 33
\bibitem[Draine(2011)]{2011ApJ...732..100D} Draine, B.~T.\ 2011, \apj, 732,100
\bibitem[Draine \& Li(2007)]{2007ApJ...657..810D} Draine, B.~T., \& Li, A.\ 2007, \apj, 657, 810
\bibitem[Elia et al.(2017)]{2017MNRAS.471..100E} Elia, D., Molinari, S., Schisano, E. et al.\ 2017, \mnras, 471, 100
\bibitem[Elmegreen(1992)]{elmegreen1992} Elmegreen, B. G. 1992, in Star Formation in Stellar Systems, ed. G. Tenorio-Tagle, M. Prieto, \& F. S\'{a}nchez (Cambridge : Cambridge Univ. Press), 383
\bibitem[Felli et al.(1993)]{1993A&AS..101..127F} Felli, M., Taylor, G. B., Catarzi, M., et al.\ 1993, \aaps, 101, 127
\bibitem[Figuer{\^e}do et al.(2008)]{2008AJ....136..221F} Figuer{\^e}do, E., Blum, R.~D., Damineli, A., Conti, P.~S., \& Barbosa, C.~L.\ 2008, \aj, 136, 221
\bibitem[Galliano et al.(2008)]{2008ApJ...672..214G} Galliano, F., Dwek, E., \& Chanial, P.\ 2008, \apj, 672, 214 
\bibitem[Gaume et al.(1993)]{1993ApJ...417..645G} Gaume, R.~A., Johnston, K.~J., \& Wilson, T.~L.\ 1993, \apj, 417, 645 
\bibitem[Ginsburg et al.(2015)]{2015A&A...573A.106G} Ginsburg, A., Bally, J., Battersby, C., et al.\ 2015, \aap, 573, 106 
\bibitem[Ginsburg et al.(2016)]{2016AnAp...595A..27G} Ginsburg, A., Goss, W.~M., Goddi, C., et al.\ 2016, \aap, 595, A27 
\bibitem[Ginsburg et al.(2017)]{2017ApJ...842...92G} Ginsburg, A., Goddi, C., Kruijssen, J.~M.~D., et al.\ 2017, \apj, 842, 92
\bibitem[Goldader \& Wynn-Williams(1994)]{1994ApJ...433..164G} Goldader, J.~D., \& Wynn-Williams, C.~G.\ 1994, \apj, 433, 164 
\bibitem[Gordon et al.(2008)]{2008ApJ...682..336G} Gordon, K.~D., Engelbracht, C.~W., Rieke, G.~H., et al.\ 2008, \apj, 682, 336 
\bibitem[Greene et al.(1994)]{1994ApJ...434..614G} Greene, T.~P., Wilking, B.~A., Andre, P., Young, E.~T., \& Lada, C.~J.\ 1994, \apj, 434, 614
\bibitem[Gutermuth et al.(2009)]{2009ApJS..184...18G} Gutermuth, R. A., Megeath, S. T., Myers, P. C. et al., 2009, ApJS, 184, 18 
\bibitem[Harvey et al.(1986)]{1986ApJ...300..737H} Harvey, P.~M., Joy, M., Lester, D.~F., \& Wilking, B.~A.\ 1986, \apj, 300, 737
\bibitem[Helou et al.(2001)]{2001ApJ...548L..73H} Helou, G., Malhotra, S., Hollenbach, D.~J., Dale, D.~A., \& Contursi, A.\ 2001, \apjl, 548, L73 
\bibitem[Herter et al.(2013)]{2013PASP...125...1393H} Herter, T. L., Vacca, W. D., Adams, J. D., et al.\ 2013, \pasp,
125, 1393
\bibitem[Hill et al.(2005)]{2005MNRAS.363..405H} Hill, T., Burton, M.~G., Minier, V., et al.\ 2005, \mnras, 363, 405
\bibitem[Hill et al.(2006)]{2006MNRAS.368.1223H} Hill, T., Thompson, M.~A., Burton, M.~G., et al.\ 2006, \mnras, 368, 1223
\bibitem[Ho et al.(1983)]{1983ApJ...266..596H} Ho, P.~T.~P., Genzel, R., \& Das, A.\ 1983, \apj, 266, 596
\bibitem[Hoare et al.(2007)]{hoare07} Hoare, M. G., Kurtz, S. E., Lizano, S., et al.\ 2007, Protostars
and Planets V (Tucson, AZ: University Arizona Press), 181
\bibitem[Hosokawa et al.(2010)]{2010ApJ...721..478H} Hosokawa, T., Yorke, H.~W., \& Omukai, K.\ 2010, \apj, 721, 478
\bibitem[Kang et al.(2009)]{2009ApJ...706...83K} Kang, M., Bieging, J. H., Povich, M. S., et al.\ 2009, \apj, 706, 83
\bibitem[Kang et al.(2010)]{2010ApJS..190...58K} Kang, M., Bieging, J. H., Kulesa, C.~A., et al.\ 2010, \apjs, 190, 58
\bibitem[Kauffmann et al.(2013)]{2013ApJ...779..185K} Kauffmann, J., Pillai, T. \& Goldsmith, P.~F.\ 2013, \apj, 779, 185
\bibitem[Koo(1997)]{1997ApJS..108..489K} Koo, B.-C.\ 1997, \apjs, 108, 489 
\bibitem[Kraemer et al.(2001)]{2001ApJ...561..282K} Kraemer, K.~E., Jackson, J.~M., Deutsch, L.~K., et al.\ 2001, \apj, 561, 282 
\bibitem[Krumholz \& Tan(2007)]{2007ApJ...654..304K} Krumholz, M.~R., \& Tan, J.~C.\ 2007, \apj, 654, 304
\bibitem[Kundu \& Velusamy(1967)]{1967AnAp...30...59K} Kundu, M.~R., \& Velusamy, T.\ 1967, Annales d'Astrophysique, 30, 59
\bibitem[Lim \& Tan(2014)]{2014ApJ...780L..29L} Lim, W. \& Tan, J. C. 2014, \apj, 780,29
\bibitem[Lim et al.(2016)]{2016ApJ...829L..19L} Lim, W., Tan, J.~C., Kainulainen, J., Ma, B., \& Butler, M.~J.\ 2016, \apjl, 829, L19 
\bibitem[Liu et al.(2010)]{2010RAA....10...67L} Liu, T., Wu, Y.-F. \& Wang, K.\ 2010, Research in Astronomy and Astrophysics, 10, 67
\bibitem[Ma et al.(2013)]{2013ApJ...779...79M} Ma, B., Tan, J.~C., \& Barnes, P.~J.\ 2013, \apj, 779, 79
\bibitem[MacLaren et al.(1988)]{1988ApJ...333..821M} MacLaren, I., Richardson, K. M. \& Wolfendale, A. W.\ 1988, \apj, 333, 821	
\bibitem[Martin(1972)]{1972MNRAS.157...31M} Martin, A.~H.~M.\ 1972, \mnras, 157, 31
\bibitem[Mason et al.(2009)]{2009AJ....137.3358M} Mason, B.~D., Hartkopf, W.~I., Gies, D.~R., et al.\ 2009, \aj, 137, 3358.
\bibitem[McKee \& Tan(2003)]{2003ApJ...585..850M} McKee, C.~F., \& Tan, J.~C.\ 2003, \apj, 585, 850
\bibitem[Mehringer(1994)]{1994ApJS...91..713M} Mehringer, D.~M.\ 1994, \apjs, 91, 713 
\bibitem[Meynet \& Maeder(2000)]{2000A&A...361..101M} Meynet, G. \& Maeder, A.\ 2000, \aap, 361, 101
\bibitem[Miller \& Scalo(1978)]{1978PASP...90..506M} Miller, G.~E., \& Scalo, J.~M.\ 1978, \pasp, 90, 506  
\bibitem[Molinari et al.(2010)]{2010A&A...518L.100M} Molinari, S., Swinyard, B., Bally, J., et al.\ 2010, \aap, 518, L100 
\bibitem[Molinari et al.(2016)]{2016A&A...591A.149M} Molinari, S., Schisano, E., Elia, D. et al.\ 2016, \aap, 591, 149
\bibitem[Morris et al.(1996)]{1996ApJ...470..597M} Morris, P.~W., Eenens, P.~R.~J., Hanson, M.~M., Conti, P.~S., \& Blum, R.~D.\ 1996, \apj, 470, 597
\bibitem[Nanda Kumar et al.(2004)]{2004MNRAS.353.1025K} Nanda Kumar, M. S., Kamath, U. S., \& Davis, C. J.\ 2004, \mnras, 353, 1025
\bibitem[Naylor(1998)]{1998MNRAS.296..339N} Naylor, T.\ 1998, \mnras, 296, 339
\bibitem[Okamoto et al.(2001)]{2001ApJ...553..254O} Okamoto, Y.~K., Kataza, H., Yamashita, T., Miyata, T., \& Onaka, T.\ 2001, \apj, 553, 254 
\bibitem[Okumura et al.(2000)]{2000ApJ...543..799O} Okumura, S.-i., Mori, A., Nishihara, E., Watanabe, E., \& Yamashita, T.\ 2000, \apj, 543, 799
\bibitem[Ossenkopf \& Henning(1994)]{1994A&A...291..943O} Ossenkopf, V., \& Henning, T.\ 1994, \aap, 291, 943
\bibitem[Penzias et al.(1971)]{1971ApJ...165..229P} Penzias, A.~A., Jefferts, K.~B., \& Wilson, R.~W.\ 1971, \apj, 165, 229
\bibitem[Robitaille et al.(2006)]{2006ApJS..167..256R} Robitaille, T.~P., Whitney, B.~A., Indebetouw, R., Wood, K., \& Denzmore, P.\ 2006, \apjs, 167, 256 
\bibitem[Robitaille et al.(2007)]{2007ApJS..169..328R} Robitaille, T.~P., Whitney, B.~A., Indebetouw, R., \& Wood, K.\ 2007, \apjs, 169, 328 
\bibitem[Saral et al.(2017)]{2017ApJ...839..108S} Saral, G., Hora, J.~L., Audard, M., et al.\ 2017, \apj, 839, 108 
\bibitem[Sato et al.(2010)]{2010ApJ...720.1055S} Sato, M., Reid, M.~J., Brunthaler, A., \& Menten, K.~M.\ 2010, \apj, 720, 1055 
\bibitem[Schaller et al.(1992)]{1992A&AS...96..269S} Schaller, G., Schaerer, D., Meynet, G., et al.\ 1992, \aaps, 96, 269
\bibitem[Schneps et al.(1981)]{1981ApJ...249..124S} Schneps, M.~H., Lane, A.~P., Downes, D., et al.\ 1981, \apj, 249, 124 
\bibitem[Scott(1978)]{1978MNRAS.183..435S} Scott, P.~F.\ 1978, \mnras, 183, 435 
\bibitem[Shi et al.(2010)]{2010ApJ...718L.181S} Shi, H., Zhao, J.-H., \& Han, J.~L.\ 2010, \apjl, 718, L181 
\bibitem[Shuping et al.(2015)]{2015ASPC..495..351S} Shuping, R.~Y., Krzaczek, R., Vacca, W.~D., et al.\ 2015, Astronomical Data Analysis Software an Systems XXIV (ADASS XXIV), 495, 351 
\bibitem[Skrutskie et al.(2006)]{2006AJ....131.1163S} Skrutskie, M.~F., Cutri, R.~M., Stiening, R., et al.\ 2006, \aj, 131, 1163
\bibitem[Stutz(2018)]{2018MNRAS.473.4890S} Stutz, A.~M.\ 2018, \mnras, 473, 4890
\bibitem[Szymczak et al.(2000)]{2000A&AS..143..269S} Szymczak, M., Hrynek, G., \& Kus, A.~J.\ 2000, \aaps, 143, 269 
\bibitem[Reach et al.(2006)]{2006AJ....131.1479R} Reach, W. T., Rho, J., Tappe, H., et al.\ 2006, \apj, 131, 147
\bibitem[Tan et al.(2013)]{2013ApJ...779...96T} Tan, J. C., Kong, S., Butler, M. J. et al.\ 2013, \apj, 779,96
\bibitem[Vacca(1994)]{1994ApJ...421..140V} Vacca, W.~D.\ 1994, \apj, 421, 140
\bibitem[Vacca \& Sandell(2011)]{2011ApJ...732....8V} Vacca, W.~D. \& Sandell, G.\ 2011, \apj, 732, 8
\bibitem[Wachter et al.(2010)]{2010AJ....139.2330W} Wachter, S., Mauerhan, J.~C., Van Dyk, S.~D., et al.\ 2010, \aj, 139, 2330
\bibitem[Westerhout(1958)]{1958BAN....14..215W} Westerhout, G.\ 1958, \bain, 14, 215
\bibitem[Wilson et al.(1970)]{1970ApL.....5...99W} Wilson, T.~L., Mezger, P.~G., Gardner, F.~F., \& Milne, D.~K.\ 1970, \aplett, 5, 99
\bibitem[Wood \& Churchwell(1989)]{1989ApJS...69..831W} Wood, D.~O.~S., \& Churchwell, E.\ 1989, \apjs, 69, 831 
\bibitem[Wu et al.(2017)]{2017ApJ...835..137W} Wu, B., Tan, J. C., Nakamura, F., et al.\ 2017, \apj, 835, 137
\bibitem[Wynn-Williams et al.(1974)]{1974ApJ...187..473W} Wynn-Williams, C.~G., Becklin, E.~E., \& Neugebauer, G.\ 1974, \apj, 187, 473 
\bibitem[Zhang \& Ho(1997)]{1997ApJ...488..241Z} Zhang, Q., \& Ho, P.~T.~P.\ 1997, \apj, 488, 241 
\bibitem[Zhang \& Tan(2011)]{2011ApJ...733...55Z} Zhang, Y., \& Tan, J.~C.\ 2011, \apj, 733, 55
\bibitem[Zhang et al.(2013)]{2013ApJ...766...86Z} Zhang, Y., Tan, J.~C., \& McKee, C.~F.\ 2013, \apj, 766, 86
\bibitem[Zhang et al.(2014)]{2014ApJ...788..166Z} Zhang, Y., Tan, J.~C., \& Hosokawa, T.\ 2014, \apj, 788, 166 
\bibitem[Zhang \& Tan(2018)]{2018ApJ...853...18Z} Zhang, Y., \& Tan, J.~C.\ 2018, \apj, 853, 18

\end{thebibliography}
\end{document}